\documentclass[aps,prl,reprint,superscriptaddress,a4paper,amsmath,amsfonts]{revtex4-2}

\usepackage[colorlinks,pdfusetitle]{hyperref}
\usepackage[english]{babel}
\usepackage[utf8]{inputenc}
\usepackage{orcidlink}
\usepackage{mathtools}
\usepackage{physics}
\usepackage[group-minimum-digits=0,table-figures-decimal=0,table-number-alignment=center]{siunitx}

\usepackage{tikz}
\usetikzlibrary{quantikz}

\newcommand{\latin}[1]{{\emph{#1}}}
\def\equationautorefname~#1\null{Eq.\,(#1)\null}
\def\figureautorefname~#1\null{Fig.\,#1\null}
\def\sectionautorefname~#1\null{App.\,#1\null}

\usepackage{dsfont}
\newcommand{\R}{\mathds{R}}

\newcommand{\Z}{\mathds{Z}}

\usepackage{fontawesome}

\newlength{\arrow}
\newcommand*{\xxmapsto}[1]{\xmapsto{\mathmakebox[\arrow]{#1}}}

\newcommand{\bigO}[1]{\ensuremath{\mathop{}\mathopen{}O\mathopen{}\left(#1\right)}}  \newcommand{\ceil}[1]{{\ensuremath{\left\lceil#1\right\rceil}}}
\newcommand{\floor}[1]{{\ensuremath{\left\lfloor#1\right\rfloor}}}

\newcommand{\Cp}{\ket{\mathcal{C}_\alpha^+}}

\newcommand{\Cm}{\ket{\mathcal{C}_\alpha^-}}

\newcommand{\Cpm}{\ket{\mathcal{C}_\alpha^\pm}}
\newcommand{\ha}{\hat{a}}
\newcommand{\had}{\hat{a}^\dagger}
\newcommand{\hb}{\hat{b}}
\newcommand{\hbd}{\hat{b}^\dagger}
\newcommand{\hc}{\hat{c}}
\newcommand{\hcd}{\hat{c}^\dagger}

\makeatletter
\DeclareExpandableDocumentCommand{\cwap}{O{}O{1.5pt}O{1.5pt}m}{|[inner sep=4pt,minimum width=#2,minimum height=#3]|\edef\n{\the\pgfmatrixcurrentrow} \edef\m{\the\pgfmatrixcurrentcolumn} \edef\options{row=\n,col=\m,#1}
	\def\DisableMinSize{0}\pgfkeys{/quantikz,wires=2,style=,label style=,braces=,cwires={1,2}}\pgfkeys{/quantikz,#1}\pgfkeysgetvalue{/quantikz/wires}{\quantwires}
	\pgfkeysgetvalue{/quantikz/style}{\a}
	\pgfkeysgetvalue{/quantikz/label style}{\b}
	\pgfkeysgetvalue{/quantikz/cwires}{\mylist}
	\pgfkeysgetvalue{/quantikz/nwires}{\nowires}
	\pgfkeysgetvalue{/quantikz/bundle}{\bundle}
\def\quantwires{2}
	\phantom{wide}
	\settowidth{\myl}{$wide$}
	\settoheight{\myh}{$wide$}
	\settodepth{\myd}{$wide$}
\IfInList{1}{\mylist}{\cw}{\IfInList{1}{\nowires}{}{\IfInList{1}{\bundle}{\qwbundle[alternate]{}}{\qw}}}\edef\k{\the\numexpr\n+\quantwires-1\relax}
	\edef\mn{\the\numexpr\m-1\relax}
	\ifthenelse{\quantwires=1}{}{\foreach \i in {\the\numexpr\n+1\relax,...,\k} {
			\edef\newcom{\noexpand\vcwhexplicit{\i-\m}{\i-\mn}}
				\edef\newcomb{\noexpand\vqwexplicit{\i-\m}{\i-\mn}}
					\edef\newcomc{\noexpand\vqbundleexplicit{\i-\m}{\i-\mn}}
			\edef\val{\the\numexpr\i+1-\n\relax}
			\IfInList{\val}{\mylist}{\newcom}{\IfInList{\val}{\nowires}{}{\IfInList{\val}{\bundle}{\newcomc}{\newcomb}}}\globaldefs=1
			\edef\dotikzset{\noexpand\tikzset{row \i\space column \m/.append style={minimum width={max(\the\myl+8pt,#2)}}}}\dotikzset \edef\undotikzset{\noexpand\tikzset{row \i\space column \m/.style={}}}\expandafter\pgfutil@g@addto@macro\expandafter\tikzcd@atendglobals\expandafter{\undotikzset}}
\globaldefs=1\edef\dotikzset{\noexpand\tikzset{row \k\space column \m/.append style={minimum height={max(\the\myh+\the\myd+8pt,#3)}}}}\dotikzset \globaldefs=0}
\expandafter\expandafter\expandafter\expandafter\expandafter\expandafter\expandafter\pgfutil@g@addto@macro\expandafter\expandafter\expandafter\expandafter\expandafter\expandafter\expandafter\tikzcd@atendsavedpaths\expandafter\expandafter\expandafter\expandafter\expandafter\expandafter\expandafter{\expandafter\expandafter\expandafter\expandafter\expandafter\expandafter\expandafter\cswap@end\expandafter\expandafter\expandafter\expandafter\expandafter\expandafter\expandafter{\expandafter\expandafter\expandafter\a\expandafter\expandafter\expandafter}\expandafter\expandafter\expandafter{\expandafter\b\expandafter}\expandafter{\options}{#4}
	}
}

\newcommand{\cswap@end}[4]{
	\def\DisableMinSize{0}
	\pgfkeys{/quantikz,#3}\pgfkeysgetvalue{/quantikz/row}{\row}
	\pgfkeysgetvalue{/quantikz/col}{\col}
	\def\quantwires{2}
	\xdef\LoopGG{}
\foreach \n in  {\row,...,\the\numexpr\row+\quantwires-1\relax} {
	\ifnodedefined{\tikzcdmatrixname-\n-\col}{
		\xdef\LoopGG{\LoopGG(\tikzcdmatrixname-\n-\col)}
		}{}
	}
\node (group\tikzcdmatrixname-\row-\col) [fit=\LoopGG,operator,inner sep=0pt,#1] {\hphantom{Wide}};
	\draw [thickness,transform canvas={yshift=0.05cm}]  (group\tikzcdmatrixname-\row-\col.west|-\tikzcdmatrixname-\row-\col.center) to[out=0,in=180] (group\tikzcdmatrixname-\row-\col.east|-\tikzcdmatrixname-\the\numexpr\row+1\relax-\col.center);
	\draw [thickness,transform canvas={yshift=-0.05cm}] (group\tikzcdmatrixname-\row-\col.west|-\tikzcdmatrixname-\row-\col.center) to[out=0,in=180] (group\tikzcdmatrixname-\row-\col.east|-\tikzcdmatrixname-\the\numexpr\row+1\relax-\col.center);
	\draw [line width=3pt,white,shorten >=0.9pt,shorten <=0.9pt] (group\tikzcdmatrixname-\row-\col.east|-\tikzcdmatrixname-\row-\col.center) to[out=180,in=0] (group\tikzcdmatrixname-\row-\col.west|-\tikzcdmatrixname-\the\numexpr\row+1\relax-\col.center);
	\draw [thickness,transform canvas={yshift=0.05cm}]  (group\tikzcdmatrixname-\row-\col.east|-\tikzcdmatrixname-\row-\col.center) to[out=180,in=0] (group\tikzcdmatrixname-\row-\col.west|-\tikzcdmatrixname-\the\numexpr\row+1\relax-\col.center);
	\draw [thickness,transform canvas={yshift=-0.05cm}] (group\tikzcdmatrixname-\row-\col.east|-\tikzcdmatrixname-\row-\col.center) to[out=180,in=0] (group\tikzcdmatrixname-\row-\col.west|-\tikzcdmatrixname-\the\numexpr\row+1\relax-\col.center);
}
\makeatother

\DeclareExpandableDocumentCommand{\ctargX}{O{}m}{|[ccrossx2,#1]| {} \cw}
\tikzset{ccrossx/.style={path picture={\draw[internal,inner sep=0pt] (path picture bounding box.south east) -- (path picture bounding box.north west) (path picture bounding box.south west) -- (path picture bounding box.north east);
  }},
	ccrossx2/.style={circle,ccrossx,minimum size=1em},
}

\usepackage{algorithm}  
\usepackage{listings}
\begin{document}

\preprint{\href{https://arxiv.org/abs/2302.06639}{arXiv:2302.06639}}
\doi{10.1103/PhysRevLett.131.040602}

\title{Performance Analysis of a Repetition Cat Code Architecture: Computing 256-bit Elliptic Curve Logarithm in 9 Hours with \num[detect-all]{126133} Cat Qubits}

\author{Élie Gouzien\,\orcidlink{0000-0002-8209-0681}}
\email[]{elie.gouzien@cea.fr}
\affiliation{Université Paris--Saclay, CNRS, CEA, Institut de physique théorique, \num[detect-all]{91191} Gif-sur-Yvette, France}
\author{Diego Ruiz\,\orcidlink{0000-0003-2003-7030}}
\affiliation{Alice\&Bob, 53 boulevard du Général Martial Valin, \num[detect-all]{75015} Paris, France}
\affiliation{Laboratoire de Physique de l'École normale supérieure, École normale supérieure, Mines Paris, Université PSL, Sorbonne Université, CNRS, Inria, \num[detect-all]{75005} Paris, France}
\author{Francois-Marie Le Régent\,\orcidlink{0000-0002-5229-7155}}
\affiliation{Alice\&Bob, 53 boulevard du Général Martial Valin, \num[detect-all]{75015} Paris, France}
\affiliation{Laboratoire de Physique de l'École normale supérieure, École normale supérieure, Mines Paris, Université PSL, Sorbonne Université, CNRS, Inria, \num[detect-all]{75005} Paris, France}
\author{Jérémie Guillaud\,\orcidlink{0000-0001-6507-9344}}
\affiliation{Alice\&Bob, 53 boulevard du Général Martial Valin, \num[detect-all]{75015} Paris, France}
\author{Nicolas Sangouard\,\orcidlink{0000-0002-3136-0266}}
\homepage[]{https://quantum.paris}
\affiliation{Université Paris--Saclay, CNRS, CEA, Institut de physique théorique, \num[detect-all]{91191} Gif-sur-Yvette, France}

\makeatletter
\hypersetup{pdfauthor={Élie Gouzien, Diego Ruiz, Francois-Marie Le Régent, Jérémie Guillaud and Nicolas Sangouard}}
\makeatother

\date{\today}

\begin{abstract}
Cat qubits provide appealing building blocks for quantum computing.
They exhibit a tunable noise bias yielding an exponential suppression of bit flips with the average photon number and a protection against the remaining phase errors can be ensured by a simple repetition code.
We here quantify the cost of a repetition code and provide valuable guidance for the choice of a large scale architecture using cat qubits by realizing a performance analysis based on the computation of discrete logarithms on an elliptic curve with Shor's algorithm.
By focusing on a 2D grid of cat qubits with neighboring connectivity, we propose to implement 2-qubit gates via lattice surgery and Toffoli gates with off-line fault-tolerant preparation of magic states through projective measurements and subsequent gate teleportations.
All-to-all connectivity between logical qubits is ensured by routing qubits.
Assuming a ratio between single- and two-photon losses of $10^{-5}$ and a cycle time of $\SI{500}{\nano\second}$, we show concretely that such an architecture can compute a $256$-bit elliptic curve logarithm in $\SI{9}{\hour}$ with \num{126133}~cat qubits and on average $19$ photons by cat state.
We give the details of the realization of Shor's algorithm so that the proposed performance analysis can be easily reused to guide the choice of architecture for others platforms.
\end{abstract}

\maketitle

\addtolength{\parskip}{\bigskipamount}

\paragraph{Introduction ---}

While quantum computing can offer substantial speedups for solving specific problems~\cite{MontanaronQI2016Quantumalgorithmsoverview}, billions of operations are typically required for implementing large scale algorithms~\cite{TroyerPotNAoS2017Elucidatingreactionmechanisms,Ekeraa2019Howfactor2048,Babbush2020CompilationFaultTolerant}.
This means that the convergence of quantum algorithms cannot realistically be ensured by requiring physical errors to occur with a probability smaller than the inverse of the number of required operations.
Instead, the concept of fault-tolerant quantum computation~\cite{VuillotN2017Roadstowardsfault} is envisioned.
It relies on the idea that, if the rate of physical errors is below a certain threshold, quantum error correction schemes suppress the logical error rate to arbitrary low levels and make possible --- at least in principle --- arbitrary long sequences of operations~\cite{Ben-OrSJoC2008FaultTolerantQuantum,KitaevRMS1997Quantumcomputationsalgorithms,ZurekPotRSoLSAMPaES1998Resilientquantumcomputation}.

With its relatively high thresholds, the surface code is one of the most popular quantum error correction codes~\cite{KitaevPRA2005Universalquantumcomputation,ClelandPRA2012SurfacecodesTowards}.
As a 2D code, the number of physical qubits per logical qubit increases quadratically with the code distance.
Their actual implementation hence comes at the price of a significant overhead in physical resources, with typically hundreds or even thousands of physical qubits per logical qubits to achieve the level of protection required for performing billions of noise-free operations~\cite{VuillotN2017Roadstowardsfault}.

Some physical platforms naturally exhibit a noise bias~\cite{HansonMB2013DiamondNVcenters} that can be exploited to increase code thresholds and hence to reduce the overhead~\cite{PreskillPRA2008Faulttolerantquantum,FlammiaPRL2018UltrahighErrorThreshold}.
Bosonic systems stabilized in a two-dimensional manifold spanned by cat states --- superpositions of coherent states with opposite phases --- with an engineered dissipation scheme combining two-photon drive and two-photon dissipation stand out in this framework.
The noise bias is indeed tunable in this case, with
bit flips that are suppressed exponentially with the mean photon number~\cite{DevoretNJoP2014Dynamicallyprotectedcat,JiangPRL2016HolonomicQuantumControl,BlaisnQI2017Engineeringquantumstates}.
The remaining phase errors can be corrected with a simple repetition code --- a 1D code with a number of physical qubits per logical qubit increasing linearly with the code distance.
Given that gate sets at the physical level preserving the noise asymmetry have been described and their use for the implementation of various universal sets at the logical level has been identified~\cite{MirrahimiPRX2019RepetitionCatQubits}, cat qubits are becoming an option for realizing a large scale quantum computer.
The gain of having a 1D over a 2D code and the details of the implementation of a large scale algorithm with cat qubits are, however, missing.

We here propose a generic tool for analyzing the performance of quantum computing architectures using Shor's algorithm~\cite{Shor1994Algorithmsquantumcomputation,ShorSJoC1997PolynomialTimeAlgorithms} for computing discrete logarithms on elliptic curves over prime fields --- a hard classical problem at the core of cryptosystems widely used for key exchange and digital signatures~\cite{Barker2020Guidelineusingcryptographic,nullITL2013DigitalSignatureStandard}.
The security level of these cryptosystems against classical attacks relies on precise knowledge of the performance of classical algorithms --- knowledge that is useful to witness a quantum advantage.
The best currently known classical algorithms to compute elliptic curve discrete logarithms are exponential in the size of the input parameters, whereas there exist subexponential algorithms for factoring.
This facilitates the achievements of a quantum advantage with respect to algorithms with subexponential speedups, including Shor's algorithm for the factorization.

We propose a concrete layout in which cat qubits are placed at the nodes of a 2D grid with physical connections to their neighboring qubits only.
All-to-all connectivity between logical qubits is ensured by means of routing qubits.
2-qubit gates are implemented by means of lattice surgery and Toffoli gates are obtained by gate teleportation through an off-line fault-tolerant magic state preparation based on projective measurements.
From detailed models of physical qubits and their manipulations, we estimate precisely the errors related to the implementation of logical operations by considering a ratio between single- and two-photon losses of $10^{-5}$.
We show concretely that such an architecture can compute a discrete logarithm on the \verb|secp256k1| curve which is used for securing signatures in Bitcoin transactions~\cite{TomamichelL2018QuantumAttacksBitcoin} in $\SI{9}{\hour}$ with \num{126133}~cat qubits and with $19$ photons per cat state on average.
Note that keeping the exponential reduction of bit flips for cat sizes up to $19$ photons on average might be experimentally challenging~\cite{LeghtasNP2020Exponentialsuppressionbit,Leghtas2022Onehundredsecond}, and our results suggest that either substantial improvements in the design of cat qubits are required, or a thin rectangular surface code~\cite{BrandaoPQ2022BuildingFaultTolerant} is more suitable for such a large computation; see \autoref{appendix:catqubits} and \autoref{appendix:tableau} for more details.
Independent of the feasibility question, the gain in using a 1D instead of a 2D code is quantified in detail.

\paragraph{Elliptic curve and discrete logarithm ---}
An elliptic curve is defined as the set of points associated with the coordinates $(x, y)$ satisfying the equation $y^2 = x^3 + ax + b$ with fixed values for $a$ and $b$.
We are interested in cryptographic relevant elliptic curves for which $x$, $y$, $a$, and $b$ belong to the field of integers modulo $p$, with $p$ a prime number ($n$ bits long).
We define a binary operation on these elliptic curves, called ``addition'' and denoted ``$+$'': for two points $P$ and $Q$, $R=P + Q$ have, in the generic case~\footnote{The following description of point addition, as well as \autoref{eq:point_addition}, are valid when $P$ and $Q$ are distinct and different from the neutral element.
	If $P=Q$, the tangent of the curve on this point is used.
	If $Q$ is the neutral element, the line to consider is the one parallel to the $y$ axis going through $P$, and the result is $P+Q = P$.
	If $Q = -P$, that is, $Q$ is symmetrical of $P$ with respect to the $x$ axis, the line joining them is vertical and the result is the neutral element: $P+Q=0$.
},
coordinates given by
\begin{subequations}\label{eq:point_addition}
\begin{align}
	x_{R} &= \lambda^2 - x_P - x_Q, \\
	y_{R} &= -y_P - \lambda (x_{R} - x_P),
\end{align}
\end{subequations}
where $\lambda = (y_Q - y_P)/(x_Q - x_P)$ is the slope of the line joining $P$ and $Q$.
A multiplication by an integer $k$ naturally arises as $k P = \underbrace{P + P + \cdots + P}_{\text{$k$ times}}$.
A cyclic subgroup can be formed from the multiples of a point $G$ (the subgroup generator) on the curve.
The security of cryptographic algorithms based on elliptic curves, such as the elliptic curve digital signature algorithm~\cite{nullITL2013DigitalSignatureStandard},
relies on the hardness to find a number $k$ (the logarithm) from the knowledge of the generator $G$ and the point $P = kG$; see \autoref{appendix:elliptic_curve} for details on elliptic curve cryptography.

\paragraph{Shor's algorithm ---}
Shor introduced an algorithm~\cite{Shor1994Algorithmsquantumcomputation,ShorPRL1996FaultTolerantError} to compute discrete logarithms on a quantum computer with a number of gates cubic in $n$.
It takes three steps and three registers.
In the first step, two registers encoding $x_1$ and $x_2$ are each prepared in a superposition of all possible integers.
In the second step, $f(x_1, x_2) = x_1 G - x_2 P$ is computed and stored in the third register.
In the last step, a quantum Fourier transform of the two registers containing $x_1$ and $x_2$ (which corresponds to a 2D quantum Fourier transform) reveals the value of $k$; see \autoref{appendix:shor_ekera} for more details and a discussion on Ekerå's version of Shor's algorithm~\cite{Ekeraa2018Quantumalgorithmscomputing}.

The preparation of registers in a superposition of all integers has a linear cost.
It is indeed obtained through the preparation of qubits in state $\ket{+} = (\ket{0} + \ket{1})/\sqrt{2}$, that is by applying a Hadamard transformation on each qubit.
Since the quantum Fourier transform precedes a measurement of each qubit, it can be implemented in a semiclassical way~\cite{NiuPRL1996SemiclassicalFourierTransform}, with a linear cost as well.
The cost of Shor's algorithm is largely dominated by the computation of $f$, which we evaluate in detail below.

\paragraph{Arithmetic circuits ---}
$f$ is the difference between the results of two scalar multiplications.
The elliptic curve scalar multiplication is implemented with a windowed approach, as in~\cite{Soeken2020ImprovedQuantumCircuits}.
The principle is to decompose the factor $k$ into groups of bits and to rewrite the multiplication as a sequence of elliptic curve point additions,
\begin{equation}
k G = \sum\limits_{\substack{i=0 \\ i \equiv 0 \mod{w_e}}}^{n_e} 2^i k_{i:i+w_e} G,
\end{equation}
with $n_e$ the number of bits in $k$, $w_e$ the width of each window, and $k_{i:i+w_e}$ the number formed from $w_e$ bits of $k$ starting at bit $i$.
For each term, the point $2^i k_{i:i+w_e} G$ is computed classically for all possible values of $k_{i:i+w_e}$ and loaded into a quantum register through a quantum table lookup circuit~\cite{NevenPRX2018EncodingElectronicSpectra,Gidney2019Windowedquantumarithmetic}, with the qubits encoding $k_{i:i+w_e}$ as controls.

Each point addition is realized from \autoref{eq:point_addition} using a quantum implementation of each operation~\footnote{\autoref{eq:point_addition} only describes the elliptic curve addition in the generic case when $P$ and $Q$ are distinct and different from the neutral element.
	Only the generic case is implemented, as this saves resources at the cost of an approximate algorithm; the exact consequences are discussed in~\cite{Lauter2017QuantumResourceEstimates}.
}.
This takes arithmetic additions, subtractions, multiplications, and divisions, modulo the prime number $p$; see \autoref{appendix:arithmetic:elliptic_curve_mult} and \autoref{appendix:arithmetic:elliptic_curve_add} for details.

A ripple-carry circuit from~\cite{Moulton2004newquantumripple} is used to perform the additions.
The basic idea is to start with the low-order bits of the inputs, compute the first carry with a Toffoli gate, take the value of the carry and the next bits of the inputs to compute the second carry, and so on up to the high-order bits.
We then work from the high-order bits back down to the low-order bits by computing the result of the addition bit by bit, store the value in the first input register, and restore the value of the second input register to get a reversible computation.
Note that the subtraction is obtained by conjugation of the addition.
To make an addition modulo $p$, the standard addition is followed by a comparison of the result with $p$, which is obtained by subtracting $p$ from the result of the addition and by checking the most significant bit of the result of this subtraction.
A subtraction of $p$ is then realized, conditioned on the result of the comparison.
In order to save resources, the comparison and subtraction, which start identically, are merged together to form a modular reduction; see the \autoref{appendix:arithmetic:add_mod} for the details on the circuits.
Note that the modular subtraction is obtained by conjugation of the modular addition.

The multiplication can be implemented with a standard double-and-add method, which can be illustrated by considering the product of two ($n$ bits) numbers $x_1$ and $x_2$.
From the binary representation of $x_1=\sum_i 2^i {[x_1]}_i$, we have $x_1 x_2=\sum_i 2^i {[x_1]}_i x_2 = {[x_1]}_0 x_2 + 2\left\lbrace{[x_1]}_1 x_2 + 2({[x_1]}_2y + \cdots)\right\rbrace$; \latin{i.e.\@}, the result of the product is obtained by first considering the last term ($x_2$ conditioned on the value of ${[x_1]}_{n-1}$), doubling the result, and adding $x_2$ conditioned on ${[x_1]}_{n-2}$, and so on up to the first term.
The multiplication modulo $p$ is then naturally obtained by performing additions and doublings modulo $p$, which takes $2n$ modular reductions.
We used a representation (compatible with the addition), the Montgomery representation, to reduce the number of reductions~\cite{Lauter2017QuantumResourceEstimates,Soeken2020ImprovedQuantumCircuits}.
It simply consists of representing a number $x_1$ by $y_1 = x_1 2^{n} \mod{p}$.
Considering the numbers $x_1$ and $x_2$, and their respective Montgomery representation $y_1$, $y_2$, the product $x_1 x_2$ is represented by $x_1 x_2 2^{n} \mod{p}$, which is obtained by computing $y_1, y_2 \mapsto y_1 y_2 2^{-n} \mod{p}$ from a double-and-add multiplier in which the doubling operation is replaced by halving.
In this case, the sums need a single modular reduction to realize a multiplication modulo $p$, hence reducing the number of modular reductions to $n+1$~\footnote{We allow the accumulation register to store a number up to $2p$ after each add and half cycle.
	The division by $2$ ensure its value not to be amplified beyond this limit, after each cycle, even when using nonmodular additions (what would not be the case when doubling).
	See \autoref{appendix:arithmetic:montgomery_mult} for the details.  }.
Note that the latter is further reduced by using a windowed version of the multiplication in the Montgomery representation; see \autoref{appendix:arithmetic:montgomery_mult} for the details of the multiplication circuit.

The modular division between two numbers $x_1$ and $x_2$ is obtained by a modular multiplication of $x_1$ and the modular inverse of $x_2$.
The modular inversion is performed with Kaliski's algorithm~\cite{KaliskiIToC1995Montgomeryinverseits}.
This algorithm is essentially a binary version of the extended Euclidean algorithm.
To make it compatible with the Montgomery representation, the result is multiplied by $2^{2n}$ such that starting from the representation $y_2 = x_2 2^{n} \mod{p}$, Kaliski's algorithm returns $x_2^{-1} 2^{n} \mod{p} = y_2^{-1} 2^{2n} \mod{p}$.
The circuit we use is inspired by~\cite{Soeken2020ImprovedQuantumCircuits}, with improvements, most notably by using subcircuits crafted for use of Toffoli gates; see \autoref{appendix:arithmetic:inversion}.

Given the number of point additions in the scalar multiplication at the core of the discrete logarithm computation, the decomposition of a point addition in elementary arithmetic operations and the number of gates that is required for implementing each of these elementary operations, we deduce that the implementation of Shor's algorithm takes $448n^3/w_e$ controlled NOT (CNOT) and $348n^3/w_e$ Toffoli gates at the leading order, with $w_e$ the size of each window for the elliptic curve multiplication; see \autoref{appendix:arithmetic:gate_count}.

\paragraph{Cat qubits with repetition code ---}
We are interested in cat qubits, in which information is encoded in two coherent states of a harmonic oscillator with the same amplitude and opposite phase $\ket{\alpha}$ and $\ket{-\alpha}$~\cite{MunroPRA1999Macroscopicallydistinctquantum,DevoretNJoP2014Dynamicallyprotectedcat}.
Here $\alpha$ is assumed real, without loss of generality.
To avoid that the state of the oscillator leaves the computation subspace $\left(\ket{\alpha}, \ket{-\alpha}\right)$ in the presence of loss and noise, a stabilization mechanism is needed.
We consider a mechanism combining a two-photon drive and an engineered two-photon dissipation, which can be implemented appropriately in a physical realization using cavity modes coupled nonlinearly by Josephson junctions.
When the corresponding stabilization rate is higher than that of typical errors, the bit-flip error rate induced by single-photon loss, thermal excitations or dephasing are exponentially suppressed with the mean number of photons in the cat size $\gamma_X \propto \exp(-2\alpha^2)$, while the phase-flip error rate typically scales linearly $\gamma_Z \propto \alpha^2$~\cite{LeghtasNP2020Exponentialsuppressionbit,Leghtas2022Onehundredsecond}.
In this work, the amplitude $\alpha$ is a free parameter.
Its value is chosen such that bit flips happen with a low probability during the run-time of Shor's algorithm and a repetition code corrects phase-flip errors only~\cite{MirrahimiPRX2019RepetitionCatQubits,BrandaoPQ2022BuildingFaultTolerant}.
Details on cat qubits and their implementation are given in \autoref{appendix:cat}.

As bit flips are not corrected, it is crucial not to introduce such errors during the algorithm execution.
At the physical level, this is obtained by using bias-preserving operations including the preparations $\mathcal{P}_{\ket{0/1}}$ and $\mathcal{P}_{\ket{\pm}}$ of the computational states $\ket{\pm\alpha}$ and cat states $\Cpm = \frac{1}{\sqrt{2(1 \pm e^{-2\alpha^2})}}(\ket{\alpha} \pm \ket{-\alpha})$, respectively, the measurements $\mathcal{M}_Z$ and $\mathcal{M}_X$, the $Z$ and $X$ gates, and CNOT and Tofolli gates; see the details in \autoref{appendix:cat}.

The principle of a distance-$d$ repetition code is to introduce redundancy in the information encoding $\ket\pm_L := \ket\pm^{\otimes d}$ and make use of $d-1$ stabilizer measurements $S_i = X_i X_{i+1}$ to identify and correct phase-flip errors after each operation.
In our case, $\ket\pm^{\otimes d} = \ket{\mathcal{C}_\alpha^\pm}^{\otimes d}$ and $S_i$ is measured from the bias-preserving operations ${\mathcal{P}_{\ket +}, \text{CNOT}, \mathcal{M}_X}$.

We consider the logical operations in the set $\mathcal{S}_L = \{\mathcal{P}_{\ket+_L},\mathcal{P}_{\ket{0}_L},\mathcal{M}_{Z_L},\mathcal{M}_{X_L},Z_L,X_L,\text{C}X^k_L,{\text{CC}X}_L\}$, where $\text{C}X^k_L$ designates the multi-target CNOT gate and ${\text{CC}X}_L$ the Toffoli gate, which can all be implemented transversally on the repetition code, except for the ${\text{CC}X}_L$ gate; see \autoref{appendix:code}.
The transversal implementation of the $\text{CNOT}_L$ gate, however, requires all-to-all couplings between the physical qubits of the processor, which is not a realistic feature of a superconducting quantum processor.
Instead, we focus on a realization based on lattice surgery using nearest-neighbor interactions only (with additional routing qubits included in the resource evaluation), as detailed in \autoref{appendix:code}.
The same idea can be extended to the multiple target $\text{C}X^k_L$ gate, which applies an $X$ gate on $k$ qubits if the control qubit is in state $\ket{1}_L$ and the identity otherwise; see also \autoref{appendix:code}.
Finally, the logical Toffoli gate ${\text{CC}X}_L$ is implemented using gate teleportation~\cite{ChuangN1999Demonstratingviabilityuniversal} from a ``Toffoli magic state'' $\ket{\text{CC}X} = \tfrac{1}{2} (\ket{000} + \ket{010} + \ket{100} + \ket{111})$, as detailed in \autoref{appendix:code:toffoli} (with the circuit depicted in \autoref{fig:toffoli_teleport}).
The fault-tolerant preparation of the Toffoli magic state at the logical level, based on a projective measurement, is discussed in \autoref{appendix:code:toffoli}.

\paragraph{Noise model ---}
We exclusively consider the single-photon loss at rate $\kappa_1$, as in the presence of two-photon dissipation, the other error mechanisms have little impact on the noise model~\cite{MirrahimiPRX2019RepetitionCatQubits}.
Our resource estimates are based on the assumptions that a two-photon dissipation rate of $\kappa_2/2\pi = \SI{1.59}{\mega\hertz}$ and a resonator lifetime of $T_1 = \SI{10}{\milli\second}$ can be achieved, which corresponds to a ratio $\kappa_1 / \kappa_2 = 10^{-5}$ and a repetition code cycle time of $T_{\text{cycle}} = 5/\kappa_2 = \SI{500}{\nano\second}$.

For a fixed gate time of $1/\kappa_2$ (assumed to be identical for state preparation, measurement, and CNOT gates), the logical error rate per cycle of a distance-$d$ repetition code is given by~\cite{MirrahimiIp2022Highperformancerepetition} (see \autoref{appendix:code:memory} for details),
\begin{equation}\label{eq:epsL_RepCode}
\epsilon_L = 5.6\times10^{-2}{\Big( \dfrac{{(\alpha^{2})}^{0.86}\kappa_1/\kappa_2}{{(\kappa_1/\kappa_2)}_{\text{th}}} \Big)}^{\frac{d+1}{2}} + 2(d-1)\times 0.50 e^{-2\alpha^2},
\end{equation}
where the first term is the logical phase-flip error rate and the second term is the logical bit-flip error rate, and ${(\kappa_1/\kappa_2)}_{\text{th}} = 1.3\times10^{-2}$.

$\epsilon_L$ corresponds to the error rate of all gates (including the identity gate), but the Toffoli gate.
For the latter, we consider two variations of the state preparations of Toffoli magic states~\cite{BrandaoPQ2022BuildingFaultTolerant} using either error detection or error correction.
The resource evaluation uses the most suitable implementation, which depends on the size of the elliptic curve logarithms to compute.

\paragraph{Methods and results ---}

As several parameters are involved, we run an exhaustive search to minimize the product of the average photon number, expected time to solution, and total number of physical qubits
(the optimization code can be found in~\cite{code}). The required resources are shown in \autoref{fig:results} as a function of the number of bits of the prime $p$; see \autoref{appendix:tableau} for the details on the optimal parameters.
We see that \num{126133}~qubits are needed to compute a 256-bit logarithm in \SI{9}{\hour} for example, with on average 19 photons as the size of each cat qubit.
The required cat size suggests significant improvement in the design of cat qubits or, alternatively, the use of thin rectangular 2D codes~\cite{BrandaoPQ2022BuildingFaultTolerant}; see \autoref{appendix:tableau}.

\begin{figure}[h]
\includegraphics[width=\linewidth]{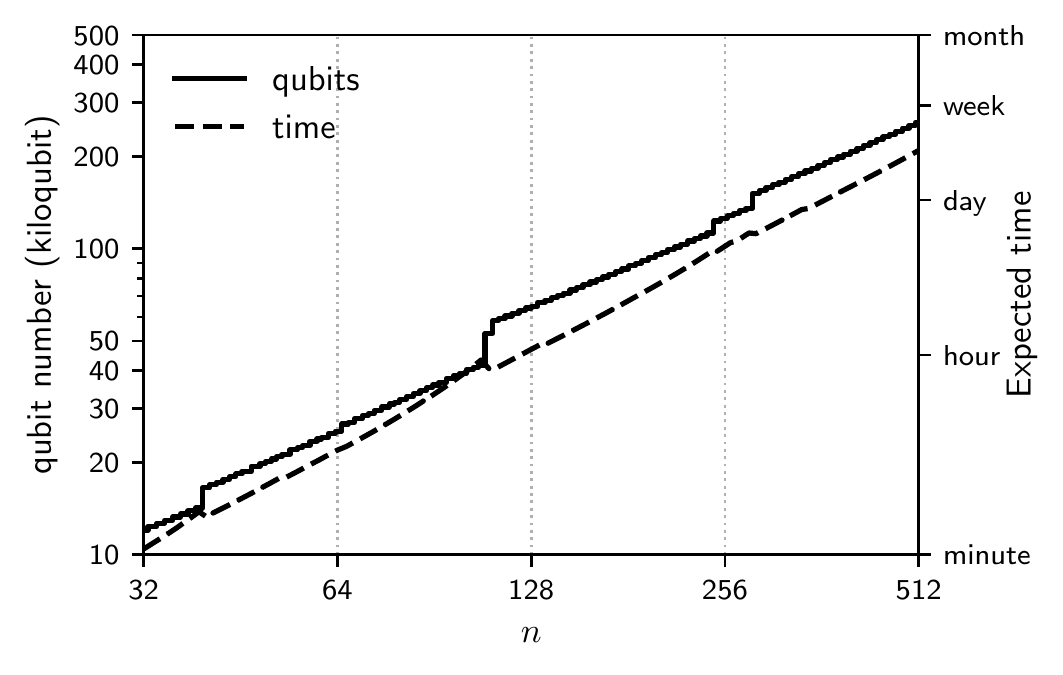}
\caption{Number of physical cat qubits and run-time for computing a discrete logarithm on an elliptic curve as a function of the number of bits in $p$.}\label{fig:results}
\end{figure}

\paragraph{Conclusion ---}
We reported on a performance analysis and provided a valuable guidance for the choice of a large scale architecture of a platform using cat qubits under the assumption that the bit flips are negligible and the remaining phase errors are corrected by a simple repetition code using Shor's algorithm.
We gave the details of an improved quantum computation of a discrete logarithm on elliptic curves, taking as an example the one used for securing signatures in Bitcoin transactions.
Assuming a ratio between single- and two-photon losses of $10^{-5}$ and a cycle time of $\SI{500}{\nano\second}$, we have shown that \num{126133} qubits are needed to compute a $256$-bit logarithm in $\SI{9}{\hour}$ and $19$ photons on average by cat state.
We also estimated that the implementation of Shor's algorithm for the factorization of \num{2048} Rivest--Shamir--Adleman (RSA) integers would take \num{349133} cat qubits and 4~days under the same assumption.
This provides a comparative analysis of the security level of two widely used cryptographic schemes.
This also favors comparisons with alternative platforms~\cite{Nickerson2022Activevolumearchitecture} and illustrates the gain in using a 1D code by comparing this cost estimation with the estimate reported in~\cite{Ekeraa2019Howfactor2048} showing that $20 \times {10}^6$ qubits and \SI{8}{\hour} would be needed for realizing the same factorization with a 2D grid of superconducting qubits and a standard surface code.
Note that, in both cases, the number of processing qubits can be substantially reduced by adding a quantum memory to the processor~\cite{SangouardPRL2021Factoring2048bit}.
Further note that parallelization has not been exploited in this work.
This could dramatically reduce the run-time, especially as the preparation of magic states is resource efficient and increasing the number of their factories would not significantly increase the total number of qubits, which would allow adequate use of look-ahead adder~\cite{Gidney2020Quantumblocklookahead}; see \autoref{appendix:ameliorations} for details.

\begin{acknowledgments}
We would like to thank Mazyar Mirrahimi for many discussions about the Toffoli magic state preparation schemes.
We are also grateful to Martin Ekerå for his comments on a first version of this manuscript.
E.~G.\@ and N.~S.\@ acknowledge funding by the Institut de Physique Théorique (IPhT), Commissariat à l'Énergie Atomique et aux Energies Alternatives (CEA), the Region Île-de-France in the framework of DIM SIRTEQ, the European Union's Horizon 2020 research and innovation program European High-Performance Computing Joint Undertaking under Grant Agreement No.\,101018180 (HPCQS), and a French national quantum initiative managed by Agence Nationale de la Recherche in the framework of France 2030 with the reference No.\,ANR-22-PETQ-0009.
D.~R.\@ and F-M.~L.~R.\@ acknowledge funding by Plan France 2030 through the project No.\,ANR-22-PETQ-0006.
\end{acknowledgments}

\appendix
\setcounter{secnumdepth}{2}

\section{Elliptic curve cryptography}\label{appendix:elliptic_curve}
This appendix introduces basic notions that are necessary to understand the discrete logarithm problem and the principles behind elliptic curve cryptographic protocols.

\subsection{Elliptic curve group}
Elliptic curve cryptographic protocols are based on elliptic curve groups which are formed from the points of an elliptic curve and an internal operation.

\subsubsection{Elliptic curve definition}
An elliptic curve is defined over a field (when the characteristic is different from $2$ and $3$) as the set of points $(x, y)$ whose coordinates satisfy an equation of the form (also refereed as Weierstrass form):
\begin{equation}\label{eq:weierstrass}
y^2 = x^3 + ax + b
\end{equation}
where $a$ and $b$ are constants.
An additional point at infinity is added to the elliptical curve.
In order to further define the group law, the curve has to be non-singular (no cusps nor self-intersections).
This holds if $4 a^3 + 27 b^2 \neq 0$.
Note that \autoref{eq:weierstrass} is unchanged under the transformation $y \mapsto -y$, hence the curve is symmetric with respect to the $x$-axis.

The coordinates and parameters of the curve can belong to any field (of characteristic different from $2$ and $3$ when defined with \autoref{eq:weierstrass}).
Elliptic curves are best represented when the chosen field is $\R$.
However, cryptographic applications typically rely on the field of integers modulo $p$, with $p$ a prime number.
The addition, multiplication and inversion over the coefficients and coordinates throughout this appendix are always to be understood as being modulo $p$.

\subsubsection{Group law}
The group law on elliptic curves is commonly referred as an addition between points of the curve, and written with the symbol $+$.
We introduce the addition as geometrical operations on the curve, and then give their translation in terms of equations on the coordinates as this is more convenient for implementations on a quantum computer.

Considering two distinct points of the elliptic curve, $P$ and $Q$, the line joining them usually intersects the curve at a third point $R'$.
The result of the addition, $R$, is the symmetric of $R'$ with respect to the $x$-axis.
We write $R = P + Q$.
Note that the described operation is commutative, which provides a justification of the designation as an addition.
A few particular cases need to be considered.
The first one arises when the line is vertical: there is no other intersection than $P$ and $Q$, and the result of the addition is the point at infinity.
The second one happens when the line joining $P$ and $Q$ is tangent to the curve at one of these two points (say the point $P$).
In that case, $R'$ is the point $P$ (it can be pictured as a double contact between the line and the curve).
The third special case corresponds to $P = Q$.
In this case, the line of interest is the tangent of the elliptic curve in $P$, and $R'$ is the intersection between this tangent and the curve (or point at infinity if the tangent is vertical).
Note that in the special case where $Q$ is at infinity, the vertical line passing through $P$ intersects the curve at the symmetric of $P$ with respect to the $x$-axis and hence $P + Q = P$.
This means that the infinite point is the neutral element.
Further note that the symmetric of $P$ with respect to the $x$-axis is its inverse, $-P$.

\medskip

Let us now see how the group law translates into equations on the coordinates of the involved points.
We study the addition of $P$ and $Q$ with coordinates $P = (x_P, y_P)$ and $Q = (x_Q, y_Q)$.
In the case where $x_P \neq x_Q$, the slope of the line joining $P$ and $Q$ is given by $\lambda = \frac{y_Q - y_P}{x_Q - x_P}$.
The division is well-defined as we work on a field (and consider the case $x_P \neq x_Q)$.
We emphasize that in the cryptographic context, the division applies on integers modulo $p$.
The third solution (apart from $P$ and $Q$) of the equation determining the intersections between the line and the curve, $R'$, has the coordinates
\begin{subequations}\label{Eq:elliadd}
\begin{align}
	x_{R'} &= \lambda^2 - x_P - x_Q \\
	y_{R'} &= y_P + \lambda (x_{R'} - x_P).
\end{align}
\end{subequations}
The result of the addition $R = P + Q = - R'$ has as coordinates $x_R = x_{R'}$ and $y_{R} = -y_{R'}$.
Note that $\lambda = \frac{y_{R'} - y_P}{x_{R'} - x_P}$ as $R'$ is also on the line joining $P$ and $Q$.
This remark is useful for uncomputing the value of $\lambda$ in the context of an in-place addition on a quantum computer.

When $x_P = x_Q$ and $y_P \neq y_Q$, necessarily $y_Q = - y_P$ and $Q = -P$.
Hence, the result of the addition is the neutral point, the one at infinity.
When $P = Q$ and $y_P = y_Q = 0$, the tangent of the curve in $P$ is vertical and the result of the addition is again the zero.
Otherwise, in the generic case for which $P = Q$, the slope of the tangent is given by $\lambda = \frac{3 x_{P}^2 + a}{2 y_{P}}$.
 The coordinates of the result are given by the formulae used in the generic case while replacing $\lambda$ by its specific value.

\subsubsection{Multiplication}
Now that the addition between two points of an elliptic curve is properly described, we can define the multiplication of a point by an integer as
\begin{equation}
k P = \underbrace{P + P + \cdots + P}_{\text{$k$ times}}.
\end{equation}
From this definition, it is clear that the scalar multiplication is associative [$(k_1 k_2) P = k_1 (k_2 P)$], distributive [$(k_1 + k_2) P = (k_1 P) + (k_2 P)$], compatible with identity ($1 \times P = P$) and hence satisfies the expected properties of a scalar multiplication.

Note that the multiplication can be computed efficiently by first decomposing the multiplier $k$ in a binary form $k = \sum_{i} 2^i k_i$.
From the distributive property, we have $k P = \sum_i 2^i k_i P$, that is the result can be obtained efficiently by computing the $2^i P$ terms through successive doubling ($2^{i+1} P = 2^i P + 2^i P$) and adding $2^i P$ into a register encoding the result when $k_i = 1$ only.
We present in \autoref{appendix:arithmetic} a windowed version of this algorithm, in the context of quantum computing.

\subsection{Diffie--Hellman key exchange}
The Diffie--Hellman protocol is a method of securely exchanging cryptographic keys over a public channel.
Prior to the algorithm, Alice and Bob agree on a specific curve, that is they agree on a given prime number $p$ such that the coordinates are numbers of the field of integers modulo $p$, and on the parameters $a$ and $b$ of \autoref{eq:weierstrass}.
They also publicly agree on a point $G$ and will work only into the cyclic sub-group generated by $G$.
For the following, we note $r$ the order of this group, and in cryptography applications, $r$ is typically a prime number.

To establish the key, Alice and Bob choose independently and randomly each one a number between $1$ and $r-1$; we write them respectively $k_a$ and $k_b$.
Those numbers are kept secret.
Then, they compute respectively $k_a G$ and $k_b G$.
The coordinates of those points are exchanged via the public channel.
Alice and Bob compute respectively $k_a (k_b G)$ and $k_b (k_a G)$, obtaining the same result due to the commutativity of the multiplication.
The $x$ coordinate of this point is the key (the other coordinate is redundant with it, as the curve is publicly known).

Note that we presented here the basic version of Diffie--Hellman algorithm.
To have a full operational protocol, authentication is required and the resulting key is usually hashed.
Also, several variants exist, for instance the cofactor Diffie--Hellman protocol helps to reduce the risk for Alice of partially revealing her private secret $k_a$ in case of a malicious Bob~\cite{Davis2018Recommendationpairwise}.

\subsection{Elliptic Curve Digital Signature Algorithm}
Suppose Alice now wants to send a signed message to Bob.
She first needs to create a pair of private and public keys.
In the elliptic curve signature algorithm, the private key of Alice is an integer $s_a$ ($< r$), while the public key is given by $P_a = s_a G$.
The message to sign is hashed down to the same number of bits than in $r$, and we write $z$ this hash.
Only this $z$ is relevant for the signing algorithm.

The signature protocol starts by randomly picking a number $k$, smaller than $r$, that must be kept secret.
Then, the point $(i, j) = k G$ is computed, and we define $x = i \mod{r}$.
In the case $x = 0$, we choose another $k$ and start again the protocol.
Then, $y = k^{-1} (z + s_a x) \mod{r}$ is computed, and if $y = 0$, another $k$ is chosen and the protocol starts again.
The signature is the couple $(x, y)$.

For verifying the signature, the point $(i, j) = (z y^{-1} \mod{r})G + (x y^{-1} \mod{r}) P_a$ is computed.
The signature is valid if $x = i \mod{r}$, invalid otherwise.
To prove the correctness of this verification, we can check that injecting the definition of $y$ in the computed point gives $(i, j) = k G$, and the result follows.

\subsubsection{Bitcoin's parameters}
The bitcoin protocol uses the elliptic curve digital signature algorithm to ensure that no one can impersonate the issuer of a transaction.
This protocol uses the \verb|secp256k1| curve, defined in~\cite{Brown2010StandardsEfficientCryptography}.
It works on the field of integers modulo $p = 2^{256} - 2^{32} - 977$ and its equation is $y^2 = x^3 + 7$.
The generator $G$ of the cyclic subgroup has the coordinates:
\begin{subequations}
\begin{align}
	x &= \begin{multlined}[t]
			55066263022277343669578718895168534326 \\ 250603453777594175500187360389116729240
		\end{multlined} \\
	y &= \begin{multlined}[t]
			32670510020758816978083085130507043184 \\ 471273380659243275938904335757337482424.
		\end{multlined}
\end{align}
\end{subequations}
Those choices result in a subgroup of order
\begin{equation}
 r = \begin{multlined}[t]
			11579208923731619542357098500868790785 \\ 2837564279074904382605163141518161494337,
		\end{multlined}
\end{equation}
which is a prime number.

\subsection{Discrete logarithm problem}
The key exchange and signature algorithm presented above are secured under the assumption that given $G$ and $kG$ for some $k < r$, finding $k$ is not possible for the adversary. This is the discrete logarithm problem.
The Pollard's rho algorithm is the best known for solving this problem with a classical computer, and a complexity $\bigO{\sqrt{r}}$, which is exponential in terms of the number of bits in $r$.
The larger discrete logarithm problem solved so far on a classical computer is based on the curve \verb|secp256k1| with a 114-bit private key~\cite{Pons2020PollardskangarooSECPK1}.

The discrete logarithm problem is tackled by Shor's algorithm, which is seen as a threat for elliptic curve cryptographic protocols.

\section{Shor's and Ekerå's algorithms for discrete logarithm}\label{appendix:shor_ekera}
Shor's algorithm for solving the discrete logarithm problem has been presented in~\cite{Shor1994Algorithmsquantumcomputation,ShorSJoC1997PolynomialTimeAlgorithms}.
It was initially designed for the multiplicative group of integers modulo $p$ but actually applies to any finite cyclic group.
Ekerå introduced a variation of this algorithm in~\cite{Ekeraa2018Quantumalgorithmscomputing}.
In this appendix, these two algorithms are presented and our choice to use Shor's algorithm is justified.

We write $G$ the generator of the cyclic subgroup of the elliptic curve we work with.
Its order is denoted by $r$, and $n$ is the number of bits of $r$ ($r < 2^n \leq 2r$).
$r$ is assumed to be a prime number.
We write $l$ the logarithm to be determined and $P$ the corresponding point, that is $P = l G$.

\subsection{Shor's algorithm}
As for Shor's algorithm for factorizing integers, the one for the discrete logarithm computation is based on the preparation of registers in a superposition of all possible integers between $0$ and a large number, followed by the application of a specific function, and the use of quantum Fourier transforms to reveal the period of this function.
We here consider the following function of two variables
\begin{equation}\label{Eq:fx1x2}
	f(x_1, x_2) = x_1 G - x_2 P.
\end{equation}
Note the alternative expression $f(x_1, x_2) = (x_1 - x_2 l)G$ holds because of the definition of the logarithm $l$.
The function $f$ is periodic in the following meaning
\begin{equation}
\forall k, f(x_1 + k l, x_2 + k) = f(x_1, x_2).
\end{equation}
This periodic feature is in essence what the quantum Fourier transform reveals, which allows one to access to $l$.

The algorithm starts by preparing two registers in a superposition of all possible numbers between $0$ and $r-1$
\begin{equation}
\frac{1}{r} \sum\limits_{x_1=0}^{r-1} \sum\limits_{x_2=0}^{r-1} \ket{x_1} \ket{x_2}.
\end{equation}
The function $f$ is computed with those registers as input and the output stored in a new register
\begin{equation}
\frac{1}{r} \sum\limits_{x_1=0}^{r-1} \sum\limits_{x_2=0}^{r-1} \ket{x_1} \ket{x_2} \ket{f(x_1, x_2)}.
\end{equation}
An inverse quantum Fourier transform is then applied to the registers containing $\ket{x_1}$ and $\ket{x_2}$.
This leads to
\begin{equation}
\frac{1}{r^2} \sum\limits_{x_1, x_2, y_1, y_2=0}^{r-1} e^{2 \pi i (x_1 y_1 + x_2 y_2)/r} \ket{y_1} \ket{y_2} \ket{f(x_1, x_2)},
\end{equation}
that we can rewrite as
\begin{equation}
\sum\limits_{y_1, y_2=0}^{r-1} \sum\limits_{k=0}^{r-1} \left[\frac{1}{r^2} \sum\limits_{\substack{x_1, x_2=0 \\ f(x_1, x_2) = k G}}^{r-1} \hspace*{-1.2em} e^{2 \pi i (x_1 y_1 + x_2 y_2)/r} \right] \ket{y_1} \ket{y_2} \ket{k G}.
\end{equation}
The part between brackets is the probability amplitude associated with the component $\ket{y_1} \ket{y_2} \ket{k G}$.
To simplify further the expression of those amplitudes, notice that the condition $f(x_1, x_2) = k G$ is equivalent to $(x_1 - x_2 l) G = k G$.
Hence $x_1 - x_2 l = k \mod{r}$, that is $x_1 = k + x_2 l \mod{r}$.
Plugging the last equality into the argument of the exponential, we obtain the following expression for the probability amplitude associated with $\ket{y_1} \ket{y_2} \ket{kG}$
\begin{equation}
\frac{1}{r^2} \sum\limits_{x_2=0}^{r-1} e^{2 \pi i (k y_1 + (y_2 + l y_1) x_2)/r}.
\end{equation}
The sum vanishes (destructive interferences) except when $y_2 + l y_1 = 0 \mod{r}$.
Hence, by measuring the registers encoding $y_1$ and $y_2$, as long as $y_1 \neq 0$ the discrete logarithm can be recovered using $l = - y_2 y_1^{-1} \mod{r}$ (from the choice of cyclic group, we imposed that $r$ is prime and hence $y_1$ is invertible modulo $r$).
Note that there are $r^2$ possible measurement outputs (as $y_2$ and $k$ can each take $r$ different values), each being produced with a probability $1/r^2$.
$r$ of them give $y_2 = 0$ and $y_1 = 0$, and can't be exploited to find the requested solution.
Hence, the probability to obtain the value of the discrete logarithm is $1-\frac{1}{r}$.

Note that when implementing the algorithm, the initial superposition of all numbers is usually extended to the next power of two, so that the register can be prepared by separately setting each qubit in the $\ket{+} = (\ket{0} + \ket{1})/\sqrt{2}$ state.
The Fourier transform also goes up to the same power of two.
This typically reduces the success probability (here not taking into account implementation errors) to an average estimated to be between $\SI{60}{\percent}$ and $\SI{82}{\percent}$~\cite{Ekeraa2019RevisitingShorsquantum}.
The success probability can be brought close to $1$ by either increasing the size of the register of $x_1$ and carrying-out a limited search on one parameter, or by performing a two-dimensional limited search post-processing, as detailed in~\cite{Ekeraa2019RevisitingShorsquantum}.
We take into account the latter into our evaluation by considering that no error comes from the algorithm itself (only from its implementation).

\subsection{Ekerå's algorithm}
Ekerå's algorithm for computing general logarithm~\cite{Ekeraa2018Quantumalgorithmscomputing} is a variation of Shor's algorithm allowing to choose a trade-off between the number of operations per run of the algorithm and the number of runs to be executed to access the desired discrete logarithm.
We introduce the trade-off parameter $s$ which is a small integer larger or equal than $1$.
$n_r$ denotes the number of bits needed to write $r$ ($2^{n_r-1} \leq r < 2^{n_r}$) and $n' = \ceil{\frac{n_r}{s}}$.
The quantum part of Ekerå's algorithm is similar to the one in Shor's algorithm (for discrete logarithm), with the difference that the two registers have the respective sizes $n_r + n'$ and $n'$.
This quantum part is typically repeated $s$ times and the outputs are fed to the classical post-processing detailed in~\cite{Ekeraa2018Quantumalgorithmscomputing}.

In Shor's algorithm, $2 n_r$ qubits are used for the two registers containing $\ket{x_1}$ and $\ket{x_2}$, and the number of elliptic curve point additions scales accordingly.
For Ekerå's algorithm, $n_r + 2 \ceil{\frac{n_r}{s}}$ qubits are used for the same purpose, giving for instance a number of $3 n_r$ qubits when $s=1$, and $n_r$ qubits in the asymptotic limit $s \to + \infty$.
We tested different values of $s$ and in any case, Shor's algorithm appeared to be a better trade-off when taking into account the overall run-time and the number of qubits~\footnote{We chose as the objective of the minimization on all parameters the product of the average run-time, the number of qubits and the average number of photons in each cat.
Focusing more on the qubit number would certainly lead to favor Ekerå's algorithm.}.
Shor's algorithm requires a prior knowledge of the order $r$ of the cyclic group, while Ekerå's algorithm doesn't need it, but can retrieve it from the measurements~\cite{Ekeraa2018Quantumalgorithmscomputing}.
This has a cost, which partially explains the differences we observed~\footnote{Also note that the analysis of the success probability of Shor's algorithm relies on an heuristic~\cite{Ekeraa2019RevisitingShorsquantum}, contrary to Ekerå's one~\cite{Ekeraa2018Quantumalgorithmscomputing}.}.
The results presented in this paper are all obtained with Shor's algorithm.

\section{Arithmetic circuits}\label{appendix:arithmetic}
We here present the circuit to implement Shor's algorithm.
More precisely, we show how to compute $f(x_1,x_2)=x_1 G - x_2 P$ (see \autoref{Eq:fx1x2}), that is to compute elliptic curve multiplications.
As we show in this appendix, an elliptic curve multiplication can be decomposed into a sequence of elliptic curve additions.
As we choose to represent the points of the elliptic curve by their coordinates, each addition is a combination of modular additions, subtractions, multiplications and divisions, see \autoref{Eq:elliadd}.

The algorithm we use is derived from the one presented in~\cite{Soeken2020ImprovedQuantumCircuits}, where integers are in Montgomery's representation, and windowed arithmetic circuits~\cite{Gidney2019Windowedquantumarithmetic} are used.
The main differences with respect to~\cite{Soeken2020ImprovedQuantumCircuits} is that the circuits are adapted for a direct use of Toffoli gates.
Moreover, subcircuits including controlled adders and comparisons are more efficiently implemented.

\subsection{Montgomery representation}
Montgomery's representation allows one to replace a reduction modulo $p$ (obtained by euclidean division) by a halving operation (a division by $2$, that is a simple bit shift) during the multiplication step~\cite{MontgomeryMoC1985Modularmultiplicationtrial,Lauter2017QuantumResourceEstimates,Chuang2018HighPerformanceQuantum}.
Here we present the Montgomery representation with parameters relevant for the computation of discrete logarithms in a cryptography context.
For a more general definition of Montgomery representation, see~\cite{MontgomeryMoC1985Modularmultiplicationtrial}.

We work with operations modulo $p$, with $p > 2$ a $n$-bit long prime number.
The basic idea of the Montgomery representation is to represent a number $x \in \Z/p\Z$ by $x' = x 2^n \mod{p}$ (still in $\Z/p\Z$).
Note that $2^n$ and $p$ are coprime, hence $2^n$ has an inverse in $\Z/p\Z$, that we write $2^{-n}$, and $x$ can be recovered as $x = x' 2^{-n} \mod{p}$.
Although the Montgomery representation is a bijection, we don't need to convert number back to the standard representation in Shor's algorithm.

\subsection{Modular addition}\label{appendix:arithmetic:add_mod}
\paragraph{Outline ---}
First, note that the standard modular addition is compatible with the Montgomery transformation
\begin{multline*}
\left[\left(x 2^{n} \mod{p}\right) + \left(y 2^{n} \mod{p}\right)\right] \mod{p} \\
	= \left(x+y\right) 2^{n} \mod{p}.
\end{multline*}

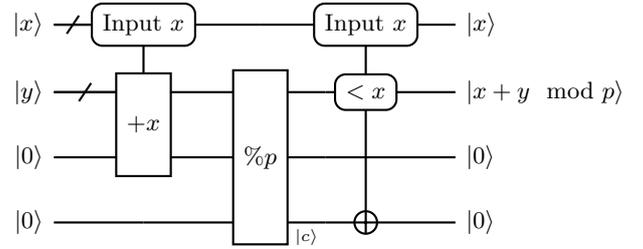
\begin{figure}[h]
\begin{quantikz}[row sep=1em]
	\lstick{$\ket{x}$}&\gate[style={rounded corners}]{\text{Input }x}\vqw{1}\qwbundle{}&\qw           &\qw          &[-2em]\gate[style={rounded corners}]{\text{Input }x}\vqw{1}&\rstick{$\ket{x}$}          \qw \\
	\lstick{$\ket{y}$}&\gate[2]{+x}\qwbundle{}                                         &\gate[3]{\% p}&\qw          &      \gate[style={rounded corners}]{< x}           \vqw{2}&\rstick{$\ket{x+y \mod p}$} \qw \\
	\lstick{$\ket{0}$}&                                                                &              &\qw          &      \qw                                                  &\rstick{$\ket{0}$}          \qw \\
	\lstick{$\ket{0}$}&\qw                                                             &              &\qw {\ket{c}}&      \targ{}                                              &\rstick{$\ket{0}$}          \qw \\
\end{quantikz}
 \caption{Addition modulo $p$: $\ket{x}\ket{y} \mapsto \ket{x}\ket{x+y \mod{p}}$.
	It is obtained by a standard nonmodular addition, a reduction modulo $p$ that also creates an ancillary qubit $\ket{c}$ indicating if the reduction happened, and the uncomputation of the ancillary qubit through comparison between the result and $x$.
	Each of those three steps is achieved with a circuit detailed in following figures.
}\label{fig:add_mod:add_mod}
\end{figure}

The modular addition is the operation $\ket{x}\ket{y} \mapsto \ket{x}\ket{y+x \mod{p}}$, where $p$ is a classically known non-negative number and $0 \leq x, y < p$.
As mentioned before, $p$ is a $n$-bit long prime number ($x$ and $y$ have the same length), hence $p < 2^{n}$.
The modular addition, shown in \autoref{fig:add_mod:add_mod}, is implemented in three steps:
\begin{enumerate}
\item a nonmodular addition of $x$ in the target register, with an additional qubit for the output ($0 \leq x + y < 2 p < 2^{n+1}$);
\item a reduction modulo $p$ of the result from first step, with creation of an ancillary qubit that indicates if the reduction happened;
\item the ancillary qubit is uncomputed by flipping it if and only if $x + y \mod{p} < x$.
\end{enumerate}
The equivalence mentioned in the third step is ensured because if the modular reduction happens, $x + y \mod{p} = x+y -p$ and as $y < p$ we have $x+y-p < x$, while if it doesn't happen, $x + y \mod{p} = x+y$ and $y \geq 0$ implies that $x+y \geq x$.

We detail each of these three steps separately.

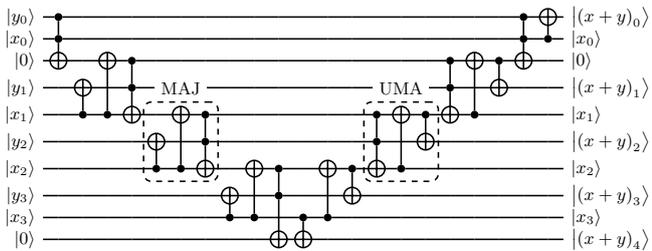
\begin{figure}[h]
\resizebox{\linewidth}{!}{\begin{quantikz}[column sep=0.35em,row sep=0.5em]
\lstick{$\ket{y_0}$}&\ctrl{1}&\qw      &\qw      &\qw     &\qw                                                                                                       &\qw      &\qw     &\qw      &\qw      &\qw     &\qw     &\qw      &\qw     &\qw                                                                                                            &\qw      &\qw     &\qw     &\qw      &\qw     &\ctrl{1}&\targ{}  &\rstick{$\ket{{(x+y)}_0}$}\qw \\
\lstick{$\ket{x_0}$}&\ctrl{1}&\qw      &\qw      &\qw     &\qw                                                                                                       &\qw      &\qw     &\qw      &\qw      &\qw     &\qw     &\qw      &\qw     &\qw                                                                                                            &\qw      &\qw     &\qw     &\qw      &\qw     &\ctrl{1}&\ctrl{-1}&\rstick{$\ket{x_0}$}      \qw \\
\lstick{$\ket{0}$}  &\targ{} &\qw      &\targ{}  &\ctrl{1}&\qw                                                                                                       &\qw      &\qw     &\qw      &\qw      &\qw     &\qw     &\qw      &\qw     &\qw                                                                                                            &\qw      &\qw     &\ctrl{1}&\targ{}  &\ctrl{1}&\targ{} &\qw      &\rstick{$\ket{0}$}        \qw \\
\lstick{$\ket{y_1}$}&\qw     &\targ{}  &\qw      &\ctrl{1}&\qw                                                                                                       &\qw      &\qw     &\qw      &\qw      &\qw     &\qw     &\qw      &\qw     &\qw                                                                                                            &\qw      &\qw     &\ctrl{1}&\qw      &\targ{} &\qw     &\qw      &\rstick{$\ket{{(x+y)}_1}$}\qw \\
\lstick{$\ket{x_1}$}&\qw     &\ctrl{-1}&\ctrl{-2}&\targ{} &\qw\gategroup[wires=3,steps=3,style={dashed,rounded corners,inner sep=-1pt},label style={fill=white}]{MAJ}&\targ{}  &\ctrl{1}&\qw      &\qw      &\qw     &\qw     &\qw      &\qw     &\ctrl{1}\gategroup[wires=3,steps=3,style={dashed,rounded corners,inner sep=-1pt},label style={fill=white}]{UMA}&\targ{}  &\ctrl{1}&\targ{} &\ctrl{-2}&\qw     &\qw     &\qw      &\rstick{$\ket{x_1}$}      \qw \\
\lstick{$\ket{y_2}$}&\qw     &\qw      &\qw      &\qw     &\targ{}                                                                                                   &\qw      &\ctrl{1}&\qw      &\qw      &\qw     &\qw     &\qw      &\qw     &\ctrl{1}                                                                                                       &\qw      &\targ{} &\qw     &\qw      &\qw     &\qw     &\qw      &\rstick{$\ket{{(x+y)}_2}$}\qw \\
\lstick{$\ket{x_2}$}&\qw     &\qw      &\qw      &\qw     &\ctrl{-1}                                                                                                 &\ctrl{-2}&\targ{} &\qw      &\targ{}  &\ctrl{1}&\qw     &\targ{}  &\ctrl{1}&\targ{}                                                                                                        &\ctrl{-2}&\qw     &\qw     &\qw      &\qw     &\qw     &\qw      &\rstick{$\ket{x_2}$}      \qw \\
\lstick{$\ket{y_3}$}&\qw     &\qw      &\qw      &\qw     &\qw                                                                                                       &\qw      &\qw     &\targ{}  &\qw      &\ctrl{2}&\qw     &\qw      &\targ{} &\qw                                                                                                            &\qw      &\qw     &\qw     &\qw      &\qw     &\qw     &\qw      &\rstick{$\ket{{(x+y)}_3}$}\qw \\
\lstick{$\ket{x_3}$}&\qw     &\qw      &\qw      &\qw     &\qw                                                                                                       &\qw      &\qw     &\ctrl{-1}&\ctrl{-2}&\qw     &\ctrl{1}&\ctrl{-2}&\qw     &\qw                                                                                                            &\qw      &\qw     &\qw     &\qw      &\qw     &\qw     &\qw      &\rstick{$\ket{x_3}$}      \qw \\
\lstick{$\ket{0}$}  &\qw     &\qw      &\qw      &\qw     &\qw                                                                                                       &\qw      &\qw     &\qw      &\qw      &\targ{} &\targ{} &\qw      &\qw     &\qw                                                                                                            &\qw      &\qw     &\qw     &\qw      &\qw     &\qw     &\qw      &\rstick{$\ket{{(x+y)}_4}$}\qw
\end{quantikz}
 }
\caption{Adder with output carry.
	It implements the operation $\ket{x}\ket{y} \mapsto \ket{x}\ket{y+x}$, where the register $y$ has one extra qubit at the output.
	Here the two inputs are represented on $4$~bits.
	The MAJ sub-operation computes from the former carry and one bit from each input the following carry (set in the bottom qubit).
	The UMA sub-operation restores the $x$ bit and the previous carry, and computes the current bit of the result.
	MAJ and UMA operations are used on every qubit pair, except the first and last for which we use a more efficient implementation.
	}\label{fig:add_mod:add}
\end{figure}

\paragraph{Nonmodular addition ---}
The first step is a nonmodular addition: $\ket{x}\ket{y} \mapsto \ket{x}\ket{x+y}$ which is implemented by the circuit presented in \autoref{fig:add_mod:add}, introduced in~\cite[Fig.\,4, modified with the optimizations~4 and~5 proposed in this reference]{Moulton2004newquantumripple}.
This circuit starts by computing successively the carry bits using repetitively the subcircuit labeled as MAJ in \autoref{fig:add_mod:add} (each carry is temporarily stored in the qubit initially containing $\ket{x_k}$).
The MAJ operations are then uncomputed while the bit encoding the result is set to its output value, with the subcircuit labeled as UMA\@.
First and last qubits are special cases with simplified and combined versions of MAJ and UMA operations.
Note that for simplicity, we did not take into account the additional optimizations proposed in~\cite{Moulton2004newquantumripple} which reduces the depth of the circuit (using time-optimal computation~\cite{Fowler2013Timeoptimalquantum,Fowler2019Flexiblelayoutsurface} would give an even shallower circuit).

\paragraph{Modular reduction ---}
The modular reduction consists of taking as input a register containing a number $z$ such that $0 \leq z < 2p$, and outputting:
\begin{equation*}
z \mod{p} =
	\begin{cases}
		z     & \text{if } z < p \\
		z - p & \text{if } z \geq p
	\end{cases}.
\end{equation*}
To ensure reversibility, another output has to be generated: a bit $c$ that indicates if the reduction occurred, and takes as value the quotient of the Euclidean division of $z$ by $p$.

The modular reduction can be implemented with various circuits, three of them are presented in \autoref{sec:modular_reduction}.
The most efficient is the third one and is the one we consider.

\begin{figure}[h]
\resizebox{\linewidth}{!}{\begin{quantikz}[column sep=0.35em,row sep=0.5em]
\lstick{$\ket{z_0}$}  &\targ{}  &\ctrl{1}&\qw     &\qw      &\qw     &\qw                                                                                                                             &\qw      &\qw     &\qw      &\qw      &\qw     &\qw     &\qw      &\qw     &\qw                                                                                                            &\qw      &\qw     &\qw     &\qw      &\qw     &\ctrl{1}&\targ{}  &\rstick{$\ket{z_0}$}\qw \\
\lstick{$\ket{x_0}$}  &\ctrl{-1}&\ctrl{1}&\qw     &\qw      &\qw     &\qw                                                                                                                             &\qw      &\qw     &\qw      &\qw      &\qw     &\qw     &\qw      &\qw     &\qw                                                                                                            &\qw      &\qw     &\qw     &\qw      &\qw     &\ctrl{1}&\ctrl{-1}&\rstick{$\ket{x_0}$}\qw \\
\lstick{$\ket{0}$}    &\qw      &\targ{} &\ctrl{1}&\targ{}  &\ctrl{1}&\qw                                                                                                                             &\qw      &\qw     &\qw      &\qw      &\qw     &\qw     &\qw      &\qw     &\qw                                                                                                            &\qw      &\qw     &\ctrl{1}&\targ{}  &\ctrl{1}&\targ{} &\qw      &\rstick{$\ket{0}$}  \qw \\
\lstick{$\ket{z_1}$}  &\qw      &\qw     &\targ{} &\qw      &\ctrl{1}&\qw                                                                                                                             &\qw      &\qw     &\qw      &\qw      &\qw     &\qw     &\qw      &\qw     &\qw                                                                                                            &\qw      &\qw     &\ctrl{1}&\qw      &\targ{} &\qw     &\qw      &\rstick{$\ket{z_1}$}\qw \\
\lstick{$\ket{x_1}$}  &\qw      &\qw     &\qw     &\ctrl{-2}&\targ{} &\ctrl{1}\gategroup[wires=3,steps=3,style={dashed,rounded corners,inner sep=-1pt},label style={fill=white}]{$\text{UMA}^\dagger$}&\targ{}  &\ctrl{1}&\qw      &\qw      &\qw     &\qw     &\qw      &\qw     &\ctrl{1}\gategroup[wires=3,steps=3,style={dashed,rounded corners,inner sep=-1pt},label style={fill=white}]{UMA}&\targ{}  &\ctrl{1}&\targ{} &\ctrl{-2}&\qw     &\qw     &\qw      &\rstick{$\ket{x_1}$}\qw \\
\lstick{$\ket{z_2}$}  &\qw      &\qw     &\qw     &\qw      &\qw     &\targ{}                                                                                                                         &\qw      &\ctrl{1}&\qw      &\qw      &\qw     &\qw     &\qw      &\qw     &\ctrl{1}                                                                                                       &\qw      &\targ{} &\qw     &\qw      &\qw     &\qw     &\qw      &\rstick{$\ket{z_2}$}\qw \\
\lstick{$\ket{x_2}$}  &\qw      &\qw     &\qw     &\qw      &\qw     &\qw                                                                                                                             &\ctrl{-2}&\targ{} &\ctrl{1} &\targ{}  &\ctrl{1}&\qw     &\targ{}  &\ctrl{1}&\targ{}                                                                                                        &\ctrl{-2}&\qw     &\qw     &\qw      &\qw     &\qw     &\qw      &\rstick{$\ket{x_2}$}\qw \\
\lstick{$\ket{z_3}$}  &\qw      &\qw     &\qw     &\qw      &\qw     &\qw                                                                                                                             &\qw      &\qw     &\targ{}  &\qw      &\ctrl{2}&\qw     &\qw      &\targ{} &\qw                                                                                                            &\qw      &\qw     &\qw     &\qw      &\qw     &\qw     &\qw      &\rstick{$\ket{z_3}$}\qw \\
\lstick{$\ket{x_3}$}  &\qw      &\qw     &\qw     &\qw      &\qw     &\qw                                                                                                                             &\qw      &\qw     &\qw      &\ctrl{-2}&\qw     &\ctrl{1}&\ctrl{-2}&\qw     &\qw                                                                                                            &\qw      &\qw     &\qw     &\qw      &\qw     &\qw     &\qw      &\rstick{$\ket{x_3}$}\qw \\
\lstick{$\ket{z < x}$}&\qw      &\qw     &\qw     &\qw      &\qw     &\qw                                                                                                                             &\qw      &\qw     &\qw      &\qw      &\targ{} &\targ{} &\qw      &\qw     &\qw                                                                                                            &\qw      &\qw     &\qw     &\qw      &\qw     &\qw     &\qw      &\rstick{$\ket{0}$}  \qw
\end{quantikz}
 }
\caption{Comparison for uncomputing the ancillary bit: $\ket{x}\ket{z}\ket{x<z} \mapsto \ket{x}\ket{z}\ket{0}$.
	This first half of this circuit is based on a subtraction modulo $2^{n+1}$ ($n$ being the number of bits of $x$ and $z$), the conjugate of an addition.
	The most significant carry is used to perform the uncomputation, and then the inputs are restored to their original values.
	The boxed $\text{UMA}^\dag$ and UMA operations (same as in \autoref{fig:add_mod:add}) compute and uncompute the carries, and are repeated for each qubit pair, except the first and last one where specific case circuits allow small optimization.
}\label{fig:add_mod:compair_uncompute}
\end{figure}
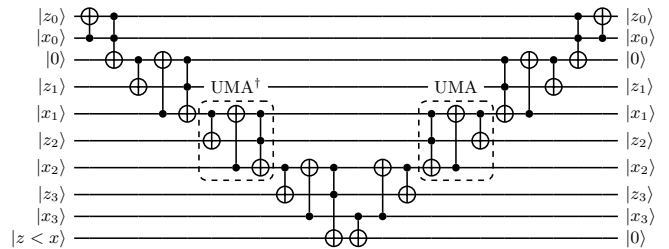

\paragraph{Comparison ---}
The last step is the uncomputation of the ancillary qubit generated by the reduction: $\ket{x}\ket{z}\ket{z < x} \mapsto \ket{x}\ket{z}\ket{0}$ ($\ket{z < x}$ is one qubit indicating if $z < x$ is true or false), which is done with a comparison between $x$ and $z$.  The corresponding circuit is presented in \autoref{fig:add_mod:compair_uncompute}.
As for the addition (\autoref{fig:add_mod:add}), it consists of propagating carries, then the most significant bit is used, and the carries are uncomputed to restore the inputs.
More precisely, the first half of the circuit is identical to the one for implementing a subtraction modulo $2^{n+1}$: $\ket{x}\ket{z} \mapsto \ket{x}\ket{z-x \mod{2^{n+1}}}$, with $n$ the number of bits of $x$ and $z$.
As $0 \leq x, z < 2^n$ and $0 \leq z-x \mod{2^{n+1}} < 2^{n+1}$, only two cases are possible:
\begin{description}
\item[$z-x \mod{2^{n+1}} = z-x$] as $z < 2^n$ and $x \geq 0$, $z-x < 2^{n}$ and we conclude that the most significant bit of the result takes value $0$.
	On the other hand, $0 \leq z-x \mod{2^{n+1}} = z-x$ implies that $x \leq z$.
\item[$z-x \mod{2^{n+1}} = z-x + 2^{n+1}$] as $x < 2^{n}$ and $z \geq 0$, $2^{n} < z-x + 2^{n+1}$ and we conclude that the most significant bit of the result takes value $1$.
	On the other hand, $z-x \mod{2^{n+1}} = z-x + 2^{n+1} < 2^{n+1}$ implies that $z < x$.
\end{description}
From those two cases, we draw the conclusion that the most significant bit takes value $1$ if and only if $z < x$.
The uncomputation of the ancillary qubit can be achieved with a single CNOT controlled by the most significant bit and targeting the qubit to uncompute.

The circuit for achieving a subtraction modulo $2^{n+1}$ is the conjugate of the one for computing an addition modulo $2^{n+1}$.
It thus starts with $\text{UMA}^\dag$ operations, as defined in \autoref{fig:add_mod:add}.
In the comparison circuit, only the first half of the subtraction is used while the last half uncomputes the carries, with UMA operations.
In \autoref{fig:add_mod:compair_uncompute}, we reduced the circuit size by directly uncomputing the target qubit instead of computing the most significant bit and then using it to control a CNOT\@.

\subsection{Modular reduction}\label{sec:modular_reduction}
A key subcircuit for the modular addition (second step) and more broadly for the discrete logarithm computation, is the modular reduction.
It consists of the following operation: $\ket{z} \mapsto \ket{z \mod{p}}\ket{c}$ where $p$ is a known integer, $0 \leq z < 2p$, and $c$ indicates if the reduction happened ($c$ is the quotient of the Euclidean division of $z$ by $p$: $c = 0 \iff z \mod{p} = z ; c = 1 \iff z \mod{p} = z - p$).
The presence of $\ket{c}$ in the outputs ensures the reversibility of the modular reduction.
We emphasize that $n$ denotes the number of bits of $p$.
$z$ is initially coded on $n+1$ qubits.
We present here separately three different implementations of the modular reduction.

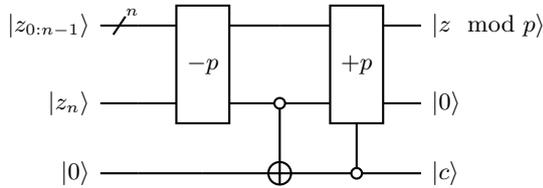
\begin{figure}[h]
\begin{quantikz}
	\lstick{$\ket{z_{0:n-1}}$}&\qwbundle{n}&\gate[2]{-p}&\qw      &\gate[2]{+p}&\rstick{$\ket{z \mod p}$}\qw \\
	\lstick{$\ket{z_{n}}$}    &\qw         &            &\octrl{1}&            &\rstick{$\ket{0}$}       \qw \\
	\lstick{$\ket{0}$}        &\qw         &\qw         &\targ{}  &\octrl{-2}  &\rstick{$\ket{c}$}       \qw \\
\end{quantikz}
 \caption{Modular reduction: first proposition.
	Apart from the reduction, it produces a qubit $\ket{c}$ indicating if the reduction happened, which is the quotient of the Euclidean division of $z$ by $p$.
	The uncomputation of this qubit state depends on the context of the modular reduction.
	The addition and subtraction are here operations modulo $2^{n+1}$.}\label{fig:modular_reduce_1}
\end{figure}

\paragraph{First implementation ---}
The most common implementation~\cite{EkertPRA1996Quantumnetworkselementary,BeauregardQIC2003CircuitShorsalgorithm,Lauter2017QuantumResourceEstimates} consists of subtracting $p$ modulo $2^{n+1}$, copying the most significant bit (that indicates the sign of the result $z-p$) and controlled by the copied bit, adding $p$.
This algorithm is depicted in \autoref{fig:modular_reduce_1}.
The subtraction and the addition are detailed in the following sub-paragraphs.

\begin{figure}[h]
\resizebox{\linewidth}{!}{\begin{quantikz}[column sep=0.3em,row sep=0.8em]
\lstick{$\ket{z_0}$}&\qw                                          &\ctrl{2}  &[1em]\qw                                                                             &[0.2em]\qw                                          &\qw       &\qw        &\qw     &\qw       &\qw       &\qw        &\qw     &\qw        &\qw     &\qw    &\qw       &\qw        &\qw     &\qw                                              &\qw     &[1em]\ctrl{2}  &\targ{}                                          &\rstick{$\ket{s_0}$} \qw \\
\lstick{$p'_0$}     &\cw                                          &\cwbend{0}&     \cw                                                                             &       \cw                                          &\cw       &\cw        &\cw     &\cw       &\cw       &\cw        &\cw     &\cw        &\cw     &\cw    &\cw       &\cw        &\cw     &\cw                                              &\cw     &     \cwbend{0}&\cwbend{-1}                                      &                     \cw \\[1ex]
                    &\lstick[label style={left=-0.4em}]{$\ket{0}$}&\targ{}   &     \qw\gategroup[wires=4,steps=17,style={dashed,rounded corners,inner xsep=.5em}]{}&       \qw                                          &\qw       &\qw        &\ctrl{3}&\qw       &\qw       &\qw        &\qw     &\qw        &\qw     &\qw    &\qw       &\targ{}    &\ctrl{1}&\targ{}                                          &\ctrl{1}&     \targ{}   &\rstick[label style={right=-0.4em}]{$\ket{0}$}\qw&                         \\
\lstick{$\ket{z_1}$}&\qw                                          &\qw       &     \qw                                                                             &       \qw                                          &\ctrl{2}  &\targ{}    &\ctrl{} &\qw       &\qw       &\qw        &\qw     &\qw        &\qw     &\qw    &\qw       &\qw        &\ctrl{2}&\qw                                              &\targ{} &     \qw       &\qw                                              &\rstick{$\ket{s_1}$} \qw \\
\lstick{$p'_1$}     &\cw                                          &\cw       &     \cw                                                                             &       \cw                                          &\cwbend{0}&\cwbend{-1}&\cw     &\cw       &\cw       &\cw        &\cw     &\cw        &\cw     &\cw    &\cwbend{1}&\cwbend{-2}&\cw     &\cwbend{-2}                                      &\cw     &     \cw       &\cw                                              &                     \cw \\
                    &                                             &          &                                                                                     &       \lstick[label style={left=-0.4em}]{$\ket{0}$}&\targ{}   &\qw        &\targ{} &\vqw{1}\qw&          &           &        &           &        &\vqw{1}&\targ{}   &\qw        &\targ{} &\rstick[label style={right=-0.4em}]{$\ket{0}$}\qw&        &               &                                                 &                         \\[1ex]
                    &                                             &          &                                                                                     &                                                    &          &           &        &          &\qw       &\qw        &\ctrl{1}&\qw        &\ctrl{1}&\qw    &          &           &        &                                                 &        &               &                                                 &                         \\
\lstick{$\ket{z_2}$}&\qw                                          &\qw       &     \qw                                                                             &       \qw                                          &\qw       &\qw        &\qw     &\qw       &\ctrl{2}  &\targ{}    &\ctrl{2}&\qw        &\targ{} &\qw    &\qw       &\qw        &\qw     &\qw                                              &\qw     &     \qw       &\qw                                              &\rstick{$\ket{s_2}$} \qw \\
\lstick{$p'_2$}     &\cw                                          &\cw       &     \cw                                                                             &       \cw                                          &\cw       &\cw        &\cw     &\cw       &\cwbend{0}&\cwbend{-1}&\cw     &\cw        &\cw     &\cw    &\cw       &\cw        &\cw     &\cw                                              &\cw     &     \cw       &\cw                                              &                     \cw \\
\lstick{$\ket{z_3}$}&\qw                                          &\qw       &     \qw                                                                             &       \qw                                          &\qw       &\qw        &\qw     &\qw       &\targ{}   &\qw        &\targ{} &\targ{}    &\qw     &\qw    &\qw       &\qw        &\qw     &\qw                                              &\qw     &     \qw       &\qw                                              &\rstick{$\ket{s_3}$} \qw \\
\lstick{$p'_3$}     &\cw                                          &\cw       &     \cw                                                                             &       \cw                                          &\cw       &\cw        &\cw     &\cw       &\cw       &\cw        &\cw     &\cwbend{-1}&\cw     &\cw    &\cw       &\cw        &\cw     &\cw                                              &\cw     &     \cw       &\cw                                              &                     \cw
\end{quantikz} }
\caption{Semiclassical subtraction performing the operation: $\ket{z} \mapsto \ket{z - p \mod{2^{n+1}}}$, with $n$ the number of bits in $p$.
	$z$ has $n+1$ qubits.
	Here we used $p' = 2^{n+1} - p$ and $s=z-p \mod{2^{n+1}}$.
	The boxed block is repeated for each qubit, except the first and last one that are handled with the depicted special cases.
	}\label{fig:semi_classical_subtract}
\end{figure}
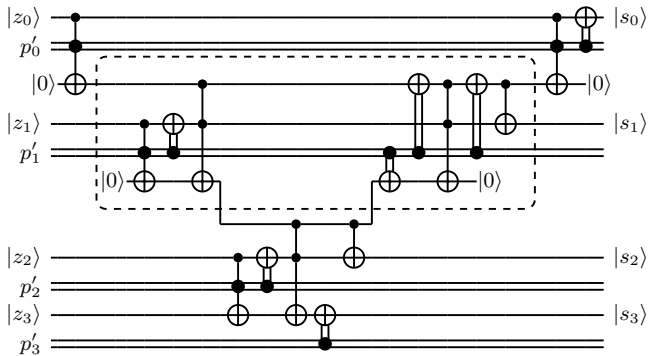

\subparagraph{Semiclassical subtraction}
We here consider the subtraction modulo $2^{n+1}$ of a classical value $p$ from a quantum register encoding $z$, see the first operation in \autoref{fig:modular_reduce_1}.
The basic idea is to add $p'=2^{n+1}-p$ modulo $2^{n+1}$ to the quantum register with the circuit presented in \autoref{fig:semi_classical_subtract}.
As in \autoref{fig:add_mod:add}, this circuit is ripple-carry based, with the main difference that one of its input is a classical number.
The circuit has been adapted to take advantage of this classical input.
It is inspired by~\cite{Moulton2004newquantumripple,Babbush2020CompilationFaultTolerant}.
In the first half of the computation, the carry is computed at each step from the previous carry, the input qubit and the classical number.
During the second half the carries are uncomputed while the result qubits are set to their desired values.
First and last qubits are treated separately as special cases for which optimizations can be used.

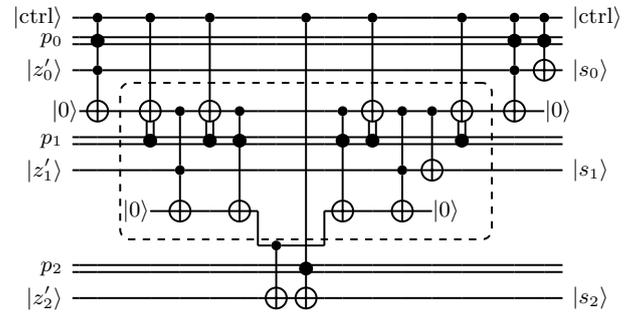
\begin{figure}[h]
\resizebox{0.95\linewidth}{!}{\begin{quantikz}[column sep=0.3em,row sep=0.8em]
  \lstick{$\ket{\text{ctrl}}$}&\qw                                          &\ctrl{2}  &[1em]\ctrl{3}                                                                       &\qw     &\ctrl{3}   &\qw       &\qw       &\qw     &\ctrl{9}  &\qw    &\qw       &\ctrl{3}   &\qw     &\qw                                              &\ctrl{3}   &[1em]\ctrl{2}&\ctrl{2}                                         &\rstick{$\ket{\text{ctrl}}$}\qw\\
  \lstick{$p_0$}              &\cw                                          &\cwbend{0}&\cw                                                                                 &\cw     &\cw        &\cw       &\cw       &\cw     &\cw       &\cw    &\cw       &\cw        &\cw     &\cw                                              &\cw        &\cwbend{0}   &\cwbend{0}                                       &\cw                            \\
  \lstick{$\ket{z'_0}$}       &\qw                                          &\ctrl{1}  &\qw                                                                                 &\qw     &\qw        &\qw       &\qw       &\qw     &\qw       &\qw    &\qw       &\qw        &\qw     &\qw                                              &\qw        &\ctrl{1}     &\targ{}                                          &\rstick{$\ket{s_0}$}\qw        \\
                              &\lstick[label style={left=-0.4em}]{$\ket{0}$}&\targ{}   &\targ{}\gategroup[wires=4,steps=13,style={dashed,rounded corners,inner xsep=.5em}]{}&\ctrl{2}&\targ{}    &\ctrl{3}  &\qw       &\qw     &\qw       &\qw    &\ctrl{3}  &\targ{}    &\ctrl{2}&\ctrl{2}                                         &\targ{}    &\targ{}      &\rstick[label style={right=-0.4em}]{$\ket{0}$}\qw&                               \\
  \lstick{$p_1$}              &\cw                                          &\cw       &\cwbend{-1}                                                                         &\cw     &\cwbend{-1}&\cwbend{0}&\cw       &\cw     &\cw       &\cw    &\cwbend{0}&\cwbend{-1}&\cw     &\cw                                              &\cwbend{-1}&\cw          &\cw                                              &\cw                            \\
  \lstick{$\ket{z'_1}$}       &\qw                                          &\qw       &\qw                                                                                 &\ctrl{1}&\qw        &\qw       &\qw       &\qw     &\qw       &\qw    &\qw       &\qw        &\ctrl{1}&\targ{}                                          &\qw        &\qw          &\qw                                              &\rstick{$\ket{s_1}$}\qw        \\
                              &                                             &          &\lstick[label style={left=-0.4em}]{$\ket{0}$}                                       &\targ{} &\qw        &\targ{}   &\vqw{1}\qw&        &          &\vqw{1}&\targ{}   &\qw        &\targ{} &\rstick[label style={right=-0.4em}]{$\ket{0}$}\qw&           &             &                                                 &                               \\
                              &                                             &          &                                                                                    &        &           &          &          &\ctrl{2}&\qw       &\qw    &          &           &        &                                                 &           &             &                                                 &                               \\
  \lstick{$p_2$}              &\cw                                          &\cw       &\cw                                                                                 &\cw     &\cw        &\cw       &\cw       &\cw     &\cwbend{0}&\cw    &\cw       &\cw        &\cw     &\cw                                              &\cw        &\cw          &\cw                                              &\cw                            \\
  \lstick{$\ket{z'_2}$}       &\qw                                          &\qw       &\qw                                                                                 &\qw     &\qw        &\qw       &\qw       &\targ{} &\targ{}   &\qw    &\qw       &\qw        &\qw     &\qw                                              &\qw        &\qw          &\qw                                              &\rstick{$\ket{s_2}$}\qw
\end{quantikz}
 }
\caption{Semiclassical controlled adder.
	$z'$ is the input number stored in a register of $n+1$ qubits.
	$p$ is known in the control software, and $s = z'+\text{ctrl}.p \mod{2^{n+1}}$.
	The boxed part takes as input the previous carry, the input qubit and classical bit, and conditioned on the control computes the following carry; it is then uncomputed and the output qubit is conditionally set.
	This boxed part is repeated for each input qubit, except the first and last ones, where optimizations are possible.
	Note that in the context of modular reduction, $p$ has $n$ bits and the most significant bit of $s$ takes value $0$.
	However, the presented circuit would work even if this is not the case.
}\label{fig:semiclassical_ctrl_add}
\end{figure}

\subparagraph{Semiclassical controlled addition}
After copying the most significant bit of $z-p \mod{2^{n+1}}$, a controlled semiclassical addition is made in the modular reduction shown in \autoref{fig:modular_reduce_1}.
For simplicity, we write $z'=z-p \mod{2^{n+1}}$ the input of this addition.
The subcircuit of interest thus performs the addition of $p$ modulo $2^{n+1}$ if the control qubit takes value~$1$: $\ket{\text{ctrl}}\ket{z'} \mapsto \ket{\text{ctrl}}\ket{z'+\text{ctrl} \times p \mod{2^{n+1}}}$.
It is realized by the circuit presented in \autoref{fig:semiclassical_ctrl_add}.
As in \autoref{fig:semi_classical_subtract}, we consider a ripple-carry based adder.
The boxed part of \autoref{fig:semiclassical_ctrl_add} conditionally computes the next carry from the previous carry, input qubit and classical bit, and then uncomputes it while setting the result qubit.
It is repeated for each qubit, with a special case for the first and last qubits, for which simplifications are possible.

\begin{figure}[h]
\begin{quantikz}
	\lstick{$\ket{z_{0:n-1}}$}&\qwbundle{n}&\gate[2,style={rounded corners}]{\geq p}\vqw{2}&\gate[2]{-p}&\rstick{$\ket{z \mod p}$}\qw \\
	\lstick{$\ket{z_{n}}$}    &\qw         &                                               &            &\rstick{$\ket{0}$}       \qw \\
	\lstick{$\ket{0}$}        &\qw         &\targ{}                                        &\ctrl{-2}   &\rstick{$\ket{c}$}       \qw \\
\end{quantikz}
 \caption{Modular reduction, second proposition.}\label{fig:modular_reduce_2}
\end{figure}
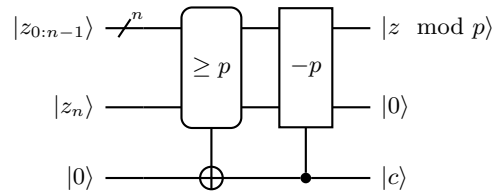

\paragraph{Second implementation ---}
A slightly simpler implementation of the modular reduction consists of comparing first $z$ and $p$, and then subtracting $p$ depending on the result of the comparison, see \autoref{fig:modular_reduce_2}.
We detail below the comparison and the controlled subtraction separately.

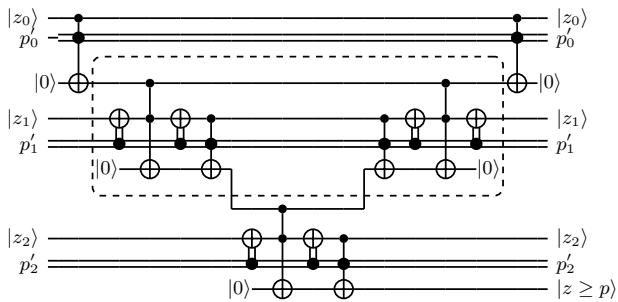
\begin{figure}[h]
\resizebox{0.95\linewidth}{!}{\begin{quantikz}[column sep=0.5em,row sep=0.75em]
	\lstick{$\ket{z_0}$}&\qw                                          &\ctrl{2}  &[0.5em]\qw                                                                             &\qw     &\qw        &\qw       &\qw     &\qw                                          &\qw     &\qw        &\qw       &\qw        &\qw       &\qw        &\qw     &\qw                                              &[0.5em]\ctrl{2}  &\qw                                              &\rstick{$\ket{z_0}$}     \qw\\
	\lstick{$p'_0$}     &\qw                                          &\cwbend{0}&       \cw                                                                             &\cw     &\cw        &\cw       &\cw     &\cw                                          &\cw     &\cw        &\cw       &\cw        &\cw       &\cw        &\cw     &\cw                                              &       \cwbend{0}&\cw                                              &\rstick{$p'_0$}          \cw\\[1em]
	                    &\lstick[label style={left=-0.4em}]{$\ket{0}$}&\targ{}   &       \qw\gategroup[wires=4,steps=14,style={dashed,rounded corners,inner sep=0.5em}]{}&\ctrl{1}&\qw        &\qw       &\qw     &\qw                                          &\qw     &\qw        &\qw       &\qw        &\qw       &\qw        &\ctrl{1}&\qw                                              &       \targ{}   &\qw\rstick[label style={right=-0.4em}]{$\ket{0}$}&                            \\
	\lstick{$\ket{z_1}$}&\qw                                          &\qw       &       \targ{}                                                                         &\ctrl{2}&\targ{}    &\ctrl{2}  &\qw     &\qw                                          &\qw     &\qw        &\qw       &\qw        &\ctrl{2}  &\targ{}    &\ctrl{2}&\targ{}                                          &       \qw       &\qw                                              &\rstick{$\ket{z_1}$}     \qw\\
	\lstick{$p'_1$}     &\cw                                          &\cw       &       \cwbend{-1}                                                                     &\cw     &\cwbend{-1}&\cwbend{0}&\cw     &\cw                                          &\cw     &\cw        &\cw       &\cw        &\cwbend{0}&\cwbend{-1}&\cw     &\cwbend{-1}                                      &       \cw       &\cw                                              &\rstick{$p'_1$}          \cw\\
	                    &                                             &          &       \lstick[label style={left=-0.4em}]{$\ket{0}$}                                   &\targ{} &\qw        &\targ{}   &\qw     &                                             &        &           &          &           &\targ{}   &\qw        &\targ{} &\rstick[label style={right=-0.4em}]{$\ket{0}$}\qw&                 &                                                 &                            \\[0.5em]
	                    &                                             &          &                                                                                       &        &           &          &\vqw{-1}&\qw                                          &\ctrl{1}&\qw        &\qw       &\qw\vqw{-1}&          &           &        &                                                 &                 &                                                 &                            \\
	\lstick{$\ket{z_2}$}&\qw                                          &\qw       &       \qw                                                                             &\qw     &\qw        &\qw       &\qw     &\targ{}                                      &\ctrl{2}&\targ{}    &\ctrl{2}  &\qw        &\qw       &\qw        &\qw     &\qw                                              &       \qw       &\qw                                              &\rstick{$\ket{z_2}$}     \qw\\
	\lstick{$p'_2$}     &\cw                                          &\cw       &       \cw                                                                             &\cw     &\cw        &\cw       &\cw     &\cwbend{-1}                                  &\cw     &\cwbend{-1}&\cwbend{0}&\cw        &\cw       &\cw        &\cw     &\cw                                              &       \cw       &\cw                                              &\rstick{$p'_2$}          \cw\\
	                    &                                             &          &                                                                                       &        &           &          &        &\lstick[label style={left=-0.4em}]{$\ket{0}$}&\targ{} &\qw        &\targ{}   &\qw        &\qw       &\qw        &\qw     &\qw                                              &       \qw       &\qw                                              &\rstick{$\ket{z \geq p}$}\qw
\end{quantikz}
 }
\caption{Semiclassical comparison circuit.
	$z$ and $p$ are compared, and the output bit takes value $1$ if and only if $z \geq p$, while the quantum register containing $z$ is restored to its initial value at the end.
	$z$ is contained in a register of $n+1$~qubits.
	$p'$ is defined as $p' = 2^{n+1} - p$ with a $n$-bit long $p$.
	The basic principle of the circuit is to compute all carries of $z + p'$ successively, the last carry taking value $1$ if and only if $z + p' \geq 2^{n+1} \iff z \geq p$.
	The boxed subcircuit computes the carry with a semiclassical version of the MAJ operation (see \autoref{fig:add_mod:add}) and then uncomputes it with the conjugated circuit; the boxed part is repeated for each bit.
	First and last bits are special cases and associated with subcircuits represented outside the box.
}\label{fig:semiclassical_comparison}
\end{figure}

\subparagraph{Semiclassical comparison}
The first step compares the value $z$ stored in the input quantum register with the number $p>0$ (known in the control software): $\ket{z}\ket{0} \mapsto \ket{z}\ket{z \geq p}$.
It is obtained with the circuit presented in \autoref{fig:semiclassical_comparison}, inspired by~\cite[Fig.\,17]{Babbush2020CompilationFaultTolerant}.
Here $z$ is written in a $n+1$~qubit register and $p$ uses $n$~bits (but the circuit we present would work identically for a $p$ with up to $n+1$ bits).
The principle is to compute the highest bit of $z + p'$ with $p' = 2^{n+1} - p$, which takes the value $1$ if and only if $z + p' \geq 2^{n+1} \iff z \geq p$.
It is obtained by computing all the carries with a semiclassical version of the MAJ operation (see \autoref{fig:add_mod:add}) and then uncomputing the carries with the conjugated circuit.

\subparagraph{Semiclassical controlled subtraction}
The semiclassical controlled subtraction is done using the circuit presented in \autoref{fig:semiclassical_ctrl_add} which implements a semiclassical controlled adder, but with a classical input $p' = 2^{n+1} -p$ (instead of $p$).

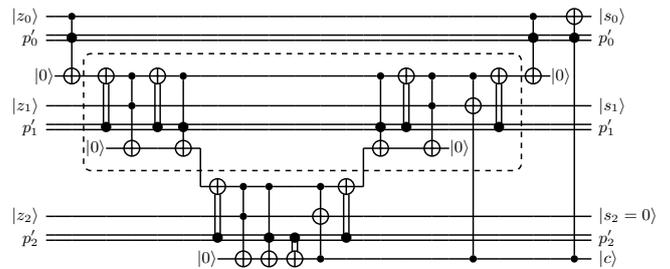
\begin{figure}[h]
\resizebox{\linewidth}{!}{\begin{quantikz}[column sep=0.5em,row sep=0.75em]
	\lstick{$\ket{z_0}$}&\qw                                          &\ctrl{2}  &[0.5em]\qw                                                                                 &\qw     &\qw        &\qw       &\qw     &\qw                                          &\qw     &\qw       &\qw       &\qw      &\qw        &\qw        &\qw       &\qw        &\qw     &\qw                                              &[0.4em]\qw      &\qw        &[0.5em]\ctrl{2}  &\qw                                              &[0.4em]\targ{}   &\rstick{$\ket{s_0}$}    \qw\\
	\lstick{$p'_0$}     &\cw                                          &\cwbend{0}&       \cw                                                                                 &\cw     &\cw        &\cw       &\cw     &\cw                                          &\cw     &\cw       &\cw       &\cw      &\cw        &\cw        &\cw       &\cw        &\cw     &\cw                                              &       \cw      &\cw        &       \cwbend{0}&\cw                                              &       \cwbend{0}&\rstick{$p'_0$}         \cw\\[1em]
	                    &\lstick[label style={left=-0.4em}]{$\ket{0}$}&\targ{}   &       \targ{}\gategroup[wires=4,steps=18,style={dashed,rounded corners,inner sep=0.5em}]{}&\ctrl{1}&\targ{}    &\ctrl{3}  &\qw     &\qw                                          &\qw     &\qw       &\qw       &\qw      &\qw        &\qw        &\ctrl{3}  &\targ{}    &\ctrl{1}&\qw                                              &       \ctrl{1} &\targ{}    &       \targ{}   &\qw\rstick[label style={right=-0.4em}]{$\ket{0}$}&                 &                           \\
	\lstick{$\ket{z_1}$}&\qw                                          &\qw       &       \qw                                                                                 &\ctrl{2}&\qw        &\qw       &\qw     &\qw                                          &\qw     &\qw       &\qw       &\qw      &\qw        &\qw        &\qw       &\qw        &\ctrl{2}&\qw                                              &       \targ{}  &\qw        &       \qw       &\qw                                              &       \qw       &\rstick{$\ket{s_1}$}    \qw\\
	\lstick{$p'_1$}     &\cw                                          &\cw       &       \cwbend{-2}                                                                         &\cw     &\cwbend{-2}&\cwbend{0}&\cw     &\cw                                          &\cw     &\cw       &\cw       &\cw      &\cw        &\cw        &\cwbend{0}&\cwbend{-2}&\cw     &\cw                                              &       \cw      &\cwbend{-2}&       \cw       &\cw                                              &       \cw       &\rstick{$p'_1$}         \cw\\
	                    &                                             &          &       \lstick[label style={left=-0.4em}]{$\ket{0}$}                                       &\targ{} &\qw        &\targ{}   &\qw     &                                             &        &          &          &         &           &           &\targ{}   &\qw        &\targ{} &\qw\rstick[label style={right=-0.4em}]{$\ket{0}$}&                &           &                 &                                                 &                 &                           \\[0.5em]
	                    &                                             &          &                                                                                           &        &           &          &\vqw{-1}&\targ{}                                      &\ctrl{1}&\ctrl{3}  &\qw       &\ctrl{1} &\targ{}    &\qw\vqw{-1}&          &           &        &                                                 &                &           &                 &                                                 &                 &                           \\
	\lstick{$\ket{z_2}$}&\qw                                          &\qw       &       \qw                                                                                 &\qw     &\qw        &\qw       &\qw     &\qw                                          &\ctrl{2}&\qw       &\qw       &\targ{}  &\qw        &\qw        &\qw       &\qw        &\qw     &\qw                                              &       \qw      &\qw        &       \qw       &\qw                                              &       \qw       &\rstick{$\ket{s_2 = 0}$}\qw\\
	\lstick{$p'_2$}     &\cw                                          &\cw       &       \cw                                                                                 &\cw     &\cw        &\cw       &\cw     &\cwbend{-2}                                  &\cw     &\cwbend{0}&\cwbend{1}&\cw      &\cwbend{-2}&\cw        &\cw       &\cw        &\cw     &\cw                                              &       \cw      &\cw        &       \cw       &\cw                                              &       \cw       &\rstick{$p'_2$}         \cw\\
	                    &                                             &          &                                                                                           &        &           &          &        &\lstick[label style={left=-0.4em}]{$\ket{0}$}&\targ{} &\targ{}   &\targ{}   &\ctrl{-2}&\qw        &\qw        &\qw       &\qw        &\qw     &\qw                                              &       \ctrl{-6}&\qw        &       \qw       &\qw                                              &       \ctrl{-9} &\rstick{$\ket{c}$}      \qw
\end{quantikz}
 }
\caption{Modular reduction, producing a garbage qubit $\ket{c}$ that indicates if the reduction happened.}\label{fig:modular_reduce_3}
\end{figure}

\paragraph{Third implementation ---}
In the previous implementation of the modular reduction, the comparison is done by first computing the carries of $z + p'$ successively, then the last carry is copied and finally the carries are uncomputed.
The subtraction is then realized by first computing again the carries of $z + p'$ (controlled on the copy qubit).
The two operations can be merged in order to avoid the computation of the same carries twice.
This requires to change the controlled semiclassical for a version where the control only applies to the UMA part.
The resulting circuit is presented in \autoref{fig:modular_reduce_3}.

This circuit being the most efficient, it is the one we use for the resource estimation.

\subsection{Multiplication}\label{appendix:arithmetic:montgomery_mult}
\subsubsection{Multiplication algorithm}
In Montgomery representation~\cite{MontgomeryMoC1985Modularmultiplicationtrial} with prime number $p$ and radix $2^{n}$, $n$ being the number of bits in $p$, the multiplication of $x$ and $y$ consists of computing $x y 2^{-n} \mod{p}$.
Note that $x$ and $y$ are here in the Montgomery representation.
Here we adopt a version that interleaves the multiplication and division by $2^{n}$ in a windowed manner~\cite{Soeken2020ImprovedQuantumCircuits,Gidney2019Windowedquantumarithmetic}.
More precisely, with $w$ the size of each window, we use the decomposition:
\begin{align*}
&x y 2^{-n} \mod{p} \\
	&= \sum\limits_{\substack{k=0 \\ k \equiv 0 \mod{w}}}^{n-1} \sum\limits_{j=0}^{w-1} x_{k+j} y 2^{j+k-n} \mod{p} \\
	&= \sum\limits_{\substack{k=0 \\ k \equiv 0 \mod{w}}}^{n-1} \left[\sum\limits_{j=0}^{w-1} x_{k+j} y 2^{j}\right]2^{k-n} \mod{p} \\
	&= \begin{multlined}[t]
	 \Big\lbrace \big[\left(x_{0} y + x_{1} 2 y + \cdots + x_{w-1}2^{w-1}y \right)2^{-w} \\
		+ x_{w}y + \cdots + x_{2w-1} 2^{w-1} y\big] 2^{-w} \\ + \cdots \Big\rbrace 2^{-(n \mod{w})} \mod{p}
	\end{multlined} \\
	&= \begin{multlined}[t]
		\Big\lbrace \big[\left(x_{0:w} y\right)2^{-w}
			+ x_{w:2w}y \big] 2^{-w} \\ + \cdots \Big\rbrace 2^{-(n \mod{w})} \mod{p}.
		\end{multlined}
\end{align*}
with $x_{a:b} = \sum_{j=a}^{b-1} x_{j} 2^{j-a}$ (we emphasize that $b$ is excluded) the number made from the slice of bits representing $x$ (from bit $a$ to bit $b-1$).
As the two inputs $x$ and $y$ are in quantum registers, there is no obvious way to use windows to group the controlled additions, and only the modular reduction is done in a windowed way.
Here we implement the multiplication in an out-of-place manner: $\ket{x}\ket{y}\ket{0} \mapsto \ket{x}\ket{y}\ket{x y 2^{-n} \mod{p}}$.
This takes controlled (nonmodular) additions, table lookups and additions (modulo $2^n$), see \autoref{fig:montgomery_mult} which is explained latter.
Before detailing the circuit implementing this multiplication algorithm, we first describe the subcircuits it uses.

\subsubsection{Subcircuits}
\begin{figure}[h]
\resizebox{\linewidth}{!}{\begin{quantikz}[column sep=0.35em,row sep=0.5em]
\lstick{$\ket{\text{ctrl}}$}&\qw     &\qw      &\qw      &\qw     &\qw                                                                                                                                                       &\qw      &\qw     &\qw      &\qw      &\qw     &\ctrl{9}&\qw     &\qw      &\ctrl{7}&\qw      &\qw                                                                                                                                                              &\qw      &\ctrl{5}&\qw      &\qw     &\qw      &\ctrl{3}&\qw      &\qw     &\ctrl{1} &\rstick{$\ket{\text{ctrl}}$}\qw \\
\lstick{$\ket{y_0}$}        &\ctrl{1}&\qw      &\qw      &\qw     &\qw                                                                                                                                                       &\qw      &\qw     &\qw      &\qw      &\qw     &\qw     &\qw     &\qw      &\qw     &\qw      &\qw                                                                                                                                                              &\qw      &\qw     &\qw      &\qw     &\qw      &\qw     &\qw      &\ctrl{1}&\targ{}  &\rstick{$\ket{z_0}$}        \qw \\
\lstick{$\ket{x_0}$}        &\ctrl{1}&\qw      &\qw      &\qw     &\qw                                                                                                                                                       &\qw      &\qw     &\qw      &\qw      &\qw     &\qw     &\qw     &\qw      &\qw     &\qw      &\qw                                                                                                                                                              &\qw      &\qw     &\qw      &\qw     &\qw      &\qw     &\qw      &\ctrl{1}&\ctrl{-1}&\rstick{$\ket{x_0}$}        \qw \\
\lstick{$\ket{0}$}          &\targ{} &\qw      &\targ{}  &\ctrl{1}&\qw                                                                                                                                                       &\qw      &\qw     &\qw      &\qw      &\qw     &\qw     &\qw     &\qw      &\qw     &\qw      &\qw                                                                                                                                                              &\qw      &\qw     &\qw      &\ctrl{1}&\qw      &\ctrl{1}&\targ{}  &\targ{} &\qw      &\rstick{$\ket{0}$}          \qw \\
\lstick{$\ket{y_1}$}        &\qw     &\targ{}  &\qw      &\ctrl{1}&\qw                                                                                                                                                       &\qw      &\qw     &\qw      &\qw      &\qw     &\qw     &\qw     &\qw      &\qw     &\qw      &\qw                                                                                                                                                              &\qw      &\qw     &\qw      &\ctrl{1}&\targ{}  &\targ{} &\qw      &\qw     &\qw      &\rstick{$\ket{z_1}$}        \qw \\
\lstick{$\ket{x_1}$}        &\qw     &\ctrl{-1}&\ctrl{-2}&\targ{} &\qw\gategroup[wires=3,steps=3,style={dashed,rounded corners,inner sep=-1pt},label style={fill=white,label position=below,anchor=north,yshift=-0.7em}]{MAJ}&\targ{}  &\ctrl{1}&\qw      &\qw      &\qw     &\qw     &\qw     &\qw      &\qw     &\qw      &\ctrl{1}\gategroup[wires=3,steps=4,style={dashed,rounded corners,inner sep=-1pt},label style={fill=white,label position=below,anchor=north,yshift=-0.7em}]{C-UMA}&\qw      &\ctrl{1}&\targ{}  &\targ{} &\ctrl{-1}&\qw     &\ctrl{-2}&\qw     &\qw      &\rstick{$\ket{x_1}$}        \qw \\
\lstick{$\ket{y_2}$}        &\qw     &\qw      &\qw      &\qw     &\targ{}                                                                                                                                                   &\qw      &\ctrl{1}&\qw      &\qw      &\qw     &\qw     &\qw     &\qw      &\qw     &\qw      &\ctrl{1}                                                                                                                                                         &\targ{}  &\targ{} &\qw      &\qw     &\qw      &\qw     &\qw      &\qw     &\qw      &\rstick{$\ket{z_2}$}        \qw \\
\lstick{$\ket{x_2}$}        &\qw     &\qw      &\qw      &\qw     &\ctrl{-1}                                                                                                                                                 &\ctrl{-2}&\targ{} &\qw      &\targ{}  &\ctrl{1}&\qw     &\ctrl{1}&\qw      &\ctrl{1}&\targ{}  &\targ{}                                                                                                                                                          &\ctrl{-1}&\qw     &\ctrl{-2}&\qw     &\qw      &\qw     &\qw      &\qw     &\qw      &\rstick{$\ket{x_2}$}        \qw \\
\lstick{$\ket{y_3}$}        &\qw     &\qw      &\qw      &\qw     &\qw                                                                                                                                                       &\qw      &\qw     &\targ{}  &\qw      &\ctrl{1}&\qw     &\ctrl{1}&\targ{}  &\targ{} &\qw      &\qw                                                                                                                                                              &\qw      &\qw     &\qw      &\qw     &\qw      &\qw     &\qw      &\qw     &\qw      &\rstick{$\ket{z_3}$}        \qw \\
\lstick{$\ket{x_3}$}        &\qw     &\qw      &\qw      &\qw     &\qw                                                                                                                                                       &\qw      &\qw     &\ctrl{-1}&\ctrl{-2}&\targ{} &\ctrl{1}&\targ{} &\ctrl{-1}&\qw     &\ctrl{-2}&\qw                                                                                                                                                              &\qw      &\qw     &\qw      &\qw     &\qw      &\qw     &\qw      &\qw     &\qw      &\rstick{$\ket{x_3}$}        \qw \\
\lstick{$\ket{0}$}          &\qw     &\qw      &\qw      &\qw     &\qw                                                                                                                                                       &\qw      &\qw     &\qw      &\qw      &\qw     &\targ{} &\qw     &\qw      &\qw     &\qw      &\qw                                                                                                                                                              &\qw      &\qw     &\qw      &\qw     &\qw      &\qw     &\qw      &\qw     &\qw      &\rstick{$\ket{z_4}$}        \qw
\end{quantikz}
 }
\caption{Controlled adder with output carry.
	It accomplishes the operation $\ket{\text{ctrl}}\ket{x}\ket{y} \mapsto \ket{\text{ctrl}}\ket{x}\ket{z = y+\text{ctrl}.x}$, with the register containing $y$ and $z$ having one extra qubit at the output.
	MAJ and UMA sub-operations have the same meaning as in \autoref{fig:add_mod:add}.
	C-UMA applies UMA when the control qubit takes value $1$ and $\text{MAJ}^{\dagger}$ otherwise.
	MAJ and C-UMA are repeated as many times as required.
}\label{fig:mult:add_ctrl_nomod}
\end{figure}
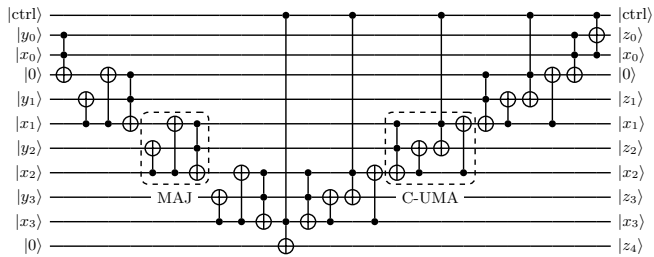

\paragraph{Controlled nonmodular addition ---}
The controlled nonmodular addition between two quantum registers, corresponding to the operation $\ket{\text{ctrl}}\ket{x}\ket{y} \mapsto \ket{\text{ctrl}}\ket{x}\ket{y+\text{ctrl}.x}$, is obtained by using the circuit presented in \autoref{fig:mult:add_ctrl_nomod}.
It works similarly to the adder circuit from \autoref{fig:add_mod:add}, but where the UMA subcircuit is modified to be controlled, that is, UMA is applied when the control qubit has value $1$ while $\text{MAJ}^{\dagger}$ is applied when the control qubit has value $0$.

\paragraph{Table lookup ---}
The table lookup operation, introduced in~\cite{NevenPRX2018EncodingElectronicSpectra} and discussed in~\cite[Fig.\,2]{Gidney2019Windowedquantumarithmetic}, consists of loading into a quantum register some value from a table, known in the control software, according to an index given by a quantum register.
For a table $T$, it performs the following operation: $\ket{k}\ket{0} \mapsto \ket{k}\ket{T[k]}$.
Note in particular that if the address register (the one specifying the index value $k$) is in a superposition, the register loaded with $T[k]$ ends up entangled with the address one, that is $\sum\limits_{k} \alpha_k \ket{k} \ket{0} \mapsto \sum\limits_{k} \alpha_k \ket{k}\ket{T[k]}$.

\begin{figure*}
\begin{quantikz}[row sep=0.5em,column sep=0.25em]
\lstick{$\ket{k_2}$}                         &\ctrl{1}&[1.5em]\qw                                                                                                     &\qw     &\qw        &\qw     &\qw        &\qw      &\qw           &[0.2em]\ctrl{2}&\qw     &\qw        &\qw     &\qw        &\qw      &\ctrl{1} &\targ{}       &\ctrl{1}&\qw     &\qw        &\qw     &\qw        &\qw      &\ctrl{2}      &\qw     &\qw        &\qw     &\qw        &\qw      &\ctrl{1}                                         &\targ{}                                          &\qw \\
\lstick{$\ket{k_1}$}                         &\ctrl{1}&\qw                                                                                                            &\qw     &\qw        &\qw     &\qw        &\qw      &\qw           &\qw            &\qw     &\qw        &\qw     &\qw        &\qw      &\octrl{1}&\qw           &\ctrl{1}&\qw     &\qw        &\qw     &\qw        &\qw      &\qw           &\qw     &\qw        &\qw     &\qw        &\qw      &\octrl{1}                                        &\qw                                              &\qw \\
\lstick[label style={left=-0.4em}]{$\ket{0}$}&\targ{} &\qw\gategroup[wires=4,steps=7,style={dashed,rounded corners,xshift=-0.6em,inner xsep=.4em,inner ysep=-0.2em}]{}&\ctrl{1}&\qw        &\ctrl{2}&\qw        &\ctrl{1} &\qw           &\targ{}        &\ctrl{1}&\qw        &\ctrl{2}&\qw        &\ctrl{1} &\targ{}  &\push{\ket{0}}&\targ{} &\ctrl{1}&\qw        &\ctrl{2}&\qw        &\ctrl{1} &\targ{}       &\ctrl{1}&\qw        &\ctrl{2}&\qw        &\ctrl{1} &\targ{}                                          &\rstick[label style={right=-0.4em}]{$\ket{0}$}\qw&    \\
\lstick{$\ket{k_0}$}                         &\qw     &\qw                                                                                                            &\ctrl{1}&\qw        &\qw     &\qw        &\octrl{1}&\qw           &\qw            &\ctrl{1}&\qw        &\qw     &\qw        &\octrl{1}&\qw      &\qw           &\qw     &\ctrl{1}&\qw        &\qw     &\qw        &\octrl{1}&\qw           &\ctrl{1}&\qw        &\qw     &\qw        &\octrl{1}&\qw                                              &\qw                                              &\qw \\
                                             &        &\lstick[label style={left=-0.4em}]{$\ket{0}$}                                                                  &\targ{} &\ctrl{7}   &\targ{} &\ctrl{7}   &\targ{}  &\push{\ket{0}}&\qw            &\targ{} &\ctrl{7}   &\targ{} &\ctrl{7}   &\targ{}  &\qw      &\push{\ket{0}}&\qw     &\targ{} &\ctrl{7}   &\targ{} &\ctrl{7}   &\targ{}  &\push{\ket{0}}&\targ{} &\ctrl{7}   &\targ{} &\ctrl{7}   &\targ{}  &\rstick[label style={right=-0.4em}]{$\ket{0}$}\qw&                                                 &    \\
                                             &        &                                                                                                               &        &           &        &           &         &              &               &        &           &        &           &         &         &              &        &        &           &        &           &         &              &        &           &        &           &         &                                                 &                                                 &    \\
                                             &        &                                                                                                               &        &           &        &           &         &              &               &        &           &        &           &         &         &              &        &        &           &        &           &         &              &        &           &        &           &         &                                                 &                                                 &    \\
\lstick{$\ket{0}$}                           &\qw     &\qw                                                                                                            &\qw     &\targ{} {?}&\qw     &\targ{} {?}&\qw      &\qw           &\qw            &\qw     &\targ{} {?}&\qw     &\targ{} {?}&\qw      &\qw      &\qw           &\qw     &\qw     &\targ{} {?}&\qw     &\targ{} {?}&\qw      &\qw           &\qw     &\targ{} {?}&\qw     &\targ{} {?}&\qw      &\qw                                              &\qw                                              &\qw \\
\lstick{$\ket{0}$}                           &\qw     &\qw                                                                                                            &\qw     &\targ{} {?}&\qw     &\targ{} {?}&\qw      &\qw           &\qw            &\qw     &\targ{} {?}&\qw     &\targ{} {?}&\qw      &\qw      &\qw           &\qw     &\qw     &\targ{} {?}&\qw     &\targ{} {?}&\qw      &\qw           &\qw     &\targ{} {?}&\qw     &\targ{} {?}&\qw      &\qw                                              &\qw                                              &\qw \\
\lstick{$\ket{0}$}                           &\qw     &\qw                                                                                                            &\qw     &\targ{} {?}&\qw     &\targ{} {?}&\qw      &\qw           &\qw            &\qw     &\targ{} {?}&\qw     &\targ{} {?}&\qw      &\qw      &\qw           &\qw     &\qw     &\targ{} {?}&\qw     &\targ{} {?}&\qw      &\qw           &\qw     &\targ{} {?}&\qw     &\targ{} {?}&\qw      &\qw                                              &\qw                                              &\qw \\
\lstick{$\ket{0}$}                           &\qw     &\qw                                                                                                            &\qw     &\targ{} {?}&\qw     &\targ{} {?}&\qw      &\qw           &\qw            &\qw     &\targ{} {?}&\qw     &\targ{} {?}&\qw      &\qw      &\qw           &\qw     &\qw     &\targ{} {?}&\qw     &\targ{} {?}&\qw      &\qw           &\qw     &\targ{} {?}&\qw     &\targ{} {?}&\qw      &\qw                                              &\qw                                              &\qw \\
\lstick{$\ket{0}$}                           &\qw     &\qw                                                                                                            &\qw     &\targ{} {?}&\qw     &\targ{} {?}&\qw      &\qw           &\qw            &\qw     &\targ{} {?}&\qw     &\targ{} {?}&\qw      &\qw      &\qw           &\qw     &\qw     &\targ{} {?}&\qw     &\targ{} {?}&\qw      &\qw           &\qw     &\targ{} {?}&\qw     &\targ{} {?}&\qw      &\qw                                              &\qw                                              &\qw \\
                                             &        &                                                                                                               &        & T[7]      &        &  T[6]     &         &              &               &        &  T[5]     &        &  T[4]     &         &         &              &        &        &  T[3]     &        &  T[2]     &         &              &        &  T[1]     &        &  T[0]     &         &                                                 &                                                 &
\end{quantikz}
 \caption{Lookup table circuit for an index composed of $3$ qubits and a value register of $5$ qubits.
	The elementary building block of the circuit is boxed; it is used recursively (by replacing the output controls by the same block acting on next qubit from $k$).
	The targets with question mark indicates that the CNOT is applied if the corresponding bit of $T[k]$ has value $1$; otherwise the gate is skipped.
	Further description of this circuit is given in the text.
}\label{fig:lookup_table}
\end{figure*}
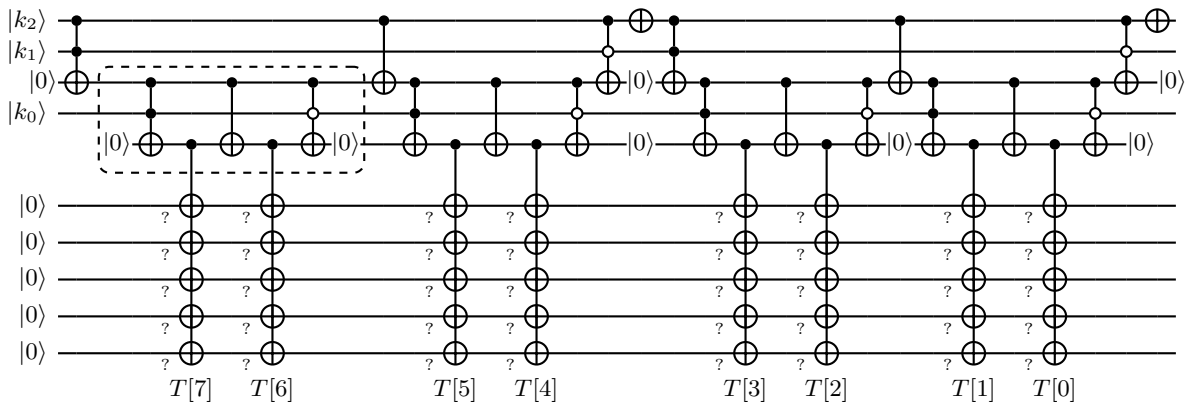

The circuit for performing this operation is presented in \autoref{fig:lookup_table}.
The part in the dashed box has an auxiliary qubit inside which is controlled by two inputs: the top one (initially in $\ket{0}$ and connected to the qubits $\ket{k_1}$ and $\ket{k_2}$ by a Toffoli gate in \autoref{fig:lookup_table}) acts as a control while the second one (in state $\ket{k_0}$ in \autoref{fig:lookup_table}) operates as a selector.
When the control qubit is in state $\ket{0}$, the ancillary qubit always stays in $\ket{0}$.
When the control qubit is in state $\ket{1}$ and the selector qubit is on state $\ket{1}$, the value of the ancillary qubit is first flipped to value $\ket{1}$ (with a first Toffoli) then back to $\ket{0}$ (with a CNOT), while if the selector qubit is on state $\ket{0}$, the ancillary qubit first takes value $\ket{0}$ and then $\ket{1}$.
By recursively using the circuit in the box (with special case for the first qubit), a circuit is obtained such that for each possible value of $k$, the last ancillary qubit takes the value $\ket{1}$ at a given step at which the target register is encoded in $\ket{T[k]}$ and stays otherwise at value $\ket{0}$.
Note that the second Toffoli gate of the boxed block can be replaced by a measurement-based uncomputing, see~\cite[Fig.\,4]{NevenPRX2018EncodingElectronicSpectra} for details.

The preparation of the target registry in $\ket{T[k]}$ (see the lower part of \autoref{fig:lookup_table}) is obtained by applying NOT gates for each bit of $T[k]$ that has the value $1$.
These NOT gates are controlled by the upper part.
Note that lattice surgery allows to merge all those NOT gates with a single control into one gate on all the involved logical qubits.

After use of $T[k]$, the register can be cleared by applying again the same circuit.
However, we consider in our resource estimation a more efficient way based on measurements which has been introduced in~\cite[Appendix~C]{BabbushQ2019QubitizationArbitraryBasis} and improved in~\cite{Gidney2019Windowedquantumarithmetic}.
For more details on the measurement-based cleanup, see~\cite[Appendix~D3]{SangouardPRL2021Factoring2048bit}.  

\begin{figure}[h]
\resizebox{\linewidth}{!}{\begin{quantikz}[column sep=0.35em,row sep=0.5em]
\lstick{$\ket{y_0}$}&\ctrl{1}&\qw      &\qw      &\qw     &\qw                                                                                                       &\qw      &\qw     &\qw      &\qw      &\qw     &\qw     &\qw      &\qw      &\qw     &\qw                                                                                                            &\qw      &\qw     &\qw     &\qw      &\qw     &\ctrl{1}&\targ{}  &\rstick{$\ket{z_0}$}\qw \\
\lstick{$\ket{x_0}$}&\ctrl{1}&\qw      &\qw      &\qw     &\qw                                                                                                       &\qw      &\qw     &\qw      &\qw      &\qw     &\qw     &\qw      &\qw      &\qw     &\qw                                                                                                            &\qw      &\qw     &\qw     &\qw      &\qw     &\ctrl{1}&\ctrl{-1}&\rstick{$\ket{x_0}$}\qw \\
\lstick{$\ket{0}$}  &\targ{} &\qw      &\targ{}  &\ctrl{1}&\qw                                                                                                       &\qw      &\qw     &\qw      &\qw      &\qw     &\qw     &\qw      &\qw      &\qw     &\qw                                                                                                            &\qw      &\qw     &\ctrl{1}&\targ{}  &\ctrl{1}&\targ{} &\qw      &\rstick{$\ket{0}$}  \qw \\
\lstick{$\ket{y_1}$}&\qw     &\targ{}  &\qw      &\ctrl{1}&\qw                                                                                                       &\qw      &\qw     &\qw      &\qw      &\qw     &\qw     &\qw      &\qw      &\qw     &\qw                                                                                                            &\qw      &\qw     &\ctrl{1}&\qw      &\targ{} &\qw     &\qw      &\rstick{$\ket{z_1}$}\qw \\
\lstick{$\ket{x_1}$}&\qw     &\ctrl{-1}&\ctrl{-2}&\targ{} &\qw\gategroup[wires=3,steps=3,style={dashed,rounded corners,inner sep=-1pt},label style={fill=white}]{MAJ}&\targ{}  &\ctrl{1}&\qw      &\qw      &\qw     &\qw     &\qw      &\qw      &\qw     &\ctrl{1}\gategroup[wires=3,steps=3,style={dashed,rounded corners,inner sep=-1pt},label style={fill=white}]{UMA}&\targ{}  &\ctrl{1}&\targ{} &\ctrl{-2}&\qw     &\qw     &\qw      &\rstick{$\ket{x_1}$}\qw \\
\lstick{$\ket{y_2}$}&\qw     &\qw      &\qw      &\qw     &\targ{}                                                                                                   &\qw      &\ctrl{1}&\qw      &\qw      &\qw     &\qw     &\qw      &\qw      &\qw     &\ctrl{1}                                                                                                       &\qw      &\targ{} &\qw     &\qw      &\qw     &\qw     &\qw      &\rstick{$\ket{z_2}$}\qw \\
\lstick{$\ket{x_2}$}&\qw     &\qw      &\qw      &\qw     &\ctrl{-1}                                                                                                 &\ctrl{-2}&\targ{} &\qw      &\targ{}  &\qw     &\ctrl{1}&\qw      &\targ{}  &\ctrl{1}&\targ{}                                                                                                        &\ctrl{-2}&\qw     &\qw     &\qw      &\qw     &\qw     &\qw      &\rstick{$\ket{x_2}$}\qw \\
\lstick{$\ket{y_3}$}&\qw     &\qw      &\qw      &\qw     &\qw                                                                                                       &\qw      &\qw     &\targ{}  &\qw      &\qw     &\ctrl{2}&\qw      &\qw      &\targ{} &\qw                                                                                                            &\qw      &\qw     &\qw     &\qw      &\qw     &\qw     &\qw      &\rstick{$\ket{z_3}$}\qw \\
\lstick{$\ket{x_3}$}&\qw     &\qw      &\qw      &\qw     &\qw                                                                                                       &\qw      &\qw     &\ctrl{-1}&\ctrl{-2}&\ctrl{1}&\qw     &\qw      &\ctrl{-2}&\qw     &\qw                                                                                                            &\qw      &\qw     &\qw     &\qw      &\qw     &\qw     &\qw      &\rstick{$\ket{x_3}$}\qw \\
\lstick{$\ket{y_4}$}&\qw     &\qw      &\qw      &\qw     &\qw                                                                                                       &\qw      &\qw     &\qw      &\qw      &\targ{} &\targ{} &\targ{}  &\qw      &\qw     &\qw                                                                                                            &\qw      &\qw     &\qw     &\qw      &\qw     &\qw     &\qw      &\rstick{$\ket{z_4}$}\qw \\
\lstick{$\ket{x_4}$}&\qw     &\qw      &\qw      &\qw     &\qw                                                                                                       &\qw      &\qw     &\qw      &\qw      &\qw     &\qw     &\ctrl{-1}&\qw      &\qw     &\qw                                                                                                            &\qw      &\qw     &\qw     &\qw      &\qw     &\qw     &\qw      &\rstick{$\ket{x_4}$}\qw
\end{quantikz}

 }
\caption{Adder modulo $2^n$ with $n$ the number of bits of the inputs.
	It accomplishes the operation $\ket{x}\ket{y} \mapsto \ket{x}\ket{z = y + x \mod{2^n}}$.
	MAJ and UMA sub-operations, which are described in the caption of \autoref{fig:add_mod:add}, are repeated for each pair of qubits except the first and two last qubits.
}\label{fig:mult:add}
\end{figure}
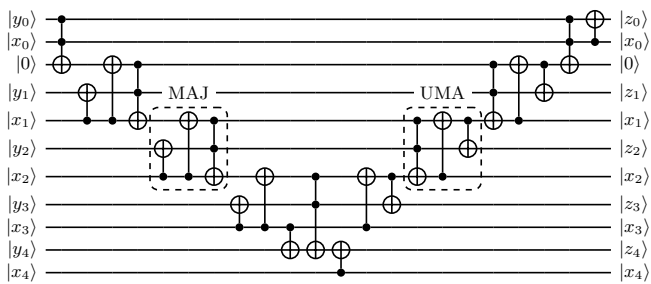

\paragraph{Addition ---}
The addition modulo $2^n$ with $n$ the number of bits in the inputs is obtained with a circuit from~\cite{Moulton2004newquantumripple} presented in \autoref{fig:mult:add}.
The repeated bloc is the same as in the nonmodular addition (\autoref{fig:add_mod:add}), and the circuits differ only for the most significant qubit.

\subsubsection{(Dirty) multiplication circuit}
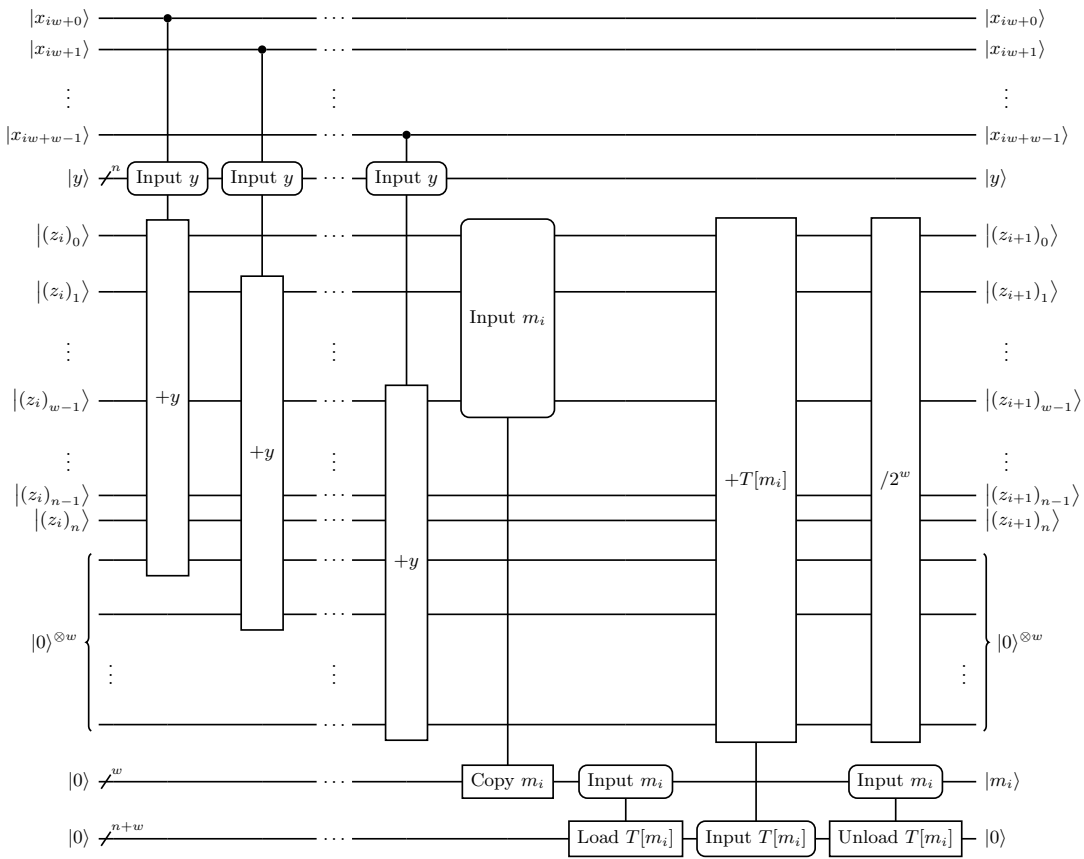
\begin{figure*}
\resizebox{0.8\linewidth}{!}{\begin{quantikz}[column sep=0.75em,row sep=1.2em]
\lstick{$\ket{x_{iw+0}}$}                   &\qw           &\ctrl{4}                                              &\qw                                                   &\push{~\dots~}&\qw                                                   &\qw                                                                  &\qw                                                    &\qw                                                        &\qw                                                    &\rstick{$\ket{x_{iw+0}}$}                    \qw\\
\lstick{$\ket{x_{iw+1}}$}                   &\qw           &\qw                                                   &\ctrl{3}                                              &\push{~\dots~}&\qw                                                   &\qw                                                                  &\qw                                                    &\qw                                                        &\qw                                                    &\rstick{$\ket{x_{iw+1}}$}                    \qw\\
\lstick[label style={left=1em}]{$\vdots$}   &              &                                                      &                                                      &\vdots        &                                                      &                                                                     &                                                       &                                                           &                                                       &\rstick[label style={right=1em}]{$\vdots$}      \\
\lstick{$\ket{x_{iw+w-1}}$}                 &\qw           &\qw                                                   &\qw                                                   &\push{~\dots~}&\ctrl{1}                                              &\qw                                                                  &\qw                                                    &\qw                                                        &\qw                                                    &\rstick{$\ket{x_{iw+w-1}}$}                  \qw\\
\lstick{$\ket{y}$}                          &\qwbundle{n}  &\gate[style={rounded corners}]{\text{Input } y}\vqw{1}&\gate[style={rounded corners}]{\text{Input } y}\vqw{2}&\push{~\dots~}&\gate[style={rounded corners}]{\text{Input } y}\vqw{4}&\qw                                                                  &\qw                                                    &\qw                                                        &\qw                                                    &\rstick{$\ket{y}$}                           \qw\\
\lstick{$\ket{{(z_{i})}_0}$}                &\qw           &\gate[8,nwires={3,5}]{+y}                             &\qw                                                   &\push{~\dots~}&\qw                                                   &\gate[4,nwires={3},style={rounded corners}]{\text{Input }m_i}\vqw{11}&\qw                                                    &\gate[11,nwires={3,5,10}]{+T[m_i]}                         &\gate[11,nwires={3,5,10}]{/2^{w}}                      &\rstick{$\ket{{(z_{i+1})}_0}$}               \qw\\
\lstick{$\ket{{(z_{i})}_1}$}                &\qw           &                                                      &\gate[8,nwires={2,4}]{+y}                             &\push{~\dots~}&\qw                                                   &                                                                     &\qw                                                    &                                                           &                                                       &\rstick{$\ket{{(z_{i+1})}_1}$}               \qw\\
\lstick[label style={left=1em}]{$\vdots$}   &              &                                                      &                                                      &\vdots        &                                                      &                                                                     &                                                       &                                                           &                                                       &\rstick[label style={right=1em}]{$\vdots$}      \\
\lstick{$\ket{{(z_{i})}_{w-1}}$}            &\qw           &                                                      &                                                      &\push{~\dots~}&\gate[8,nwires={2,7}]{+y}                             &                                                                     &\qw                                                    &                                                           &                                                       &\rstick{$\ket{{(z_{i+1})}_{w-1}}$}           \qw\\
\lstick[label style={left=1em}]{$\vdots$}   &              &                                                      &                                                      &\vdots        &                                                      &                                                                     &                                                       &                                                           &                                                       &\rstick[label style={right=1em}]{$\vdots$}      \\
\lstick{$\ket{{(z_{i})}_{n-1}}$}            &\qw           &                                                      &                                                      &\push{~\dots~}&                                                      &\qw                                                                  &\qw                                                    &                                                           &                                                       &\rstick{$\ket{{(z_{i+1})}_{n-1}}$}           \qw\\
\lstick{$\ket{{(z_{i})}_{n}}$}              &\qw           &                                                      &                                                      &\push{~\dots~}&                                                      &\qw                                                                  &\qw                                                    &                                                           &                                                       &\rstick{$\ket{{(z_{i+1})}_{n}}$}             \qw\\
\lstick[4]{$\ket{0}^{\otimes w}$}           &\qw           &                                                      &                                                      &\push{~\dots~}&                                                      &\qw                                                                  &\qw                                                    &                                                           &                                                       &\rstick[4]{$\ket{0}^{\otimes w}$}            \qw\\
                                            &\qw           &\qw                                                   &                                                      &\push{~\dots~}&                                                      &\qw                                                                  &\qw                                                    &                                                           &                                                       &                                             \qw\\
\lstick[label style={left=-1.2em}]{$\vdots$}&              &                                                      &                                                      &\vdots        &                                                      &                                                                     &                                                       &                                                           &                                                       &\rstick[label style={right=-1.2em}]{$\vdots$}   \\
                                            &\qw           &\qw                                                   &\qw                                                   &\push{~\dots~}&                                                      &\qw                                                                  &\qw                                                    &                                                           &                                                       &                                             \qw\\
\lstick{$\ket{0}$}                          &\qwbundle{w}  &\qw                                                   &\qw                                                   &\push{~\dots~}&\qw                                                   &\gate{\text{Copy }m_i}                                               &\gate[style={rounded corners}]{\text{Input }m_i}\vqw{1}&\qw                                                        &\gate[style={rounded corners}]{\text{Input }m_i}\vqw{1}&\rstick{$\ket{m_i}$}                         \qw\\
\lstick{$\ket{0}$}                          &\qwbundle{n+w}&\qw                                                   &\qw                                                   &\push{~\dots~}&\qw                                                   &\qw                                                                  &\gate{\text{Load }T[m_i]}                              &\gate[style={rounded corners}]{\text{Input }T[m_i]}\vqw{-2}&\gate{\text{Unload }T[m_i]}                            &\rstick{$\ket{0}$}                           \qw
\end{quantikz}
 }
\caption{Circuit for one window of the Montgomery multiplication.
	The full multiplication is obtained by repeating it over all the blocks of the windowed representation of $x$, leading to the operation $\ket{x}\ket{y}\ket{0} \mapsto \ket{x}\ket{y}\ket{x y 2^{-n}}$.
	$w$ is the size of the window, $n$ the number of qubits in $x$ and $y$, and $T[m_i] = \left(-m_i p^{-1} \mod{2^{w}}\right) \times p$.
	At the first iteration, the register containing $\ket{z_i}$ starts at $\ket{z_0 = 0}$.
	After the last iteration, a modular reduction is applied to the register of $z$ to obtain the result.
}\label{fig:montgomery_mult}
\end{figure*}

We emphasize that the goal is to compute $x y 2^{-n} \mod{p}$ out-of-place from the following sum of blocks $\Big\lbrace \big[\left(x_{0:w} y\right)2^{-w} + x_{w:2w}y \big] 2^{-w} + \cdots \Big\rbrace 2^{-(n \mod{w})} \mod{p}$.
This is done by repeatedly applying the circuit presented in \autoref{fig:montgomery_mult} (taken from~\cite[Fig.\,4]{Soeken2020ImprovedQuantumCircuits}) for each block $x_{k:k+w}$, with $k \in \lbrace 0, w, 2w, \ldots, \floor{\frac{n}{w}}w \rbrace$, of the windowed decomposition of the input $x$.
In \autoref{fig:montgomery_mult}, we consider a labelling of windows according to $i \in \lbrace 0, 1, 2, \ldots, \floor{\frac{n}{w}} \rbrace$ such that $k = i w$.
Thus, the blocks of $x$ are written $x_{iw:iw+w}$ (keep in mind that the last index $iw+w$ is excluded from the sum; \latin{i.e.\@}, $x_{iw:iw+w} = \sum_{j=iw}^{iw+w-1} x_j 2^{j-iw}$).
The circuit starts by the nonmodular addition of $x_{iw:iw+w}y$ to a target register (which includes the register $\ket{z}$ and $w$ auxiliary qubits as shown in \autoref{fig:montgomery_mult}, all initialized to $\ket{0}$ for $i=0$), by using $w$ shifted additions.
The division of $z_i+x_{iw:iw+w}y$ by $2^{w}$ modulo $p$ is obtained by first adding to $z_i+x_{iw:iw+w}y$ the smallest multiple of $p$ canceling its first $w$ bits and by then performing a standard (nonmodular) division by $2^w$, by forgetting the first $w$ bits with a bit re-labellization.
This is done in five steps.
First, the $w$ less significant bits of the target register, which encode a number we call $m_i$ is copied in a garbage register (second to last register in \autoref{fig:montgomery_mult}).
Second, a table lookup is used to load
$T[m_i] = \left(-m_i p^{-1} \mod{2^{w}}\right) \times p$ in an auxiliary register (last register in \autoref{fig:montgomery_mult}).
$T[m_i]$ is the smallest multiple of $p$ such that $m_i + T[m_i] = 0 \mod{2^{w}}$.
Third, $T[m_i]$ is added to the target register, setting its $w$ lowest bits to $0$.
Fourth, the division by $2^{w}$ is realized by forgetting the first $w$ zeros of the target register and by relabelling the $w+1$th bit as the first bit, the $w+2$th bit as the second bit and so on up to the last bit.  Fifth, the register containing $T[m_i]$ is cleaned.
Once the circuit is applied on each block of the windowed decomposition of $x$, the last step consists of realizing a modular reduction; see \autoref{sec:modular_reduction} for its implementation.
This creates an additional garbage qubit, which is stored in the same way as the $\ket{m_i}$.
Note that for the last iteration, the size of the last window is reduced to the number of remaining qubits.
Further note that the garbage qubits are entangled with the outputs of the circuit.
They are typically cleared by latter applying the conjugate of the whole multiplication circuit, \latin{cf.\@} next subsubsection for an example.

We now discuss the size of the different registers.
By definition, we consider $0 \leq x, y < p $ and since $p < 2^n$, we have $0 \leq x, y < 2^n$.
Let $z$ be the initial value encoded in the target register, as in \autoref{fig:montgomery_mult}.
Let us prove by induction that $0 \leq z < 2p$.
For the first iteration $z=0$ and the bound is satisfied.
Let us assume that $z < 2p$ and prove the same bound after the circuit represented in \autoref{fig:montgomery_mult}.
As $x_{iw:iw+w} \leq 2^{w} - 1$ and $y < p$, we have $x_{iw:iw+w} y < 2^{w}p - p$.
We also have $-m_i p^{-1} \mod{2^{w}} \leq 2^{w} - 1$, hence $T[m_i] \leq 2^{w}p - p$.
We can then conclude
\begin{equation*}
z + x_{iw:iw+w} y + T[m_i] < 2p + p 2^{w} - p + p 2^{w} - p = p 2^{w+1}.
\end{equation*}
After the division by $2^{w}$, we are back to a number strictly smaller than $2p$.
This concludes the proof that the value of the target register satisfies $0 \leq z < 2p$; \latin{i.e.\@}, $n+1$ qubits are sufficient.

Note that in our case, the multiplication circuit is used to compute $x y 2^{-n} \mod{p}$; \latin{i.e.\@}, a final modular reduction is required to go back to a number strictly smaller than $p$.

Further note that the advantage of the Montgomery representation with respect to a standard double-and-add method lies in the fact that it uses modular divisions by two instead of modular doubling.
The former is less expensive to compute as long as we don't clean the ancillary qubits, because the need for a reduction can be assessed only by probing the less significant bit value, while the doubling requires a full comparison (which is costly, even when merged with the reduction as in \autoref{fig:modular_reduce_3}).

\subsubsection{Clean multiplication}
In some situations, a multiplication circuit that does not create garbage qubits is required.
The garbage register in the dirty multiplication circuit presented before can be reset to its initial value $\ket{0}^{\otimes w}$ by copying the result encoded in the target register into another register with CNOT gates, and then applying the conjugate of the multiplication circuit~\cite{BennettIJoRaD1973LogicalReversibilityComputation}, that is
\begin{alignat*}{7}
&\ket{x}\ket{y}\ket{0}\ket{0}\ket{0}
	 &&\xxmapsto{\text{mul}}         &&\ket{x} &&\ket{y} &&\ket{0}        &&\ket{xy2^{-n}} &&\ket{\text{\footnotesize \faTrashO}_{\times}^{(x,y)}} \\
	&&&\xxmapsto{\text{copy}}        &&\ket{x} &&\ket{y} &&\ket{xy2^{-n}} &&\ket{xy2^{-n}} &&\ket{\text{\footnotesize \faTrashO}_{\times}^{(x,y)}} \\
	&&&\xxmapsto{\text{mul}^\dagger} &&\ket{x} &&\ket{y} &&\ket{xy2^{-n}} &&\ket{0}        &&\ket{0}
\end{alignat*}
Note that in some cases where the garbage qubits can be kept temporary, the dirty multiplication is enough (the copy is not needed).

\subsubsection{Squaring}
For implementing an out-of-place squaring operation, note that the circuit of \autoref{fig:montgomery_mult} performs an out-of-place multiplication which does not modify $y$.
Hence, the squaring is obtained by copying the bits of $y$ into the register encoding $x$ with CNOT gates, then applying the multiplication, and finally cleaning the $x$ register with CNOT gates.
This can be done window by window in order to limit the size of the ``$x$'' register to $w$ qubits.
A clean version of this squaring operation is obtained using the strategy used to get the clean multiplication.
To subtract the squaring result from another register, the copy sub-operation in the clean squaring is replaced by a subtraction, obtained by conjugation of the adder circuit from \autoref{fig:mult:add}.

\subsection{Modular inversion}\label{appendix:arithmetic:inversion}
During the elliptic curve group operation, the computation of a slope is necessary.
This takes a modular division, which is presented in this section.

As we use the Montgomery representation to simplify the multiplications, we need to consider an implementation of the inversion which is compatible with this representation.
Concretely, given a number $y$ with a Montgomery representation $x=y 2^{n} \mod{p}$, an in-place version of the modular inversion operates as
$\ket{x} \mapsto \ket{x^{-1} 2^{2n} \mod{p}}$.
As in~\cite{Lauter2017QuantumResourceEstimates,Soeken2020ImprovedQuantumCircuits}, we consider an implementation of the modular inversion in Montgomery representation using Kaliski's algorithm~\cite{KaliskiIToC1995Montgomeryinverseits}.

\paragraph{Kaliski Algorithm ---}

A python ``pseudo-code'' of the algorithm is given by
\begin{algorithm}[H]
\begin{lstlisting}[]
def kaliski(x: int, p: int, n: int):
    u, v, r, s = p, x, 0, 1
    for i in range(0, 2*n):
        # Note: u and v even impossible
        if v == 0:
            r = 2*r elif u v = v//2
            r = 2*r
        elif u u = u//2
            s = 2*s
        # From here, u and v both odd
        elif u > v:
            u = u - v
            r = r + s
            u = u//2
            s = 2*s
        else:
            v = v - u
            s = s + r
            v = v//2
            r = 2*r
    assert u == 1
    assert v == 0
    assert s == p
    if r >= p:
        r = r - p
    return p - r
\end{lstlisting}
\caption{Kaliski algorithm~\cite{python}}\label{algo:kaliski}
\end{algorithm}

This algorithm runs on $u$, $v$, $r$ and $s$.
They are initialized as $u=p$, $v=x$, $r=0$ and $s=1$, with $x < p$.
At the end of the algorithm $u=1$, $v=0$, $s=p$ and $r$ is such that $p-r = x^{-1} 2^{2n} \mod{p}$ is the desired result, as we show below.
The number of iterations $k$ which is required to reach $v=0$ is in between $n$ and $2n$ depending on the input values.
Stopping the algorithm after $2n$ iterations is thus sufficient (note that on a classical computer, we would stop right after $v=0$ and group the remaining modular doublings of $r$).

We now present the main ideas of how to bound the size of registers, compute the number of iterations for the convergence of the algorithm and prove the values taken by the outputs.
More detailed proofs can be found in~\cite{KaliskiIToC1995Montgomeryinverseits}.
In this paragraph, we consider the $k$ first iterations of the algorithm, which lead to $v=0$ (only modular doubling of $r$ happens from the $(k+1)$th iteration, see the first \texttt{if} of the \texttt{for} loop in \autoref{algo:kaliski}).
First note that the equality
\begin{equation}\label{eq:conservation_inversion}
p = us + vr
\end{equation}
is initially satisfied and we can easily check that it holds at each iteration.
We can also prove by induction that $1 \leq s$, $1 \leq u$, $0 \leq v \leq x$, $\gcd(u, v) = \gcd(x, p) = 1$ and that after $i$ iterations of the loop (while the condition $v=0$ is not yet reached):
\begin{subequations}\label{eq:Kaliski_conserv}
\begin{align}
	x r &\equiv -u 2^{i} \mod{p} \label{eq:Kaliski_conserv:r} \\
	x s &\equiv \phantom{-}v 2^{i} \mod{p}.\label{eq:Kaliski_conserv:s}
\end{align}
\end{subequations}
Due to the non-negativity of all integers involved in \autoref{eq:conservation_inversion}, it is clear that for all iterations $i<k$, we have $0 \leq u, v, r, s \leq p$, the upper bound being even strict for $v$ and $r$ (because $1 \leq us$ and $v$ is decreasing).
After the $k$th iteration resulting in $v=0$, we have $0 < r < 2p$.
Those bounds allow us to correctly size the registers containing the different variables.

Next, we can check that the product $uv$ is at least divided by $2$ at each step.
This implies that after $i$ iterations, $uv \leq \frac{px}{2^i} < 2^{2n - i}$.
We can conclude that after $2n$ steps, $uv = 0$ hence $v=0$.
On the other hand, $u+v$ is at most divided by $2$ at each step, hence after $i$ iterations, $u+v \geq \frac{x+p}{2^i} > \frac{p}{2^i} \geq \frac{2^{n-1}}{2^i}$.
Since at the penultimate ($k-1$) iteration $u=v=1$, we can conclude that the number of iterations before reaching $v=0$ is at least $n$.

To prove the terminal values, first note that at the end, $v=0$.
Using $\gcd(u, v) = 1$, we conclude that at the end $u=1$.
Given \autoref{eq:Kaliski_conserv:s} and $1 \leq s \leq p$, we get $s=p$ when $v=0$.
Finally, for the $k$ such that $v=0$ we deduce from \autoref{eq:Kaliski_conserv:r} and $u=1$ that $r \equiv -x^{-1} 2^k \mod{p}$.
The $2n-k$ following loop executions consist in modular doubling, and result in $r = -x^{-1} 2^{2n} \mod{p}$.

Note that numerical tests we performed suggest that the maximum possible value of $k$ could be equal to $2n-1$ instead of $2n$, as presented above.
We did not investigate this further, as this would insignificantly change the cost of the overall algorithm.

Note that in \autoref{algo:kaliski}, the two operations performed in the first two \texttt{else if} are identical.
They are applied on $v$ and $r$ for the first \texttt{else if} and on $u$ and $s$ for the second \texttt{else if}.
Similarly, the four operations in the last two \texttt{else if} are the same and are applied on $u,v,r,s$ and $v,u,s,r$ respectively.
Moreover, in each \texttt{else if}, one division and one multiplication by $2$ are applied.
Instead of performing these operations in a conditional way, we can apply them systematically and make use of controlled swap operations to apply them on the proper register, as proposed in~\cite[Fig.\,7]{Soeken2020ImprovedQuantumCircuits}.
This reduces the number of operations and make their implementation easier as they are not controlled anymore.
The resulting classical formulation of the algorithm is:
\vspace{-1ex}
\begin{algorithm}[H]
\begin{lstlisting}[]
def kaliski_swaps(x: int, p: int, n: int):
    u, v, r, s = p, x, 0, 1
    for i in range(0, 2*n):
        if v == 0:
            r = r*2 continue
        swap = False
        if (u or (u and u > v):
            u, v = v, u
            r, s = s, r
            swap = True
        if u v = v - u
            s = s + r
        v = v//2
        r = 2*r
        if swap:
            u, v = v, u
            r, s = s, r
    return u, v, r, s
\end{lstlisting}
\caption{Kaliski algorithm with swaps~\cite{python}}\label{algo:kaliski_swaps}
\end{algorithm}

\begin{figure*}
\resizebox{0.95\linewidth}{!}{\begin{quantikz}[column sep=1em,row sep=1em]
\lstick{$\ket{u}$}    &\qwbundle{n}&\qw                                      &\qw\slice{1}&\octrl{5}&\qw      &\qw      &\qw\slice{2}&\gate[2,style={rounded corners}]{u > v}\vqw{4}\slice{3}&\swap{1} &\qw\slice{4}&\gate[style={rounded corners}]{\text{Input }u}\vqw{1}&\qw\slice{5}                                         &\qw      &\qw      &\qw      &\qw                   &\qw      &\swap{1} &\qw      &\rstick{$\ket{u}$}  \qw \\
\lstick{$\ket{v}$}    &\qwbundle{n}&\gate[style={rounded corners}]{=0}\vqw{5}&\qw         &\qw      &\octrl{4}&\qw      &\qw         &                                                       &\targX{} &\qw         &\gate{-u}                                            &\qw                                                  &\qw      &\qw      &\gate{/2}&\qw                   &\qw      &\targX{} &\qw      &\rstick{$\ket{v}$}  \qw \\
\lstick{$\ket{r}$}    &\qwbundle{n}&\qw                                      &\qw         &\qw      &\qw      &\qw      &\qw         &\qw                                                    &\qw      &\swap{1}    &\qw                                                  &\gate[style={rounded corners}]{\text{Input }r}\vqw{1}&\qw      &\qw      &\qw      &\gate{\times 2 \mod p}&\swap{1} &\qw      &\qw      &\rstick{$\ket{r}$}  \qw \\
\lstick{$\ket{s}$}    &\qwbundle{n}&\qw                                      &\qw         &\qw      &\qw      &\qw      &\qw         &\qw                                                    &\qw      &\targX{}    &\qw                                                  &\gate{+r}                                            &\qw      &\qw      &\qw      &\qw                   &\targX{} &\qw      &\octrl{2}&\rstick{$\ket{s}$}  \qw \\
\lstick{$\ket{b=0}$}  &\qw         &\qw                                      &\qw         &\qw      &\qw      &\targ{}  &\targ{}     &\octrl{2}                                              &\qw      &\qw         &\octrl{-3}                                           &\octrl{-1}                                           &\targ{}  &\targ{}  &\qw      &\qw                   &\qw      &\qw      &\qw      &\rstick{$\ket{0}$}  \qw \\
\lstick{$\ket{a=0}$}  &\qw         &\qw                                      &\qw         &\targ{}  &\octrl{1}&\ctrl{-1}&\qw         &\targ{}                                                &\ctrl{-5}&\ctrl{-3}   &\qw                                                  &\qw                                                  &\qw      &\ctrl{-1}&\qw      &\qw                   &\ctrl{-3}&\ctrl{-5}&\targ{}  &\rstick{$\ket{0}$}  \qw \\
\lstick{$\ket{m_i=0}$}&\qw         &\targ{}                                  &\ctrl{1}    &\qw      &\targ{}  &\qw      &\ctrl{-2}   &\targ{}                                                &\qw      &\qw         &\qw                                                  &\qw                                                  &\ctrl{-2}&\qw      &\qw      &\qw                   &\qw      &\qw      &\qw      &\rstick{$\ket{m_i}$}\qw \\
\lstick{$\ket{f}$}    &\qw         &\ctrl{-1}                                &\targ{}     &\ctrl{-2}&\ctrl{-1}&\qw      &\qw         &\ctrl{-1}                                              &\qw      &\qw         &\ctrl{-3}                                            &\ctrl{-3}                                            &\qw      &\qw      &\ctrl{-6}&\qw                   &\qw      &\qw      &\qw      &\rstick{$\ket{f}$}  \qw
\end{quantikz}
 }
\caption{One iteration of Kaliski algorithm for modular inversion, with swaps.
	This figure is inspired by~\cite[Fig.\,6b]{Soeken2020ImprovedQuantumCircuits}.
	Controls on a register indicate a control by its least significant bit.
	At the first step, $f=1$.
	The different $\ket{m_i}$ are kept as garbage qubits, and are cleared by applying the conjugate of this circuit.
	See the text for the description of the algorithm and circuit.
}\label{fig:kaliski}
\end{figure*}
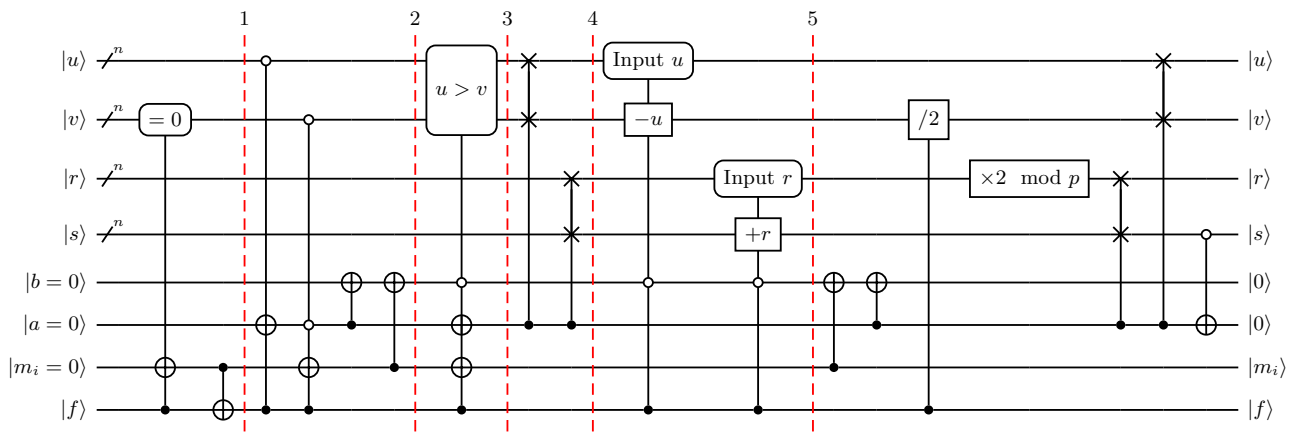

\paragraph{Quantum circuit ---}
The quantum circuit corresponding to one iteration of \autoref{algo:kaliski_swaps} is given in \autoref{fig:kaliski}, which is largely inspired by~\cite[Fig.\,6b]{Soeken2020ImprovedQuantumCircuits}.

Note that $u$, $v$, $r$ and $s$ have the same meaning that in \autoref{algo:kaliski} and must be initialized accordingly.
The last qubit, in state $\ket{f}$, indicates when the ``stop'' condition $v=0$ is reached.
It starts with the value $1$ and is switched to $0$ at the beginning of the first round for which $v=0$, see the first two gates, before step~1.
Once it is set to $0$, only the modular doubling is applied.

Qubits $\ket{a}$, $\ket{b}$ and $\ket{m_i}$ are used to decide which branch of the algorithm is run.
More precisely, as long as $f=1$, those qubits take the values $a=(\text{$u$ is even})$, $b=(\text{$u$ even or $v$ even})$ and $m_i=(\text{$u$ odd and $v$ even})$ before the comparison, see step~2.
Note that the control by a register corresponds to a control by the least significant qubit of the register (control through the parity of the integer encoded in the register).
After the comparison, the values become $a=(\text{$u$ even or [$u$ and $v$ odd and $u>v$]})$, $b=(\text{$u$ even or $v$ even})$ and $m_i=(\text{[$u$ odd and $v$ even] or [$u$ and $v$ odd and $u>v$]})$, see step~3.
Since $u$ and $v$ can't be simultaneously even (\autoref{eq:conservation_inversion} tells us that if $u$ and $v$ are both even, $p$ is even meaning that it is not a prime integer), $a$ corresponds exactly to the condition for applying the swaps in \autoref{algo:kaliski_swaps} (second \texttt{if}).
The logical complement of $b$ is the condition for applying the subtraction and addition, see the gates between step~4 and step~5.

$b$ is uncomputed right after the subtraction and addition by means of the two CNOT gates with $a$ and $m_i$ as controls and $b$ as target.
Note that the contributions to $a$ and $m_i$ from the comparison (between steps~2 and~3) cancel each other when clearing $b$.
The uncomputation of $a$ relies on the swap operations that $\ket{a=1}$ generates.
More precisely, the register $r$ and $s$ are swapped before the subtraction and after the modular multiplication by $2$.
Note that the bound $0 \leq s < p$ has been proven with nonmodular multiplication by $2$, hence the modular reduction never happens.
In addition, $r$ and $s$ can't be simultaneously even; see \autoref{eq:conservation_inversion}.
This guarantees that if and only if the swap happened $s$ is even right before the very last NOT operation on $a$ conditioned on $s=0$, hence, cleaning $\ket{a}$.

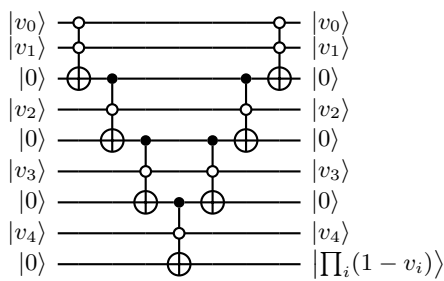
\begin{figure}[h]
\begin{quantikz}[column sep=0.35em,row sep=0.5em]
\lstick{$\ket{v_0}$} & \octrl{1} & \qw       & \qw       & \qw       & \qw       & \qw       & \octrl{1} & \rstick{$\ket{v_0}$}                 \qw \\
\lstick{$\ket{v_1}$} & \octrl{1} & \qw       & \qw       & \qw       & \qw       & \qw       & \octrl{1} & \rstick{$\ket{v_1}$}                 \qw \\
\lstick{$\ket{0}$}   & \targ{}   & \ctrl{1}  & \qw       & \qw       & \qw       & \ctrl{1}  & \targ{}   & \rstick{$\ket{0}$}                   \qw \\
\lstick{$\ket{v_2}$} & \qw       & \octrl{1} & \qw       & \qw       & \qw       & \octrl{1} & \qw       & \rstick{$\ket{v_2}$}                 \qw \\
\lstick{$\ket{0}$}   & \qw       & \targ{}   & \ctrl{1}  & \qw       & \ctrl{1}  & \targ{}   & \qw       & \rstick{$\ket{0}$}                   \qw \\
\lstick{$\ket{v_3}$} & \qw       & \qw       & \octrl{1} & \qw       & \octrl{1} & \qw       & \qw       & \rstick{$\ket{v_3}$}                 \qw \\
\lstick{$\ket{0}$}   & \qw       & \qw       & \targ{}   & \ctrl{1}  & \targ{}   & \qw       & \qw       & \rstick{$\ket{0}$}                   \qw \\
\lstick{$\ket{v_4}$} & \qw       & \qw       & \qw       & \octrl{1} & \qw       & \qw       & \qw       & \rstick{$\ket{v_4}$}                 \qw \\
\lstick{$\ket{0}$}   & \qw       & \qw       & \qw       & \targ{}   & \qw       & \qw       & \qw       & \rstick{$\ket{\prod_{i} (1 - v_i)}$} \qw
\end{quantikz}
 \caption{Multiple-controlled-NOT, with negative control.
	The small white circles designate controls on state $\ket{0}$.
}\label{fig:ccccccnot}
\end{figure}

In \autoref{fig:kaliski}, the first operation is a check of equality with $0$, that is a multiple-controlled-NOT gate.
It can be implemented by applying successively AND gates: a Toffoli gate targeting a clean ancillary qubit which is later uncomputed with another Toffoli gate (or via a measurement and correction as in~\cite[Fig.\,4]{NevenPRX2018EncodingElectronicSpectra}), as depicted in \autoref{fig:ccccccnot}.
Alternatively, it is possible to use borrowed qubits (ancillary qubits which don't need to be initialized and are restored into their original state) as described in~\cite[Lemma~7.2]{WeinfurterPRA1995Elementarygatesquantum}.
Further intuition on how to build multiple-controlled-NOT gates can be found in~\cite{Gidney2015ConstructingLargeControlled}.
As ancillary qubits are available at that point of the algorithm, we consider the successive Toffoli method presented in \autoref{fig:ccccccnot} for the resource evaluation.

The comparison operation between steps~2 and~3 (\autoref{fig:kaliski}) is done with the circuit already presented in \autoref{fig:add_mod:compair_uncompute}.

\begin{figure}[h]
\resizebox{0.95\linewidth}{!}{\begin{quantikz}[column sep=0.35em,row sep=0.5em]
\lstick{$\ket{\text{ctrl}}$}&\qw     &\qw      &\qw      &\qw     &\qw                                                                                                                                                       &\qw      &\qw     &\qw     &\ctrl{8} &\qw     &\qw                                                                                                                                                              &\qw      &\ctrl{5}&\qw      &\qw     &\qw      &\ctrl{3}&\qw      &\qw     &\ctrl{1} &\rstick{$\ket{\text{ctrl}}$}\qw \\
\lstick{$\ket{y_0}$}        &\ctrl{1}&\qw      &\qw      &\qw     &\qw                                                                                                                                                       &\qw      &\qw     &\qw     &\qw      &\qw     &\qw                                                                                                                                                              &\qw      &\qw     &\qw      &\qw     &\qw      &\qw     &\qw      &\ctrl{1}&\targ{}  &\rstick{$\ket{z_0}$}        \qw \\
\lstick{$\ket{x_0}$}        &\ctrl{1}&\qw      &\qw      &\qw     &\qw                                                                                                                                                       &\qw      &\qw     &\qw     &\qw      &\qw     &\qw                                                                                                                                                              &\qw      &\qw     &\qw      &\qw     &\qw      &\qw     &\qw      &\ctrl{1}&\ctrl{-1}&\rstick{$\ket{x_0}$}        \qw \\
\lstick{$\ket{0}$}          &\targ{} &\qw      &\targ{}  &\ctrl{1}&\qw                                                                                                                                                       &\qw      &\qw     &\qw     &\qw      &\qw     &\qw                                                                                                                                                              &\qw      &\qw     &\qw      &\ctrl{1}&\qw      &\ctrl{1}&\targ{}  &\targ{} &\qw      &\rstick{$\ket{0}$}          \qw \\
\lstick{$\ket{y_1}$}        &\qw     &\targ{}  &\qw      &\ctrl{1}&\qw                                                                                                                                                       &\qw      &\qw     &\qw     &\qw      &\qw     &\qw                                                                                                                                                              &\qw      &\qw     &\qw      &\ctrl{1}&\targ{}  &\targ{} &\qw      &\qw     &\qw      &\rstick{$\ket{z_1}$}        \qw \\
\lstick{$\ket{x_1}$}        &\qw     &\ctrl{-1}&\ctrl{-2}&\targ{} &\qw\gategroup[wires=3,steps=3,style={dashed,rounded corners,inner sep=-1pt},label style={fill=white,label position=below,anchor=north,yshift=-0.7em}]{MAJ}&\targ{}  &\ctrl{1}&\qw     &\qw      &\qw     &\ctrl{1}\gategroup[wires=3,steps=4,style={dashed,rounded corners,inner sep=-1pt},label style={fill=white,label position=below,anchor=north,yshift=-0.7em}]{C-UMA}&\qw      &\ctrl{1}&\targ{}  &\targ{} &\ctrl{-1}&\qw     &\ctrl{-2}&\qw     &\qw      &\rstick{$\ket{x_1}$}        \qw \\
\lstick{$\ket{y_2}$}        &\qw     &\qw      &\qw      &\qw     &\targ{}                                                                                                                                                   &\qw      &\ctrl{1}&\qw     &\qw      &\qw     &\ctrl{1}                                                                                                                                                         &\targ{}  &\targ{} &\qw      &\qw     &\qw      &\qw     &\qw      &\qw     &\qw      &\rstick{$\ket{z_2}$}        \qw \\
\lstick{$\ket{x_2}$}        &\qw     &\qw      &\qw      &\qw     &\ctrl{-1}                                                                                                                                                 &\ctrl{-2}&\targ{} &\ctrl{2}&\qw      &\ctrl{2}&\targ{}                                                                                                                                                          &\ctrl{-1}&\qw     &\ctrl{-2}&\qw     &\qw      &\qw     &\qw      &\qw     &\qw      &\rstick{$\ket{x_2}$}        \qw \\
\lstick{$\ket{y_3}$}        &\qw     &\qw      &\qw      &\qw     &\qw                                                                                                                                                       &\qw      &\qw     &\qw     &\targ{}  &\qw     &\qw                                                                                                                                                              &\qw      &\qw     &\qw      &\qw     &\qw      &\qw     &\qw      &\qw     &\qw      &\rstick{$\ket{z_3}$}        \qw \\
\lstick{$\ket{x_3}$}        &\qw     &\qw      &\qw      &\qw     &\qw                                                                                                                                                       &\qw      &\qw     &\targ{} &\ctrl{-1}&\targ{} &\qw                                                                                                                                                              &\qw      &\qw     &\qw      &\qw     &\qw      &\qw     &\qw      &\qw     &\qw      &\rstick{$\ket{x_3}$}        \qw
\end{quantikz}
 }
\caption{Controlled addition modulo $2^n$.
	The output is $z = x + \text{ctrl}.y \mod{2^{n}}$, with $n$ the number of bits in $x$ and $y$ ($n=4$ here).
	MAJ and C-UMA operation are described in \autoref{fig:mult:add_ctrl_nomod}.
	MAJ and C-UMA are repeated as many times as required.
	Note that the controlled addition modulo $2^n$ is simplified compared with the controlled addition presented in \autoref{fig:mult:add_ctrl_nomod} as it does not take an additional qubit for the output (the result of the desired sum is strictly smaller than $2^n$) and the number of operations is reduced.
}\label{fig:controlled_add}
\end{figure}
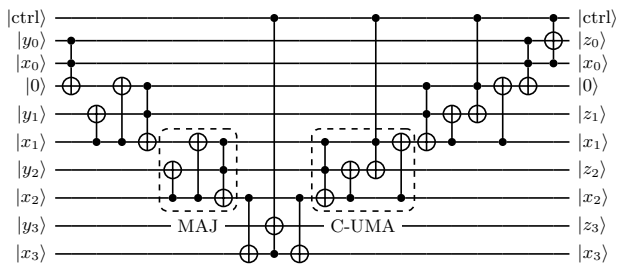

The controlled addition between steps~4 and~5 is performed according to \autoref{fig:controlled_add}, while the subtraction is obtained through the conjugation of this circuit.
It is similar to \autoref{fig:mult:add_ctrl_nomod} except that in \autoref{fig:controlled_add}, the controlled addition is performed modulo $2^n$ with $n$ the number of qubits of the inputs.
The desired addition between $s$ and $r$ is not modular, but we have proven by induction that at the end of any iteration $s, r \leq p$ (except for $r$ after the $k$th iteration, but the corresponding branch does not apply the addition to $r$).
Given that $s$ or $r$ at the iteration $i$ take the value of the sum $s+r$ at round $i-1$, we conclude that $s+r \leq p <2^n$.
Although in \autoref{fig:controlled_add} the addition and subtraction are performed modulo $2^{n}$, we deduce that the properties of \autoref{algo:kaliski_swaps} ensure that no overflow can occur.

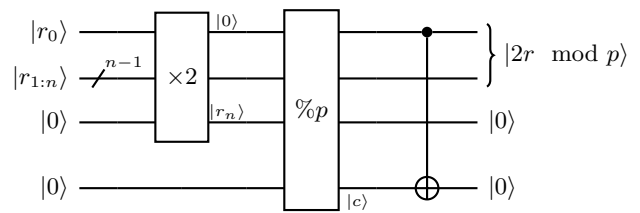
\begin{figure}[h]
\begin{quantikz}[row sep=1em]
\lstick{$\ket{r_0}$}    &\qw           &\gate[3]{\times 2}&\qw [yshift=1em] {\ket{0}}  &\gate[4]{\% p}&\qw           &\ctrl{3}&\rstick[2]{$\ket{2r \mod p}$}  \qw \\
\lstick{$\ket{r_{1:n}}$}&\qwbundle{n-1}&                  &\qw                         &              &\qw           &\qw     &                               \qw \\
\lstick{$\ket{0}$}      &\qw           &                  &\qw [yshift=1em] {\ket{r_n}}&              &\qw           &\qw     &\rstick{$\ket{0}$}             \qw \\
\lstick{$\ket{0}$}      &\qw           &\qw               &\qw                         &              &\qw {\ket{c}} &\targ{} &\rstick{$\ket{0}$}             \qw \\
\end{quantikz} \caption{Doubling modulo $p$, with $p$ odd.
	The multiplication by two is obtained by a simple register shift.
	The modular reduction, represented by $\%p$ operator is done according to \autoref{fig:modular_reduce_3}, while the parity of the result is sufficient to uncompute the qubit $\ket{c}$ indicating if a reduction has been applied.
}\label{fig:double_mod}
\end{figure}

The modular doubling applied to $r$ once $v=0$ in \autoref{fig:kaliski} is implemented with the scheme presented in \autoref{fig:double_mod}.

\begin{figure}[h]
\resizebox{0.95\linewidth}{!}{\begin{quantikz}[column sep=0.3em,row sep=0.8em]
  \lstick{$x_0$}      &\cw                                          &\cwbend{0}\vqw{1}&[1em]\cw                                                                                       &\cw     &\cw       &\cw              &\cw       &\cw       &\cw       &\cw              &\cw     &\cw       &\cw    &\cw              &\cw       &\cw     &\cw                                              &[1em]\cwbend{0}\vqw{1}&\cwbend{1}                                       &\cw                            \\
  \lstick{$\ket{y_0}$}&\qw                                          &\ctrl{1}         &\qw                                                                                            &\qw     &\qw       &\qw              &\qw       &\qw       &\qw       &\qw              &\qw     &\qw       &\qw    &\qw              &\qw       &\qw     &\qw                                              &     \ctrl{1}         &\targ{}                                          &\rstick{$\ket{z_0}$}\qw        \\
                      &\lstick[label style={left=-0.4em}]{$\ket{0}$}&\targ{}          &\qw\gategroup[wires=4,steps=15,style={dashed,rounded corners,inner xsep=.5em}]{}               &\ctrl{2}&\qw       &\qw              &\qw       &\qw       &\qw       &\qw              &\qw     &\qw       &\qw    &\qw              &\qw       &\ctrl{2}&\ctrl{2}                                         &     \targ{}          &\rstick[label style={right=-0.4em}]{$\ket{0}$}\qw&                               \\
  \lstick{$x_1$}      &\cw                                          &\cw              &\cwbend{1}                                                                                     &\cw     &\cwbend{1}&\cwbend{0}\vqw{1}&\cw       &\cw       &\cw       &\cw              &\cw     &\cw       &\cw    &\cwbend{0}\vqw{1}&\cwbend{1}&\cw     &\cw                                              &     \cw              &\cw                                              &\cw                            \\
  \lstick{$\ket{y_1}$}&\qw                                          &\qw              &\targ{}                                                                                        &\ctrl{1}&\targ{}   &\ctrl{1}         &\qw       &\qw       &\qw       &\qw              &\qw     &\qw       &\qw    &\ctrl{1}         &\targ{}   &\ctrl{1}&\targ{}                                          &     \qw              &\qw                                              &\rstick{$\ket{z_1}$}\qw        \\
                      &                                             &                 &\lstick[label style={left=-0.4em}]{$\ket{0}$}                                                  &\targ{} &\qw       &\targ{}          &\vqw{1}\qw&          &          &                 &        &          &\vqw{1}&\targ{}          &\qw       &\targ{} &\rstick[label style={right=-0.4em}]{$\ket{0}$}\qw&                      &                                                 &                               \\
                      &                                             &                 &                                                                                               &        &          &                 &          &\qw       &\qw       &\qw              &\ctrl{2}&\ctrl{2}  &\qw    &                 &          &        &                                                 &                      &                                                 &                               \\
  \lstick{$x_2$}      &\cw                                          &\cw              &\cw                                                                                            &\cw     &\cw       &\cw              &\cw       &\cwbend{1}&\cwbend{4}&\cwbend{0}\vqw{1}&\cw     &\cw       &\cw    &\cw              &\cw       &\cw     &\cw                                              &     \cw              &\cw                                              &\cw                            \\
  \lstick{$\ket{y_2}$}&\qw                                          &\qw              &\qw                                                                                            &\qw     &\qw       &\qw              &\qw       &\targ{}   &\qw       &\ctrl{3}         &\ctrl{3}&\targ{}   &\qw    &\qw              &\qw       &\qw     &\qw                                              &     \qw              &\qw                                              &\rstick{$\ket{z_2}$}\qw        \\
                      &                                             &                 &                                                                                               &        &          &                 &          &          &          &                 &        &          &       &                 &          &        &                                                 &                      &                                                 &                               \\
  \lstick{$x_3$}      &\cw                                          &\cw              &\cw                                                                                            &\cw     &\cw       &\cw              &\cw       &\cw       &\cw       &\cw              &\cw     &\cwbend{1}&\cw    &\cw              &\cw       &\cw     &\cw                                              &     \cw              &\cw                                              &\cw                            \\
  \lstick{$\ket{y_3}$}&\qw                                          &\qw              &\qw                                                                                            &\qw     &\qw       &\qw              &\qw       &\qw       &\targ{}   &\targ{}          &\targ{} &\targ{}   &\qw    &\qw              &\qw       &\qw     &\qw                                              &     \qw              &\qw                                              &\rstick{$\ket{z_3}$}\qw
\end{quantikz}
 }
\caption{Semiclassical addition.
	The result is $z = x + y \mod{2^{n}}$ with $n$ the number of bits in $x$ and $y$.
	The boxed part computes the next carry, and then uncompute it while setting the result qubit to the appropriate value.
	It is repeated for each qubit, except the first and two last (special cases presented in the scheme).
	}\label{fig:semiclassical-add}
\end{figure}
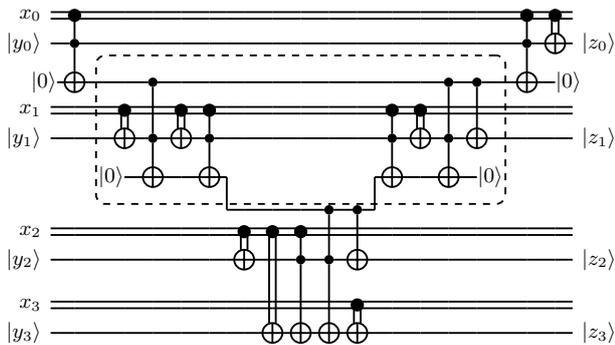

After the circuit presented in \autoref{fig:kaliski} is repeated $2n$ times, the transformation $\ket{r} \mapsto \ket{p-r}$ is applied.
The negation $r \mapsto -r$ is obtained by flipping each qubit and adding $+1$ (which is merged with the addition of $p$), as for a negation in two's complement.
Given that $p$ is a known integer, the addition is obtained by the semiclassical addition circuit presented in \autoref{fig:semiclassical-add}.

The $2n$ repetition of the circuit presented in \autoref{fig:kaliski} together with the transformation $\ket{r} \mapsto \ket{p-r}$ clarifies how to perform the inversion presented in~\cite[Fig.\,6b]{Soeken2020ImprovedQuantumCircuits}.
Note that here we don't need to keep track of the current number of iterations $i$ with a counter as together with $f$ the values $m_i$ are sufficient to ensure reversibility (to indicate if the iteration $k$ has been reached) and the $2n$ repetition of the circuit in \autoref{fig:kaliski} ensures that the doubling is applied $2^{n-k}$ times.
Further note that the modular multiplication by $2$ is merged with the doubling required in the main branch (according to the proven bounds, the modular reduction can happen only once $v=0$, thus the original algorithm is not disturbed).
The comparison step is also simplified, as its result can be used before uncomputing the carries, by using a scheme similar with \autoref{fig:add_mod:compair_uncompute} (with use of the comparison result instead of using it for uncomputation).

\subsection{Division}\label{annexe:arithmetic:division}
Following~\cite{Soeken2020ImprovedQuantumCircuits}, a full out-of-place division $\ket{x}\ket{y} \mapsto \ket{x}\ket{y}\ket{y/x \mod{p}}$ is obtained by combining the inversion and multiplication according to the following protocol
\begin{multline*}
\ket{x}\ket{y}\ket{0}\ket{0}\ket{0}\ket{0} \\
\begin{alignedat}{7}
	&\xxmapsto{\text{inv}}         &&\ket{x^{-1}} &&\ket{y} &&\ket{0}   &&\ket{0}   &&\ket{0}                                                    &&\ket{\text{\footnotesize \faTrashO}_{/}^{(x)}} \\
	&\xxmapsto{\text{mul}}         &&\ket{x^{-1}} &&\ket{y} &&\ket{0}   &&\ket{y/x} &&\ket{\text{\footnotesize \faTrashO}_{\times}^{(x^{-1},y)}} &&\ket{\text{\footnotesize \faTrashO}_{/}^{(x)}} \\
	&\xxmapsto{\text{copy}}        &&\ket{x^{-1}} &&\ket{y} &&\ket{y/x} &&\ket{y/x} &&\ket{\text{\footnotesize \faTrashO}_{\times}^{(x^{-1},y)}} &&\ket{\text{\footnotesize \faTrashO}_{/}^{(x)}} \\
	&\xxmapsto{\text{mul}^\dagger} &&\ket{x^{-1}} &&\ket{y} &&\ket{y/x} &&\ket{0}   &&\ket{0}                                                    &&\ket{\text{\footnotesize \faTrashO}_{/}^{(x)}} \\
	&\xxmapsto{\text{inv}^\dagger} &&\ket{x}      &&\ket{y} &&\ket{y/x} &&\ket{0}   &&\ket{0}                                                    &&\ket{0}
\end{alignedat}
\end{multline*}
where all the operations are modulo $p$, and all the numbers in Montgomery representation.
Note that $\ket{y/x}$ is the step before the last one is the state of the copy register.
$\ket{\text{\footnotesize \faTrashO}_{/}^{(x)}}$ indicates garbage qubits created during the inversion of $x$ (they depend on the value $x$), and $\ket{\text{\footnotesize \faTrashO}_{\times}^{(x^{-1},y)}}$ those generated by the multiplication of $x^{-1}$ by $y$.
The copy operation is done by applying for each qubit a CNOT gate targeting a qubit initialized in $\ket{0}$.
At the end of the process, all ancillary qubits are restored to their initial state.
A controlled version of the division is obtained by controlling the copy, hence changing the CNOT gates into Toffoli gates.

The reader might be interested to look at the circuit presented in~\cite[Fig.\,8b]{Soeken2020ImprovedQuantumCircuits} for an illustration of the division (note that the operation $\times 2^{2n-k}$ presented in~\cite[Fig.\,8b]{Soeken2020ImprovedQuantumCircuits} does not have to be applied since it is already included in the inversion).

\subsection{Elliptic curve scalar multiplication}\label{appendix:arithmetic:elliptic_curve_mult}
The elliptic curve scalar multiplication of an arbitrary integer $k$ by an arbitrary point $P$ is performed in a windowed manner, where $k$ is stored in a quantum register and $P$ is known in the classical control software.
Writing $n_e$ the number of bit in the multiplying factor $k$, we start with a decomposition of $k$ into windows of $w_e$ bits
\begin{equation*}
k = \sum\limits_{\substack{i=0 \\ j \equiv 0 \mod{w_e}}}^{n_e} 2^j k_{j:j+w_e}.
\end{equation*}
where $k_{j:j+w_e}$ is the number composed of the $w_e$ bits of $k$ starting at index $j$ (from least to most significant bits).
Then we can decompose the multiplication operation into a sequence of additions
\begin{equation*}
k P = \sum\limits_{\substack{j=0 \\ j \equiv 0 \mod{w_e}}}^{n_e} 2^j k_{j:j+w_e} P
\end{equation*}
hence the multiplication can be done with $\ceil{\frac{n_e}{w_e}}$ additions into an accumulation register of a point depending on $k_{j:j+w_e}$ loaded from the table associating $i \mapsto 2^j i P$, that contains $2^{w_e}$ different values.
More precisely, $\floor{\frac{n_e}{w_e}}$ additions from a table indexed by $w_e$ qubits are required and eventually one addition from a smaller table indexed by $(n_e \mod{w_e})$ qubits.

An algorithm for performing elliptical curve addition between one point stored in a quantum register and one point loaded from a table into another quantum register is required.
The corresponding circuit is detailed in the following sub-section.
Note that it is also possible to perform the elliptic curve scalar multiplication with controlled semiclassical elliptic curve additions, for details see~\cite{Lauter2017QuantumResourceEstimates,Soeken2020ImprovedQuantumCircuits}.

As the multiplication is performed with successive additions into an accumulation register, if the latter originally contains a point $P_0$, the performed operation is
$\ket{k}\ket{P_0} \mapsto \ket{k} \ket{P_0 + kP}$.

\subsection{Elliptic curve addition}\label{appendix:arithmetic:elliptic_curve_add}
Here we detail how to add into a quantum register a point $P[i]$ loaded from a classical table $i \mapsto P[i]$, with an index $i$ contained in a quantum register: $\ket{i}\ket{Q} \mapsto \ket{i}\ket{Q+P[i]}$, where $Q$ is the point initially contained in the target register.

\begin{figure*}
\resizebox{\linewidth}{!}{\begin{quantikz}[column sep=0.5em,row sep=2em]
\lstick{$\ket{i}$}  &\qwbundle{w_e}    &\qw         &\gate[style={rounded corners}]{\text{Input $i$}}\vqw{3}&\qw                                                            &\qw                                                            &\gate[style={rounded corners}]{\text{Input $i$}}\vqw{3}&\qw                                                    &\qw                                                           &\gate[style={rounded corners}]{\text{Input $i$}}\vqw{4}&\qw                                                     &\gate[style={rounded corners}]{\text{Input $i$}}\vqw{4}&\qw                                                           &\qw                                                           &\qw                                                     &\gate[style={rounded corners}]{\text{Input $i$}}\vqw{3}&\qw                                                            &\qw                                                            &\gate[style={rounded corners}]{\text{Input $i$}}\vqw{3}&\qw               &\qw\\
\lstick{$\ket{x_Q}$}&\qwbundle{n}      &\qw         &\qw                                                    &\gate{-x_{P[i]}}                                               &\qw                                                            &\qw                                                    &\gate[style={rounded corners}]{\text{Input $x$}}\vqw{1}&\gate[style={rounded corners}]{\text{Input $x$}}\vqw{1}       &\qw                                                    &\gate{+a}                                               &\qw                                                    &\gate{-\lambda^2}                                             &\gate[style={rounded corners}]{\text{Input x}}\vqw{1}         &\gate[style={rounded corners}]{\text{Input $x$}}\vqw{1} &\qw                                                    &\gate{-x_{P[i]}}                                               &\qw                                                            &\qw                                                    &\gate{\times (-1)}&\qw\\
\lstick{$\ket{y_Q}$}&\qwbundle{n}      &\qw         &\qw                                                    &\qw                                                            &\gate{-y_{P[i]}}                                               &\qw                                                    &\gate[style={rounded corners}]{\text{Input $y$}}\vqw{1}&\gate{\oplus x \times \lambda}                                &                                                       &                                                        &                                                       &                                                              &\gate[nwires={1}]{\lambda \times x}                           &\gate[style={rounded corners}]{\text{Input $y$}}\vqw{1} &\qw                                                    &\qw                                                            &\gate{-y_{P[i]}}                                               &\qw                                                    &\qw               &\qw\\
                    &\lstick{$\ket{0}$}&\qwbundle{n}&\gate[2]{\text{Load $P[i]$}}                           &\gate[style={rounded corners}]{\text{Input $x_{P[i]}$}}\vqw{-2}&\qw                                                            &\gate[2]{\text{Unload $P[i]$}}                         &\gate[nwires={1}]{\frac{y}{x}}                         &\gate[style={rounded corners}]{\text{Input $\lambda$}}\vqw{-1}&\qw                                                    &\qw                                                     &\qw                                                    &\gate[style={rounded corners}]{\text{Input $\lambda$}}\vqw{-2}&\gate[style={rounded corners}]{\text{Input $\lambda$}}\vqw{-1}&\gate{\oplus \frac{y}{x}}                               &\gate[2,nwires={1,2}]{\text{Load $P[i]$}}              &\gate[style={rounded corners}]{\text{Input $x_{P[i]}$}}\vqw{-2}&\qw                                                            &\gate[2]{\text{Unload $P[i]$}}                         &                  &   \\
                    &\lstick{$\ket{0}$}&\qwbundle{n}&                                                       &\qw                                                            &\gate[style={rounded corners}]{\text{Input $y_{P[i]}$}}\vqw{-2}&                                                       &                                                       &                                                              &\gate[nwires={1}]{\text{Load $3 x_{P[i]}$}}            &\gate[style={rounded corners}]{\text{Input $a$}}\vqw{-3}&\gate{\text{Unload $3 x_{P[i]}$}}                      &                                                              &                                                              &                                                        &                                                       &\qw                                                            &\gate[style={rounded corners}]{\text{Input $y_{P[i]}$}}\vqw{-2}&                                                       &                  &
\end{quantikz} }
\caption{Elliptic curve lookup-addition $\ket{i}\ket{Q} \mapsto \ket{i}\ket{Q+P[i]}$.
	This scheme summarizes \autoref{eq:elliptic_lookup_add}.
	It is very close to the circuit of~\cite[Fig.\,10]{Soeken2020ImprovedQuantumCircuits}.
	Note that the ancillary qubits used by the suboperations are not represented.
}\label{fig:elliptic_lookup_add}
\end{figure*}

Following~\cite{Lauter2017QuantumResourceEstimates}, we only implement the elliptic curve addition for the generic case, where the two points are distinct, not the inverse of each other and none is the neutral element.
As detailed in~\cite{Lauter2017QuantumResourceEstimates}, these exceptional cases happen in a negligible number unless the accumulation register is initialized in zero, which can be avoided by initializing the accumulation register with another point than the neutral element (which has no impact on the measurement statistics after the Fourier transform).

The addition $\ket{i}\ket{Q} \mapsto \ket{i}\ket{Q+P[i]}$ is realized with the following procedure
\begin{widetext}
\settowidth{\arrow}{\scriptsize unlookup}
\begin{equation}\label{eq:elliptic_lookup_add}
\begin{aligned}
\ket{i}\ket{x_Q}\ket{y_Q}
	&\xxmapsto{\text{lookup}} \ket{i}\ket{x_Q}\ket{y_Q}\ket{x_{P[i]}}\ket{y_{P[i]}} \\
	&\xxmapsto{\text{substr}} \ket{i}\ket{x_Q-x_{P[i]}}\ket{y_Q-y_{P[i]}}\ket{x_{P[i]}}\ket{y_{P[i]}} \\
	&\xxmapsto{\text{unlookup}} \ket{i}\ket{x_Q-x_{P[i]}}\ket{y_Q-y_{P[i]}} \\
	&\xxmapsto{\text{div}} \ket{i}\ket{x_Q-x_{P[i]}}\ket{y_Q-y_{P[i]}}\ket{\lambda = \frac{y_Q-y_{P[i]}}{x_Q-x_{P[i]}}} \\
	&\xxmapsto{\text{mul}^\dagger} \ket{i}\ket{x_Q-x_{P[i]}}\ket{0 = (y_Q-y_{P[i]}) - \lambda \times (x_Q-x_{P[i]})}\ket{\lambda} \\
	&\xxmapsto{\text{lookup}} \ket{i}\ket{x_Q-x_{P[i]}}\ket{0}\ket{\lambda}\ket{3 x_{P[i]}} \\
	&\xxmapsto{\text{add}} \ket{i}\ket{x_Q+2x_{P[i]}}\ket{0}\ket{\lambda}\ket{3 x_{P[i]}} \\
	&\xxmapsto{\text{unlookup}} \ket{i}\ket{x_Q+2x_{P[i]}}\ket{0}\ket{\lambda} \\
	&\xxmapsto{\text{square-}} \ket{i}\ket{x_Q+2x_{P[i]}-\lambda^2 = x_{P[i]} - x_{P[i]+Q}}\ket{0}\ket{\lambda} \\
	&\xxmapsto{\text{mul}} \ket{i}\ket{x_{P[i]} - x_{P[i]+Q}}\ket{\lambda \times (x_{P[i]} - x_{P[i]+Q}) = y_{P[i]+Q} + y_{P[i]}}\ket{\lambda} \\
	&\xxmapsto{\text{div}^\dagger} \ket{i}\ket{x_{P[i]} - x_{P[i]+Q}}\ket{y_{P[i]+Q} + y_{P[i]}} \\
	&\xxmapsto{\text{lookup}} \ket{i}\ket{x_{P[i]} - x_{P[i]+Q}}\ket{y_{P[i]+Q} + y_{P[i]}}\ket{x_{P[i]}}\ket{y_{P[i]}} \\
	&\xxmapsto{\text{substr}} \ket{i}\ket{-x_{P[i]+Q}}\ket{y_{P[i]+Q}}\ket{x_{P[i]}}\ket{y_{P[i]}} \\
	&\xxmapsto{\text{unlookup}} \ket{i}\ket{-x_{P[i]+Q}}\ket{y_{P[i]+Q}} \\
	&\xxmapsto{\text{neg}} \ket{i}\ket{x_{P[i]+Q}}\ket{y_{P[i]+Q}}.
\end{aligned}
\end{equation}
\end{widetext}
The process is also represented in \autoref{fig:elliptic_lookup_add}.
Each sub-operation has been previously described.
Note that with the usage of the clean multiplication and division techniques previously presented (see \autoref{appendix:arithmetic:montgomery_mult} and \autoref{annexe:arithmetic:division}), no garbage qubits are created during the elliptic curve addition.

When used as a sub-operation of the elliptic curve scalar multiplication for the $j$th iteration, the specific choice $P[i] = i P 2^{j} = i P'$ with $P' = 2^{j} P$ is made.
This allows to use an additional improvement introduced in~\cite[Sec.\,5.1]{Soeken2020ImprovedQuantumCircuits}.
The idea is to replace $iP'$ by $(i - 2^{w_e - 1})P'$, with the advantage of saving one bit for the control of each lookup, hence saving a factor $2$ for the upper part of \autoref{fig:lookup_table}.
As this modification only subtracts the point $2^{w_e-1} P'$, independently of $i$, it introduces a bijection in the accumulation register that does not modify the interferences exploited by the quantum Fourier transforms.
When $i \geq 2^{w_e-1}$, the most significant bit takes value $1$ while the others encode the number $i - 2^{w_e - 1}$, and using those qubits for controlling the table lookup operation allows loading the correct point.
In the other case, $i < 2^{w_e-1}$ and the most significant bit takes value is $0$.
To prepare the correct point in this case, we start by applying the transformation $i \mapsto 2^{w_e - 1} - i$, then use the same table lookup circuit as in the other case to load $(2^{w_e - 1} - i)P'$, and change the sign of the $y$ coordinate modulo $p$ to negate the point and obtain $(i - 2^{w_e - 1})P'$.
As $i$ is stored in a quantum register, we distinguish the two cases by controlling the modular inversions by the most significant qubit (control on value $\ket{0}$).
This improvement is taken into account in our resource estimation code~\cite{code}.

\subsection{Shor's algorithm}\label{appendix:arithmetic:shor}
As detailed in \autoref{appendix:shor_ekera}, Shor's algorithm for computing the discrete logarithm of $P$ in the cyclic group generated by $G$ relies on the preparation of a superposition of all possible numbers in two registers, the computation of $f(x_1, x_2) = x_1 G - x_2 P$ (with $x_1$ and $x_2$ taken from the two registers) and a Fourier transform.
The more expensive operation is the computation of $f(x_1, x_2)$, which is obtained according to the following procedure
\begin{align*}
\ket{x_1}\ket{x_2}\ket{P_0}
	&\xmapsto{\text{scal mult}} \ket{x_1}\ket{x_2}\ket{P_0 + x_1 G} \\
	&\xmapsto{\text{scal mult}} \ket{x_1}\ket{x_2}\ket{P_0 + x_1 G + x_2 (-P)}
\end{align*}
where the elliptic curve scalar multiplication implementation is detailed in \autoref{appendix:arithmetic:elliptic_curve_mult}.

Note that to correctly implement $f(x_1, x_2)$, $P_0$ should be chosen as the elliptic curve neutral element.
However, as detailed in \autoref{appendix:arithmetic:elliptic_curve_add}, it is preferable to take another value to avoid exceptional cases during the elliptic curve addition (as a translation as no influence on the outcome after the Fourier transform).

\subsection{Gate count}\label{appendix:arithmetic:gate_count}
In this subsection, we present an estimation of the number of CNOT and Toffoli gates in the implementation of elliptic curve discrete logarithm computation.
To keep it simple, we show the dominant order in $n$ (which is the number of bits in the prime $p$).
Note, however, that the resource estimates given in the main text come from an exact resource estimation obtained via a numerical counting (the corresponding code is available at~\cite{code}).

\paragraph{Multiplication ---}
The cost of \autoref{fig:montgomery_mult} is asymptotically dominated by $n$ controlled additions when the choice $w_m \sim \log_2(n)$ is made.
Each addition implemented with \autoref{fig:mult:add_ctrl_nomod} requires at the leading order $4n$~CNOT gates and $3n$~Toffoli.
Hence, Montgomery multiplication uses essentially $4n^2$~CNOT gates and $3n^2$~Toffoli.

A clean multiplication is implemented by running the Montgomery multiplication, coping the result, and running the conjugate of Montgomery multiplication.
Hence, it requires asymptotically $8n^2$~CNOT and $6n^2$~Toffoli.
Note that the squaring operation (eventually including a subtraction) has the same asymptotic cost; \latin{i.e.\@}, it takes about $8n^2$~CNOT and $6n^2$~Toffoli.

\paragraph{Division ---}
We start by evaluating the asymptotic number of gates involved in \autoref{fig:kaliski}.
The first step is a comparison with $0$.
When implemented according to \autoref{fig:ccccccnot}, it uses $2n$~Toffoli gates.
The comparison is implemented following \autoref{fig:add_mod:compair_uncompute} and uses $4n$~CNOT and $2n$~Toffoli.
Each controlled swap (also known as Fredkin gate) can be implemented with $2$~CNOT gates and $1$~Toffoli~\cite{Chuang2010QuantumComputationQuantum}.
Hence, a full register controlled swap involves $2n$~CNOT and $n$~Toffoli.
Each controlled addition or subtraction (\autoref{fig:controlled_add}) uses $4n$~CNOT and $3n$~Toffoli.
The controlled division by two is obtained by repeating controlled swaps and costs $2n$~CNOT and $n$~Toffoli.
The modular multiplication by two (\autoref{fig:double_mod}) is dominated by the modular reduction (\autoref{fig:modular_reduce_3}), and uses asymptotically $n$~CNOT and $3n$~Toffoli on average (we don't count the NOT gates).
By summing those contributions, we conclude that \autoref{fig:kaliski} involves asymptotically $23n$~CNOT and $18n$~Toffoli gates.

The modular inversion is obtained by repeating $2n$ times the \autoref{fig:kaliski}, hence it involves $46n^2$~CNOT and $36n^2$~Toffoli.

The division is obtained by running the inversion, a clean multiplication and the conjugate of the inversion.
Hence, it involves $100n^2$~CNOT and $78n^2$~Toffoli gates.

\paragraph{Elliptic curve addition ---}
In the elliptic curve addition protocol (with one point taken from a lookup circuit), only the two divisions, the two multiplications and the squaring contribute in the asymptotic cost (with the typical choice $w_e = \log_2(n)$ the table lookup does not contribute at the first order).
Hence, it takes asymptotically $224 n^2$~CNOT and $174 n^2$~Toffoli gates.
Note that the choice $w_e = \log_2(n)$ is illustrative, but not optimal (for instance $w_e = 2\log_2(n)$ is a better choice).
In our analysis, the best value for $w_e$ is automatically computed for a given $n$~\cite{code}.

\paragraph{Discrete logarithm computation ---}
As previously seen, the whole algorithm is dominated by the scalar multiplications on elliptic curve.
For Shor's algorithm $x_1$ and $x_2$ have both typically $\approx n$ bits and the two elliptic curve multiplications are almost equivalent to a multiplication of a number containing $n_e \approx 2n$ bits, obtained from $\ceil{\frac{n_e}{w_e}}$ elliptic curve additions.
Hence, the asymptotic number of gates for the whole algorithm is $448 n^3/w_e$~CNOT and $348 n^3/w_e$~Toffoli gates.

\subsection{Qubit count}\label{appendix:arithmetic:qubit_count}
The maximum number of required logical qubits during the discrete logarithm computation is reached at the division step, and more exactly for the modular inversion.

$w_e$ logical qubits are used for the current window of the scalar factor of the elliptic curve multiplication (written $i$ in \autoref{appendix:arithmetic:elliptic_curve_add}).
The two coordinates of the current point of the elliptic curve ($x$ and $y$) each uses $n$ logical qubits.
The register associated with the slope ($\lambda$) also takes $n$ logical qubits.
To summarize, we need $3n+w_e$ logical qubits at this level.

In the modular inversion represented in \autoref{fig:kaliski} $a$, $b$ and $f$ use each one logical qubit, $u$, $v$, $r$ and $s$ are $n$ qubits registers, and we need to store $2n$ different $m_i$ values.
At the first round the input is used as of $v$, so we don't count the corresponding qubits twice (it is an in-place inversion).
Hence, the registers represented in \autoref{fig:kaliski} require $5n+3$ additional qubits.

We also have to take into account the ancillary qubits used by the sub-operations of the inversion.
The more demanding one is the modular multiplication by two, and more precisely the modular reduction.
As shown in \autoref{fig:double_mod}, the extension of $r$ for the nonmodular doubling uses one qubit and another one is required for indicating if the reduction has been applied ($c$).
The circuit shown in \autoref{fig:modular_reduce_3} uses by itself $n$ ancillary qubits.
Note that at this stage of the modular inversion, the register $b$ is available and can be reused.
Hence, the modular doubling adds $n+1$ logical qubits to the count.

By summing up those numbers, we obtain that $9n + w_e + 4$ logical qubits are required for the elliptic curve logarithm computation.
Note that by using a modular reduction based on the semiclassical adder involving only $2$ ancillary qubits presented in~\cite{SvoreQIC2017Factoringusing2n2}, a reduction to $8n + w_e + 6$ logical qubits could be achieved.

\section{Cat qubits, physical gates and noise models}\label{appendix:cat}
\subsection{Cat qubits}\label{appendix:catqubits}
\begin{figure}[h!]
	\resizebox{0.5\textwidth}{!}{\includegraphics[width=\linewidth]{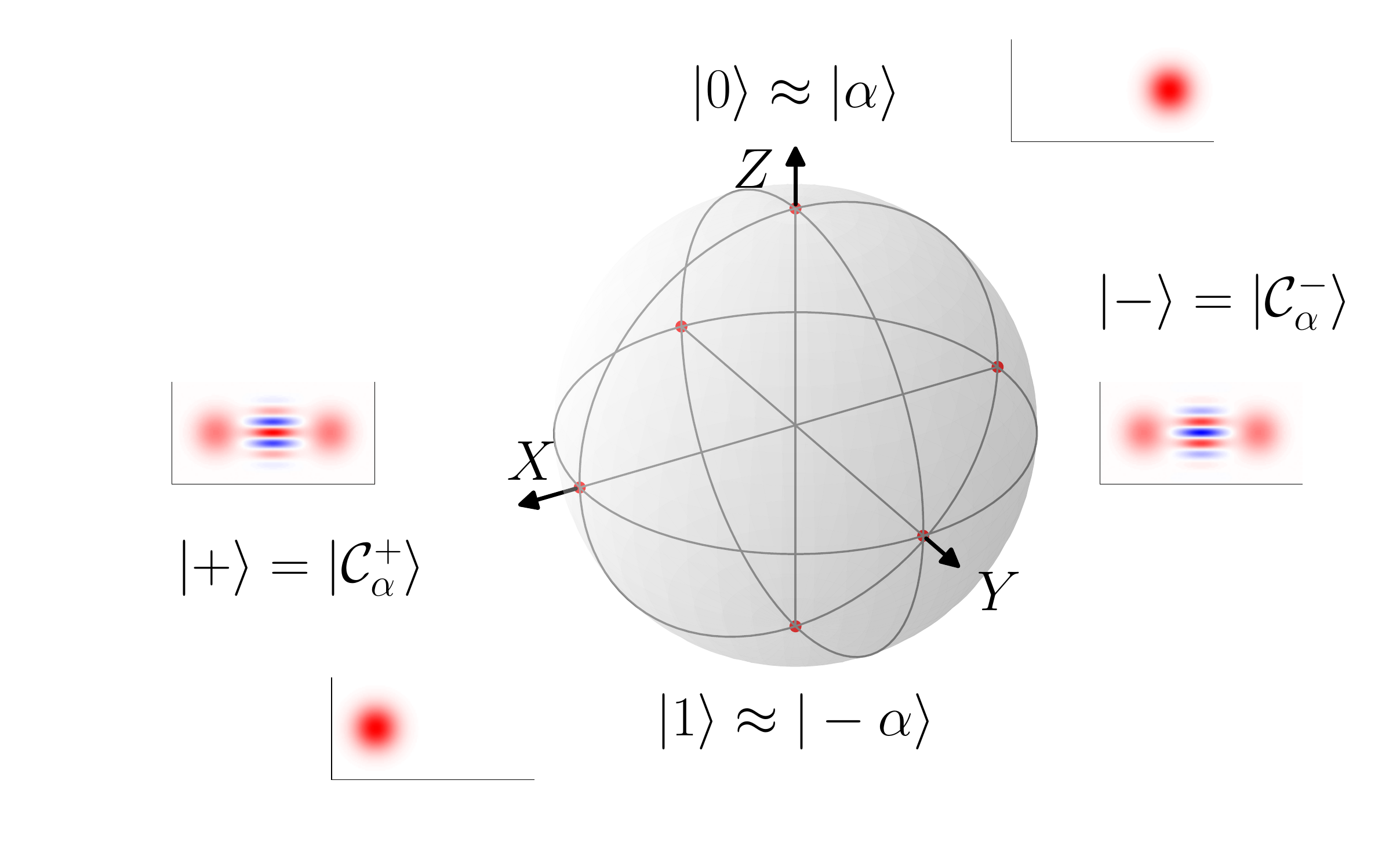}}
	\caption{Bloch sphere representation of a cat qubit.}\label{fig:Bloch_sphere}
\end{figure}

The cat qubit belongs to the family of bosonic qubits~\cite{GaoQSaT2021Quantuminformationprocessing}, for which quantum information is encoded in a subspace of the infinite dimensional Hilbert space of a quantum harmonic oscillator.
The specific choice of the encoding is referred to as the bosonic code.
The (two-component) cat code is defined by two coherent states of same amplitude and opposite phase $\ket{\alpha}$ and $\ket{-\alpha}$~\cite{MunroPRA1999Macroscopicallydistinctquantum,DevoretNJoP2014Dynamicallyprotectedcat}, where $\alpha$ is assumed real without loss of generality.
The amplitude of the cat is related to the average number of photons in the coherent state, $\expval{\had \ha} = \alpha^2$.
An orthonormal basis for the computational subspace --- the \emph{codespace} --- is the superpositions of these two states, the so-called Schrödinger cat states
\begin{align*}
	\Cp &= \frac{1}{\sqrt{2(1+e^{-2\alpha^2})}}(\ket{\alpha} + \ket{-\alpha}) \\
	&= \frac{e^{-0.5\alpha^2}}{\sqrt{2(1+e^{-2\alpha^2})}} \sum \limits_{n} \frac{\alpha^{2n}}{\sqrt{(2n)!}} \ket{2n} \\
	\Cm &= \frac{1}{\sqrt{2(1-e^{-2\alpha^2})}}(\ket{\alpha} - \ket{-\alpha})\\
	&= \frac{e^{-0.5\alpha^2}}{\sqrt{2(1-e^{-2\alpha^2})}} \sum \limits_{n} \frac{\alpha^{2n+1}}{\sqrt{(2n+1)!}} \ket{2n+1},
\end{align*}
which are chosen to be the eigenstates of the Pauli $X$ operator of the cat qubit with eigenvalues $+1$ and $-1$, respectively; see \autoref{fig:Bloch_sphere}.
Note that the computational states $\ket{0}$, $\ket{1}$ are exponentially close (in $\alpha^2$) to the coherent states $\ket{\alpha}$, $\ket{-\alpha}$ as the normalization factor of the superpositions of coherent states are not exactly identical.
To ensure the state of the quantum harmonic oscillator remains in the code space throughout the computation, a stabilization mechanism is required.
In the case of cat qubits, this stabilization is realized by implementing a two-photon dissipation~\cite{DevoretS2015Confiningstatelight,DevoretPRX2018CoherentOscillationsQuantum,LeghtasNP2020Exponentialsuppressionbit,Leghtas2022Onehundredsecond} modelled by the Lindblad master equation
\begin{equation}\label{eq:2photonpumping}
	\dot{\rho} = \kappa_2 \mathcal{D}[\ha^2 - \alpha^2]\rho
\end{equation}
where the superoperator $\mathcal{D}$ is given by
\begin{equation*}
	\mathcal{D}[\hat L]\bullet = \hat{L} \bullet \hat{L}^\dagger - \tfrac{1}{2} \hat{L}^\dagger \hat{L} \bullet - \tfrac{1}{2} \bullet \hat{L}^\dagger \hat{L}
\end{equation*}
and $\kappa_2$ is the rate of the engineered two-photon dissipation.

A major interest of cat qubits lies in the bias of the noise: when the stabilization rate is higher than that of typical errors, the bit-flip error rate induced by the errors of the quantum harmonic oscillator (single-photon loss, thermal excitations, dephasing, \latin{etc.\@}) is exponentially suppressed with the mean number of photon in the cat size $\gamma_X \propto \exp(-2\alpha^2)$, while the phase-flip error rate typically scales linearly $\gamma_Z \propto \alpha^2$~\cite{LeghtasNP2020Exponentialsuppressionbit,Leghtas2022Onehundredsecond}.
As a result, even for moderate cat sizes, the noise bias $\gamma_Z/\gamma_X$ of the cat qubit can be extremely large.

A large noise bias is a highly desirable feature for quantum error correction~\cite{FlammiaPRL2018UltrahighErrorThreshold,FlammiaPRX2019TailoringSurfaceCodes,BrownPRL2020FaultTolerantThresholds}, lowering the requirements on the fault-tolerance threshold, and many recent works have been addressing how to best tailor the codes to leverage the noise bias~\cite{Jiang2022TailoredXZZXcodes,BrownNC2021XZZXsurfacecode,Campbell2022Biastailoredquantum}.
In this work, we assume that the exponential suppression of bit flips in the cat qubits will be sufficient to produce extremely low bit-flip error rate, such that the quantum error correcting code can be used to correct phase-flip errors only~\cite{MirrahimiPRX2019RepetitionCatQubits,BrandaoPQ2022BuildingFaultTolerant}.
In practice however, this assumption may not be realistic, as there could be error mechanisms causing the exponential suppression of bit flips to saturate for some photon number, as has been typically observed experimentally~\cite{LeghtasNP2020Exponentialsuppressionbit,Leghtas2022Onehundredsecond}.
Thus, it may be required in practice to use a code that can correct for a certain amount of bit flips, such as a thin rectangular surface code~\cite{BrandaoPQ2022BuildingFaultTolerant}, but quantifying precisely the extra-overhead induced in this case is beyond the scope of this work (we however provide a rough estimate of this overhead in \autoref{appendix:tableau}).
In the two next paragraphs, we recall for self-completeness the physical implementations of operations on cat qubits.
A thorough discussion of the code operation can be found in~\cite{MirrahimiPRX2019RepetitionCatQubits,BrandaoPQ2022BuildingFaultTolerant}.

\subsection{Physical realization of cat qubits with superconducting circuits}
The physical platform considered in this work for the realization of cat qubits is superconducting circuits in the context of circuit quantum electrodynamics (cQED)~\cite{SchoelkopfN2004Strongcouplingsingle,WallraffRoMP2021Circuitquantumelectrodynamics}.
The bosonic mode used to store quantum information is a mode of a superconducting co-planar waveguide microwave resonator or of a three-dimensional microwave cavity.
The resonators are typically realized using a superconducting material like aluminium or niobium on a dielectric substrate, usually sapphire or silicon.
To operate microwave resonators in the quantum regime, the chips are cooled at a temperature around \SI{10}{\milli\kelvin} in dilution refrigerators.

Universal bosonic quantum information processing requires non-linear dynamics.
In circuit QED, this non-linearity is inherited from the Josephson junction, a non-dissipative non-linear circuit element.
The Hamiltonian of a Josephson junction is given by $\hat H = -E_J \cos(\hat\varphi)$ where $\hat\varphi$ is the phase of the superconducting order parameter across the junction.

The non-linear dynamics required to stabilize cat qubits and process quantum information are realized using the mixing capabilities of the Josephson junction~\cite{MartinisS1988QuantumMechanicsMacroscopic}.
In practice, a cat qubit processor is made of many (linear) bosonic modes hosting the cat qubits; coupled together using non-linear circuit elements built from Josephson junctions.
Quantum information processing is carried out using the non-linear wave-mixing capabilities of Josephson junctions, via the activation of parametric microwave drives and pumps at well-chosen frequencies.

We now briefly illustrate these principles by focusing on the realization of the two-photon dissipation $\mathcal{D}[\hat{a}^2]$, following closely the derivation given in~\cite{LeghtasNP2020Exponentialsuppressionbit}. The realization of this dissipation has been demonstrated experimentally in~\cite{DevoretS2015Confiningstatelight,DevoretPRX2018CoherentOscillationsQuantum,LeghtasNP2020Exponentialsuppressionbit}.
For detailed derivations of more complex operations (weak Hamiltonians required for the Zeno gates~\cite{DevoretPRX2018CoherentOscillationsQuantum} and topological gates), we refer the reader to~\cite{MirrahimiPRX2019RepetitionCatQubits,BrandaoPQ2022BuildingFaultTolerant}.

\begin{figure}[h!]
	\resizebox{0.5\textwidth}{!}{\includegraphics[width=0.9\linewidth]{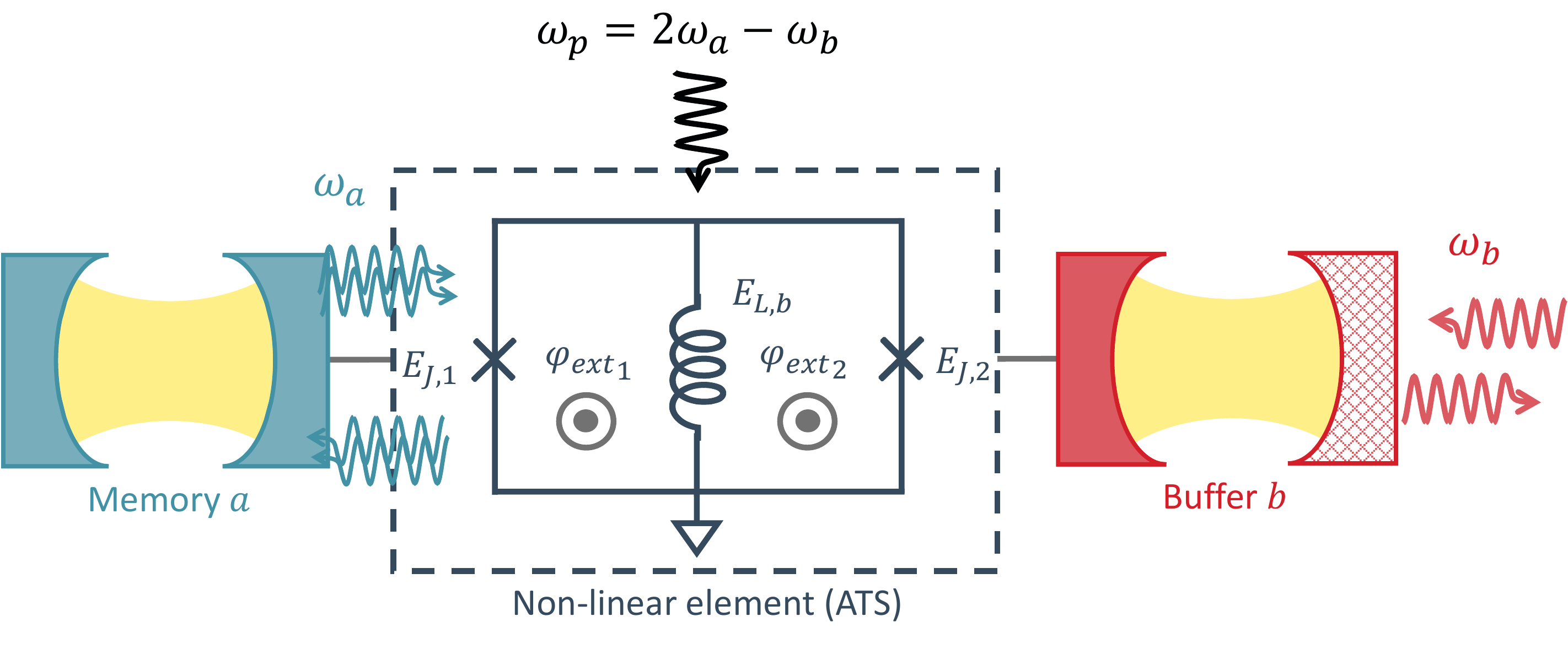}}
	\caption{The cat qubit memory (blue) is coupled to the buffer (red).
	Pumping the ATS at frequency $\omega_p = 2\omega_a - \omega_b$ (black arrow), where $\omega_a$ and $\omega_b$ are the cat qubit and buffer frequencies, activates the conversion of two photons of the cat qubit (blue arrows) into one photon of the buffer (red arrows) and one photon of the pump, and vice-versa.
	}\label{fig:cat_qubit_circuit}
\end{figure}

To realize this dissipation on a high-Q mode $\hat{a}$, called the \emph{memory mode}, a dissipative (low-Q) auxiliary mode $\hat{b}$ called \emph{the buffer mode} is required.
The memory and the buffer mode are coupled using a non-linear circuit element.
In the first experimental realizations of cat qubits, this non-linear circuit element was a simple Josephson junction~\cite{DevoretS2015Confiningstatelight,DevoretPRX2018CoherentOscillationsQuantum}.
Later, a more refined circuit containing more Josephson junctions called the Asymmetrically Threaded SQUID (ATS)~\cite{LeghtasNP2020Exponentialsuppressionbit} was used.
The advantage of this more elaborate circuit is that it allows to engineer the desired two-to-one photon conversion without the creation of many spurious terms that spoil the cat qubit.
The first experimental demonstration of the exponential suppression of bit flips was realized using this circuit, which is why we illustrate the principles using this specific realization, as depicted in \autoref{fig:cat_qubit_circuit}.
An ATS is a SQUID shunted in its center by a large inductance, thus forming two loops with flux $\varphi_{\mathrm{ext}, 1}$ and $\varphi_{\mathrm{ext}, 2}$.
Its potential energy is a function of the phase $\varphi$ across the inductor
\begin{multline}
	U(\varphi)=\frac{1}{2} E_{L, b} \varphi^2
		-E_{J, 1} \cos\left(\varphi+\varphi_{\mathrm{ext}, 1}\right) \\
		-E_{J, 2} \cos\left(\varphi+\varphi_{\mathrm{ext}, 2}\right).
\end{multline}

By writing the half-sum $\varphi_{\Sigma}=\frac{1}{2}\left(\varphi_{\mathrm{ext}, 1}+\varphi_{\mathrm{ext}, 2}\right)$ and the half-difference $\varphi_{\Delta}=\frac{1}{2}\left(\varphi_{\mathrm{ext}, 1}-\varphi_{\mathrm{ext}, 2}\right)$ and setting them such that $\varphi_{\Sigma}=\pi / 2 + \epsilon(t)=\pi / 2+\epsilon_0 \cos \left(\omega_p t\right)$ and $\varphi_{\Delta}=\pi / 2$ with $\epsilon_0$ small, and assuming $E_{J, 1}=E_{J, 2}=E_J$, the time-dependent potential energy becomes
\begin{equation}\label{eq:sine}
U(\varphi)=\frac{1}{2} E_{L, b} \varphi^2-2 E_J \epsilon(t) \sin(\varphi).
\end{equation}

The non-linear coupling of the memory mode $\hat a$ to the buffer mode $\hat b$ by an ATS is described by the Hamiltonian ($\hbar = 1$)
\begin{equation*}\label{eq:ATS_coupling_Hamiltonian}
	\hat{H}=\omega_{a,0}\hat{a}^\dagger\hat{a} + \omega_{b,0}\hat{b}^\dagger\hat{b} - 2E_J \epsilon(t)\sin(\hat\varphi_a + \hat\varphi_b)
\end{equation*}
where $\omega_{a/b,0}$ are the bare resonance frequencies of the modes, $\hat\varphi_a = \varphi_a(\hat{a} + \hat{a}^\dagger)$, $\hat\varphi_b = \varphi_b(\hat{b} + \hat{b}^\dagger)$ and $\varphi_{a,b}$ are the zero point fluctuations of the modes across the ATS dipole, related to the geometry of the circuit.

Setting the pump frequency to be $\omega_p = 2\omega_a - \omega_b$, where $\omega_{a,b}$ are the renormalized (AC-Stark shifted) frequencies of the mode due to the presence of the pump, expanding the sine in~\autoref{eq:ATS_coupling_Hamiltonian} up to third order, and neglecting the fast-oscillating terms (the so-called RWA approximation), one finds\cite{LeghtasNP2020Exponentialsuppressionbit} that the above Hamiltonian reduces to
\begin{equation*}
\hat{H}=g_2^* \hat{a}^2 \hat{b}^{\dagger}+g_2 \hat{a}^{\dagger 2} \hat{b}
\end{equation*}
where $g_2 = E_J\epsilon_0\varphi_a^2\varphi_b/2$.
Finally, the buffer mode $\hat{b}$ being highly dissipative, its fast relaxation dynamics can be adiabatically eliminated to obtain an effective single-mode description of the slower dynamics induced on the memory mode, which takes the form of the effective two-photon dissipation $\mathcal{D}[\hat{a}^2]$.

These principles can be used to implement all the pieces of dynamics of the next subsection.
By using resonant pumps at well-chosen frequencies, one can parametrically engineer more complex interaction Hamiltonians to perform gates on the cat qubits~\cite{MirrahimiPRX2019RepetitionCatQubits,BrandaoPQ2022BuildingFaultTolerant}.

\subsection{Physical gates on cat qubits}
The architecture relies by design on the noise bias of the cat qubit.
It is thus of crucial importance to maintain this bias in the noise, even during the execution of the gates.
Operations that preserve the noise bias are called bias-preserving, and only such operations may be used at the physical level.

\paragraph{State preparation and measurement ---}
The eigenstates of the $X$ operator having a well-defined photon-number parity, ${(-1)}^{\had\ha}\Cpm = \pm \Cpm$, the measurement of $X$ can be realized with a photon-number parity measurement, routinely achieved in circuit QED~\cite{HarochePRL2002DirectMeasurementWigner}.
The preparation of $\Cp$ may be simply realized by preparing the oscillator in the vacuum state and by turning on the cat qubit stabilization.
Alternatively, a faster state preparation can be realized using optimal control.
The $Z$ measurement distinguishes the coherent states $\ket{\pm \alpha}$, which can be done \latin{e.g.\@} with a homodyne measurement.
The eigenstates of the $Z$ operator $\ket{\pm \alpha}$ can be reliably prepared by applying a strong resonant microwave pulse to the oscillator prepared in the vacuum state to displace the oscillator and by turning on the stabilization immediately after the displacement.

\paragraph{Single qubit gates ---}
The implementation of the single qubit $Z$ gate is realized using the quantum Zeno effect, by applying a weak resonant drive of complex amplitude $\epsilon_Z$ to the cat qubit in presence of the stabilization (at rate $\kappa_2 \gg \abs{\epsilon_Z}$), as was first proposed in~\cite{DevoretNJoP2014Dynamicallyprotectedcat}.
The dynamics implementing this gate is modelled by the master equation
\begin{equation}\label{eq:zeno_Z}
	\dot{\rho} = -i \comm{\epsilon_Z \ha + \epsilon_Z^*\had}{\rho} + \kappa_2 \mathcal{D}[\ha^2 - \alpha^2]\rho.
\end{equation}

A $\pi$-rotation around the $Z$ axis is realized in time $T = \pi/[4\alpha \Re(\epsilon_Z)]$.
Note that arbitrary rotations around the $Z$-axis $Z(\theta)$ can be realized using this implementation, giving access to interesting gates (S gate for $\theta = \pi/2$, T gate for $\pi/4$); but it is not clear how to use these gates at the logical level.

The implementation of the $X$ gate is realized by making the two-photon stabilization time-dependent,
\begin{equation}\label{eq:Xgate}
\kappa_2 \mathcal{D}[\ha^2 - \alpha^2] \rightarrow \kappa_2 \mathcal{D}[\ha^2 - {(\alpha e^{i\pi t /T})}^2]
\end{equation}
such that the two states $\ket{\pm\alpha}$ are swapped adiabatically.
Note that the fact that the $X$ gate can be implemented in a bias-preserving manner may be counter-intuitive at first sight.
Indeed, as noted in~\cite{PreskillPRA2008Faulttolerantquantum} in the case of regular qubits (two-level systems), it may be natural to assume that the imperfections of the physical process implementing the $X$ will lead to an $X$ component in the error model of the gate, \latin{e.g.\@} if a slight over or under rotation is implemented, the corresponding error should have an $X$ component.
Indeed, in the case of regular two-level systems, it has been shown~\cite{MirrahimiPRX2019RepetitionCatQubits} that it is impossible to implement an $X$ gate in a bias-preserving manner on two-level systems.
In the case of cat qubits, the bias-preserving implementation is possible by exploiting the infinite dimensional Hilbert space of the underlying harmonic oscillator to implement a continuous code deformation.

The fully dissipative implementation of \autoref{eq:Xgate} needs to be adiabatic with respect to the dissipative time-scale $\kappa_2^{-1}$.
As a consequence, this may led to low fidelity gates as the cat qubit suffers from decoherence during the execution of the gate.
To accelerate the gate, a ``feedforward'' Hamiltonian $\hat{H}_{X} = -\pi/T \had\ha$ can be added.
The role of this Hamiltonian is to ensure that the state of the quantum harmonic oscillator remains in the code space during the gate, instead of relying only on the dissipative stabilization.
Loosely speaking, in presence of this Hamiltonian, the role of the dissipative stabilization is no longer to ``drag'' the quantum state to implement the gate but merely to ensure that the errors do not distort the state, while the actual dynamics implementing the gate is realized by the Hamiltonian part of the dynamics.

\paragraph{CNOT gate ---}
Combining the ideas of the $X$ gate and the $Z$ gate, the CNOT gate is realized by making the rotation of the pumping of the target qubit implementing the $X$ gate that depends on the state of the control qubit, by modifying the dissipative stabilization channels of the control cat qubit $\mathcal{L}_{\ha} = \kappa_2 \mathcal{D}[L_{\ha}]$ and target cat qubit $\mathcal{L}_{\hb} = \kappa_2 \mathcal{D}[L_{\hb}(t)]$ respectively as
\begin{subequations}\label{eq:CNOT_ME}
\begin{align}
	\hat L_{\ha} &= \ha^2 - \alpha^2 \\
	\hat L_{\hb}(t) &= \hat{b}^2 - \tfrac{1}{2}\alpha(\hat{a}+\alpha) + \tfrac{1}{2}\alpha {e^{2 i \frac{\pi}{T} t}}(\hat{a}-\alpha)
\end{align}
\end{subequations}
where $\ha$ and $\hb$ denote the annihilation operators of the control and target cat qubit respectively, and $T$ denotes the CNOT gate time.
Similarly to the case of the $X$ gate, the gate speed can be dramatically increased, thus leading to higher gate fidelity, by adding a ``longitudinal coupling'' Hamiltonian $\hat{H}_{\text{C}X} = \frac{\pi}{4\alpha T}(\ha + \had - 2\alpha)(\hbd \hb - \alpha^2)$ to the gate dynamics.

\paragraph{Toffoli gate ---}
Finally, the CNOT gate can be generalized to a Toffoli gate by adding another control cat qubit.
The dynamics implementing the adiabatic Toffoli gate is realized with
the stabilization channels $\mathcal{L}_{\ha} = \kappa_2 \mathcal{D}[L_{\ha}]$, $\mathcal{L}_{\hb} = \kappa_2 \mathcal{D}[L_{\hb}]$, $\mathcal{L}_{\hc} = \kappa_2 \mathcal{D}[L_{\hc}(t)]$ (respectively the first control, the second control and the target)
\begin{subequations}\label{eq:CCX_ME}
\begin{align}
	\hat L_{\hat a} &= \hat{a}^2 - \alpha^2\\
	\hat L_{\hat b} &= \hat{b}^2 -\alpha^2\\
	\hat L_{\hat c}(t) &=
		\begin{multlined}[t]
			\hat{c}^2 -\tfrac{1}{4} (\hat a + \alpha)(\hat b + \alpha)
				+\tfrac{1}{4} (\hat a + \alpha)(\hat b - \alpha) \\
			+\tfrac{1}{4} (\hat a - \alpha)(\hat b + \alpha)- \tfrac{1}{4} e^{2 i \frac{\pi}{T} t} (\hat a - \alpha)(\hat b - \alpha),
		\end{multlined}
\end{align}
\end{subequations}
together with the Hamiltonian $\hat{H}_{\text{CC}X} = \frac{\pi}{16\alpha^2 T}(\ha + \had - 2\alpha)(\hb + \hbd - 2\alpha)(\hcd \hc - \alpha^2)$.

\subsection{Physical noise model}\label{appendix:code:physical_noise}
We now detail the error models associated with the physical implementations of the gates described above, with errors coming from two different sources: the single-photon loss of the quantum harmonic oscillator at a rate $\kappa_1$, and the non-adiabaticity of the gates.

\begin{table}[t!]
	\begin{tabular}{ccc}
&                                                                                 &$\alpha^2=19$, $\kappa_1/\kappa_2 = 10^{-5}$ \\ \hline
	                       &                                                                                 &                                             \\
	$\mathcal{P}_{\ket{+}}$&                                                                                 &$T_{\text{prep}} = 1/\kappa_2$               \\ \hline
	$1-\mathcal{F}$        &$\alpha^2 \kappa_1 T_{\text{prep}}$                                              &$1.9\times 10^{-4}$                          \\ \hline
	                       &                                                                                 &                                             \\
	$\mathcal{M}_{X}$      &                                                                                 &$T_{\text{meas}} = 1/\kappa_2$               \\ \hline
	$1-\mathcal{F}$        &$\alpha^2 \kappa_1 T_{\text{meas}}$                                              &$1.9\times 10^{-4}$                          \\ \hline
	                       &                                                                                 &                                             \\
	C$X$                   &                                                                                 &$T_{\text{C}X} = 1/\kappa_2$                 \\ \hline
	$1-\mathcal{F}$        &                                                                                 &$8.4 \times 10^{-3}$                         \\
	$Z_1$                  &$\alpha^2 \kappa_1 T_{\text{C}X} + \frac{\pi^2}{64\alpha^2\kappa_2T_{\text{C}X}}$&$8.3\times 10^{-3}$                          \\
	$Z_2 = Z_1Z_2$         &$\tfrac{1}{2}\alpha^2 \kappa_1 T_{\text{C}X}$                                    &$9.5\times 10^{-5}$                          \\ \hline
	\end{tabular}
	\caption{Pauli phase-flip error models used in this section.
		For the dissipative implementations considered in this work, we summarize the analytical errors models derived in~\cite{BrandaoPQ2022BuildingFaultTolerant}.
		The equalities in the left column, \latin{e.g.\@} $Z_2 = Z_1Z_2$, means that these errors occur with equal probability.
	}\label{tab:error_models}
\end{table}

\paragraph{Physical phase-flip errors ---}
Using the ``shifted Fock basis'', a subsystem decomposition of the quantum harmonic oscillator well suited to the cat qubit encoding~\cite{BrandaoPQ2022BuildingFaultTolerant}, the (dominant) phase-flip error probabilities of the single and multi-qubit gates can be obtained analytically.
The corresponding Pauli error model is summarized in \autoref{tab:error_models}.
Because the phase-flip errors induced by the natural losses of the quantum harmonic oscillator increase with the gate time, while the phase-flip errors induced by non-adiabaticity decrease with the gate time, the combination of these two sources of errors gives rise to an optimal finite gate time that minimizes the phase-flip errors.

As previously discussed~\cite{MirrahimiPRX2019RepetitionCatQubits,BrandaoPQ2022BuildingFaultTolerant}, the timescale $1/\kappa_2$ sets the typical time of quantum operations on a cat qubit, and hence the clock cycle time of the architecture, and the ratio $\kappa_1/\kappa_2$ sets the typical fidelity of quantum operations.
For the specific value of $\kappa_1/\kappa_2 = 10^{-5}$ considered here, and an average number of photons $\alpha^2 = 19$, we obtain a total state preparation and measurement error probability of $p_{\text{SPAM}} = p_{\text{prep}} + p_{\text{meas}} = 3.8\times 10^{-4}$ and a CNOT gate infidelity of $8.4\times 10^{-3}$.
The noise model associated with the Toffoli gate is later described in \autoref{tab:error_models_ccx}.

Note that the CNOT gate time $T_{\text{C}X} = 1/\kappa_2$ is sub-optimal for the gate fidelity, but nonetheless achieves a lower repetition code cycle error rate.
This somewhat counter-intuitive fact is detailed in the recent work~\cite{MirrahimiIp2022Highperformancerepetition}.

\begin{figure}[h!]
	\resizebox{0.5\textwidth}{!}{\includegraphics[width=0.9\linewidth]{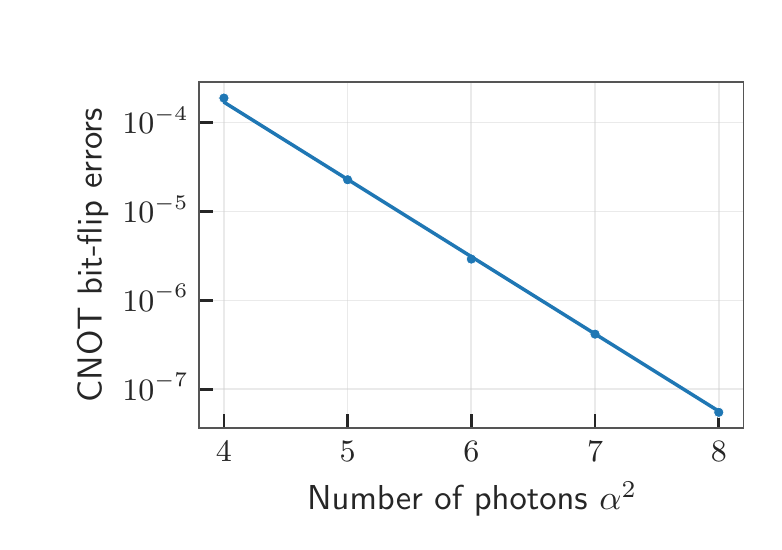}}
	\caption{Total bit-flip error probability induced during the implementation of a CNOT gate versus the average photon number of the cat qubit, obtained through a numerical process tomography of \autoref{eq:CNOT_ME}.
	}\label{fig:pX_CX}
\end{figure}

\paragraph{Physical bit-flip errors ---}
In order to estimate the (exponentially suppressed) physical bit-flip error probability, we perform numerical process tomography of all the gates described in \autoref{appendix:cat}, at the exception of the CC$X$ gate.
Indeed, resolving accurately the bit-flip error probability requires to simulate the gate over a typical range of photon number $\alpha^2 = 2 \textendash{} 20$ photons, and to use an appropriate truncation in the Fock basis.
The current simulation method does not allow to perform a sufficiently accurate simulation of the 3-mode CC$X$ gate due to the size of the Hilbert space required for the simulation to converge.
We thus assume that the bit-flip error rate of the CC$X$ gate to be similar to that of the CNOT gate.

We find that, while all operations have a bit-flip error probability scaling as $e^{-2\alpha^2}$, the operation that dominates the physical bit-flip error probability is the CNOT gate.
Due to the fast implementation of this gate, we find that the dominant bit flips are those induced by non-adiabaticity.
We numerically estimate the value of the bit-flip errors by performing a full process tomography~\cite{Chuang2010QuantumComputationQuantum} of the dynamics implementing a CNOT gate.
The sum of the probabilities of all the 12 bit-flip type errors (\latin{e.g.\@} $X_1$, $Z_1\otimes Y_2$, \latin{etc.\@}) is depicted in \autoref{fig:pX_CX}.
To extrapolate to larger number of photons, we fit this curve and find
\begin{equation}
	p_X^{\text{C}X} = 0.50 e^{-2\alpha^2}.
\end{equation}

\begin{figure*}[t!]
	\resizebox{\textwidth}{!}{\includegraphics[width=\linewidth]{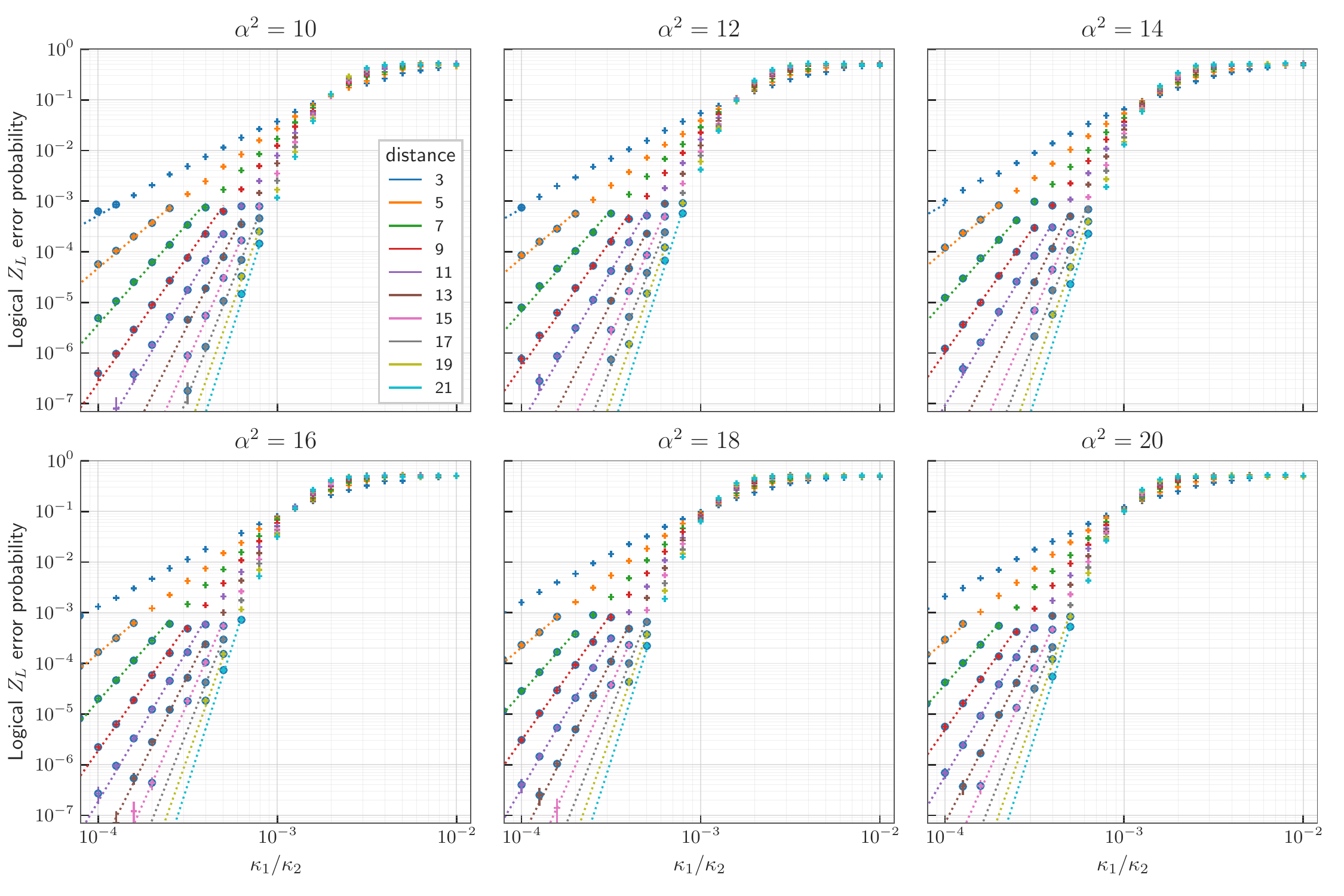}}
	\caption{Logical phase-flip error probability for various values of the average photon number $\alpha^2$, calculated using Monte Carlo sampling of the repetition code circuit under circuit-level noise.
		The error bars display the statistical error.
		The dotted lines correspond to a leading order fit in the regime where errors are sparse, $\kappa_1/\kappa_2 \ll 1$, the data points used in the fit are circled.
	}\label{fig:QEC_error_probability}
\end{figure*}

\section{Repetition code operation}\label{appendix:code}
In this appendix, we detail the architecture of our cat qubit quantum computer at the logical (repetition code) level.

\subsection{Layout}\label{appendix:repetition_code:layout}
\begin{figure}[h]
	\includegraphics[width=0.7\linewidth]{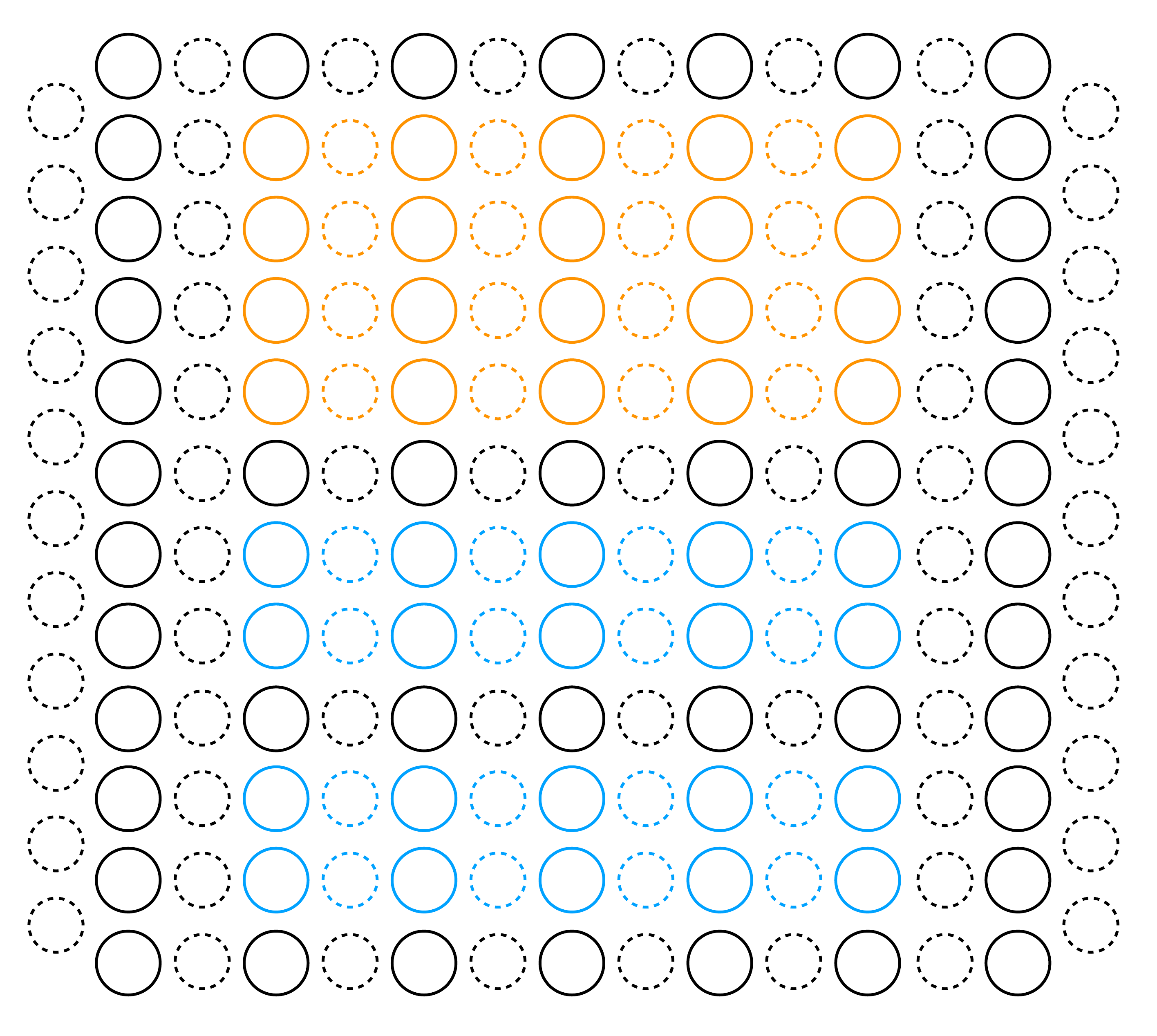}
	\caption{Layout for a quantum processor using cat qubits, with 4~logical qubits and a code distance of 5.
		The data qubits are in plain line, the ancilla qubits are dashed, the qubits used to perform the computation are in blue, the qubits used in the magic state factory are orange, and the routing qubits are black.
		}\label{fig:architecture}
	\end{figure}

With currently available technology, cat qubits can only be built with planar connections.
However, it is convenient to benefit from an all-to-all connectivity between the logical qubits, as it prevents proliferation of logical swaps.
This is achieved by laying the physical qubits as presented in \autoref{fig:architecture}.
The proposed layout consists of organizing data busses for applying 2-qubit gates through lattice surgery.
Each logical qubit corresponds to a horizontal line of data qubits, and ancillary qubits between them allow the measurements of the code stabilizers.
To apply CNOT gates between far away logical qubits with lattice surgery, the logical qubits are accessed through the routing qubits either on their extremity or with each of its physical qubits involved.
Toffoli gate is realized by preparation of a magic state in a dedicated factory (yellow) and by a teleportation, that requires only 2-qubit gates as detailed in \autoref{appendix:code:toffoli}.

For a code distance $d$, each logical qubit requires $d$ physical qubits for the data and $d-1$ for the ancillary qubits that allow measurements of the stabilizers.
Each magic state factory involves $4$ logical qubits; we adjust the number of magic state factory to deliver the states at a similar speed at which they can be consumed (we do not consider any parallelization, but ensure to have on average one magic state ready for each time interval corresponding to a gate teleportation).
With $\text{nb}_{\text{log}}$ and $\text{nb}_\text{factory}$ respectively the numbers of logical qubits and factories, we count $\text{nb}_{\text{rout}} = \ceil{(\text{nb}_{\text{log}} + \text{nb}_{\text{factory}})/2} + 1$ horizontal lines of routing qubits.
On each side, there is $3(\text{nb}_{\text{log}} + \text{nb}_{\text{rout}} + 4 \text{nb}_{\text{factory}}) - 1$ additional routing qubits, including the corresponding ancillaries.
Those numbers are added in the code used to evaluate the resources~\cite{code}.

Note that with the proposed layout only up to two CNOT gates can be realized in parallel.
Large parallelization would require modifications, for instance by placing the logical qubits on several columns, as considered for surface codes~\cite{Gidney2019Lowoverheadquantum,LitinskiQ2019GameSurfaceCodes,CampbellPQ2022UniversalQuantumComputing}.

\subsection{Repetition code performance}\label{appendix:code:memory}
In this work, we follow~\cite{MirrahimiPRX2019RepetitionCatQubits,BrandaoPQ2022BuildingFaultTolerant} and assume that the bit-flip error rate can be made sufficiently low to run large-scale algorithms while relying only on a simple repetition code against phase flips to produce a logical qubit.
The logical $\ket{+}_L$ and $\ket{-}_L$ states of a distance-d repetition code are $\ket{\pm}_L := \ket\pm^{\otimes d} = \ket{\mathcal{C}_\alpha^\pm}^{\otimes d}$.
Note that the logical $\ket{0/1}_L = \tfrac{1}{\sqrt{2}}(\ket{+}_L \pm \ket{-}_L)$ state corresponds to the equal superposition of all the $2^{d-1}$ states that have an even/odd number of $\ket{1}$ states.
The $d-1$ stabilizers of the code $S_i = X_i X_{i+1}$ can be measured using bias-preserving operations ${\mathcal{P}_{\ket +}, \text{CNOT}, \mathcal{M}_X}$.

We now detail how logical error models (at the level of the repetition code) can be obtained from the physical error models described in \autoref{appendix:code:physical_noise}, beginning this subsection with the repetition code error rates (memory).
Here, we follow the recent proposal~\cite{MirrahimiIp2022Highperformancerepetition} that optimizes the repetition code performance.
We proceed in two-steps: as the repetition code only corrects for phase flips, we estimate the logical phase-flip error probability $p_{Z_L}$ and the logical bit-flip error probability $p_{X_L}$ separately, and the logical error rate is then estimated as $\epsilon_L = p_{Z_L} + p_{X_L}$.

\paragraph{Repetition code cycle time ---}
As discussed in the main text, in this work we assume that a two-photon stabilization rate of $\kappa_2/2\pi = \SI{1.59}{\mega\hertz}$.
Following~\cite{MirrahimiIp2022Highperformancerepetition}, a full repetition code cycle is executed in five time steps (preparation of the ancilla cat qubits in the $\ket{+}$ state, first round of CNOT gates, idling time to suppress cat qubit leakage, second round of CNOT gates, and ancilla measurement), each of which is assumed to be realized in time $1/\kappa_2 = \SI{100}{\nano\second}$.
Therefore, the total repetition code cycle time is $T_{\text{cycle}} = \SI{500}{\nano\second}$.

\paragraph{Logical phase flip ---}
The logical phase-flip error probability is numerically estimated using circuit-level noise simulations of the repetition code circuit, where every operation in the circuit (including idle locations) is noisy.
More precisely, every noisy operation is modelled as a perfect operation followed by the stochastic application of a Pauli phase-flip noise model, given in \autoref{tab:error_models}.
For the value of $\kappa_1/\kappa_2 = 10^{-5}$ and $\alpha^2=19$ considered in the main text, this translates to a preparation and measurement infidelity of $p_{\text{prep}} = p_{\text{prep}} = 1.9\times 10^{-4}$, and a CNOT gate infidelity of $p_{\text{C}X} = 8.4\times 10^{-3}$.
To estimate the per cycle logical phase-flip error rate, the state of the circuit is initialized in $\ket{+}_L$, and the stabilizers are measured $d$ times, followed by a final perfect round of stabilizer measurements.
This final perfect round of stabilizer measurements emulates the fact that, in practice, the measurement of the stabilizers continues.
This ensures the convergence of the decoding algorithm as if the data were buried under additional data as the processor keeps running.
The estimate of the per cycle logical error rate is accurate as long as the temporal window used in the simulation (here, $d$ code cycles) is sufficiently long~\cite{PreskillJoMP2002Topologicalquantummemory}.
Note that, at the end of the quantum algorithm, the data qubits are measured, and the finale value of the stabilizers may be inferred from this measurement.
The history of syndrome outcomes is decoded using the open source minimum-weight perfect matching decoder pyMatching~\cite{HiggottAToQC2022PyMatchingPythonPackage}.
Decoding the syndromes yields the correction to apply to the state.
If the state after (perfect correction) is $\ket{+}_L$, the quantum error correction circuit correctly interpreted the errors that occurred, while if the state after correction is $\ket{-}_L$, a logical phase-flip error $Z_L$ occurred.

\begin{figure*}[t!]\includegraphics[width=0.45\linewidth]{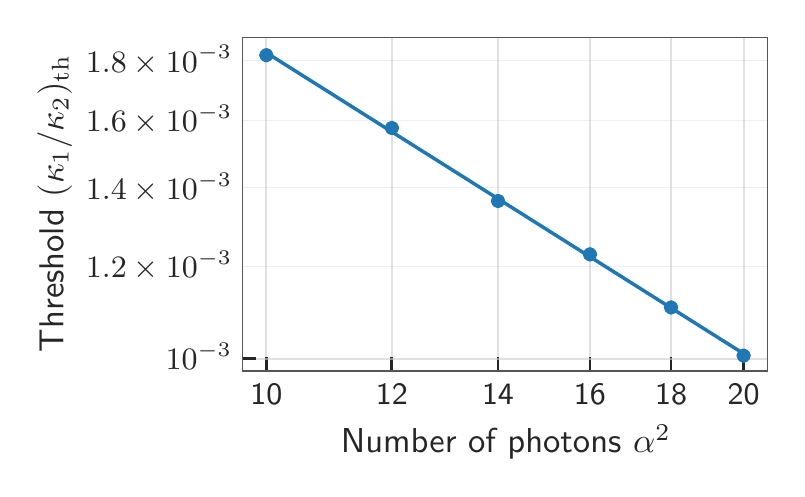}
	\qquad
	\includegraphics[width=0.45\linewidth]{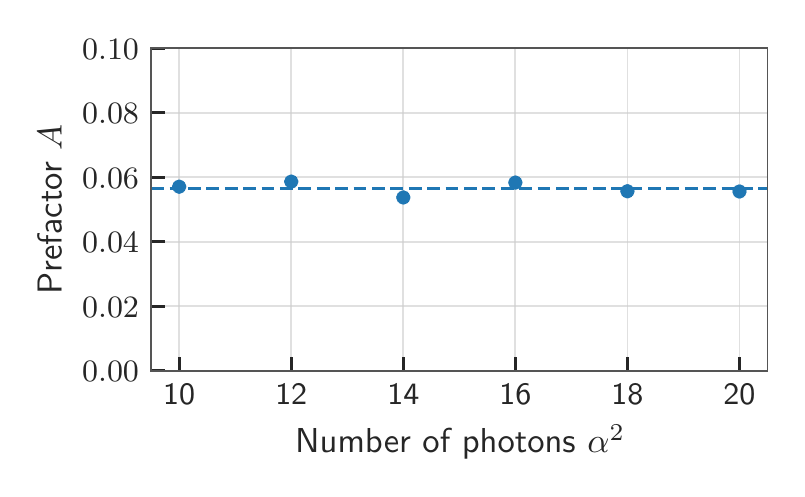}
	\caption{Numerical fit of the prefactor and value of the threshold in \autoref{eq:pZL_ansatz}.
		We find that the prefactor is independent of $\alpha^2$, while the threshold decreases almost linearly with $\alpha^2$, as one expects from the fact that the photon-loss induced phase flips increase linearly with $\alpha^2$.
	}\label{fig:pZL_numerical_fits}
\end{figure*}

The probability of such a failure is then estimated by repeating the above procedure, and is depicted in \autoref{fig:QEC_error_probability} for various values of $\kappa_1/\kappa_2$, average photon number $\alpha^2$ and code distance $d$.
In the regime where errors are rare, $\kappa_1/\kappa_2 \ll 1$, we expect the logical error rate to be dominated by the contributions of trajectories where exactly $(d+1)/2$ errors occurred (as any combination of less than $(d+1)/2$ errors cannot lead to a logical error, and combinations of more than $(d+1)/2$ errors are unlikely to occur in this regime).
Thus, we fit the logical error probability after $d$ rounds of stabilizer measurements to the ansatz
\begin{equation}\label{eq:pZL_ansatz}
	p_{Z_L} = A(\alpha^2)d{\Big( \dfrac{\kappa_1/\kappa_2}{{(\kappa_1/\kappa_2)}_{\text{th}}(\alpha^2)} \Big)}^{\frac{d+1}{2}}.
\end{equation}
We find numerically (see \autoref{fig:pZL_numerical_fits}) that $A(\alpha^2)$ is independent of $\alpha^2$, $A(\alpha^2) \approx 5.6 \times 10^{-2}$, and ${(\kappa_1/\kappa_2)}_{\text{th}}(\alpha^2)$ is well approximated by
\begin{equation*}
	{(\kappa_1/\kappa_2)}_{\text{th}}(\alpha^2) = \frac{1.3\times 10^{-2}}{{(\alpha^2)}^{0.86}}.
\end{equation*}

We interpret the fact that ${(\kappa_1/\kappa_2)}_{\text{th}}(\alpha^2)$ is not exactly inversely proportional to $\alpha^2$ as the effect of non-adiabatic (measurement) errors on the control.
Indeed, in the absence of such errors, the only source of errors considered in our model is the single-photon loss which scales linearly with $\alpha^2$, which would produce a threshold inversely proportional to $\alpha^2$.

\paragraph{Logical bit flip ---}
From the characterization of the physical bit-flip error probabilities in \autoref{appendix:code:physical_noise}, dominated by the bit-flip errors induced by the CNOT gate, we estimate the probability of logical bit-flip error per repetition code cycle to be $p_{X_L} = n_{\text{CX}}\times p_{X}^{\text{C}X} = 2(d-1)\times 0.50 e^{-2\alpha^2}$, where $n_{\text{CX}}$ is the number of CNOT gates in a repetition code cycle.

\paragraph{Repetition code performance ---}
Taking into account both errors, the per cycle error probability of the repetition code is given by (presented as \autoref{eq:epsL_RepCode} in main text)
\begin{equation*}
	\epsilon_L= 5.6\times10^{-2}{\Big( \dfrac{{(\alpha^{2})}^{0.86}\kappa_1/\kappa_2}{{(\kappa_1/\kappa_2)}_{\text{th}}} \Big)}^{\frac{d+1}{2}} + 2(d-1)\times 0.50 e^{-2\alpha^2}
\end{equation*}
with ${(\kappa_1/\kappa_2)}_{\text{th}} = 1.3\times10^{-2}$.

Note that this architecture does not have a threshold, as for any finite value of the noise $\kappa_1/\kappa_2$, there is an optimal value for the mean number of photons $\alpha^2$ and the repetition code distance $d$ that minimizes the logical error rate.
However, in this work we argue that the absence of a threshold may not be problematic as long as the minimal error rate that can be achieved is sufficiently low for the targeted algorithm.

\subsection{Transversal logical gates implementation}
We now show how the logical operations of the set $\{\mathcal{P}_{\ket+_L},\mathcal{P}_{\ket{0}_L},\mathcal{M}_{Z_L},\mathcal{M}_{X_L},Z_L,X_L,\text{C}X^k_L\}$ can be implemented on the repetition code, postponing the more involved implementation of the ${\text{CC}X}_L$ gate to the next subsection.

Note that the logical gate $\text{C}X^k_L$ is not required for universality as it can be synthesized from ${\text{C}X}_L$ gates, but here we show how it can be directly implemented as it is widely used in the table lookup circuit (\autoref{fig:lookup_table}).

All of the operations in $\mathcal{S}_L = \{\mathcal{P}_{\ket+_L},\mathcal{P}_{\ket{0}_L},\mathcal{M}_{Z_L},\mathcal{M}_{X_L},Z_L,X_L,\text{C}X^k_L,{\text{CC}X}_L\}$ can be implemented transversally on the repetition code, except for the ${\text{CC}X}_L$ gate, the implementation of the latter being discussed in the next subsection.

\begin{figure*}[t!]
	\resizebox{\textwidth}{!}{\includegraphics[width=\linewidth]{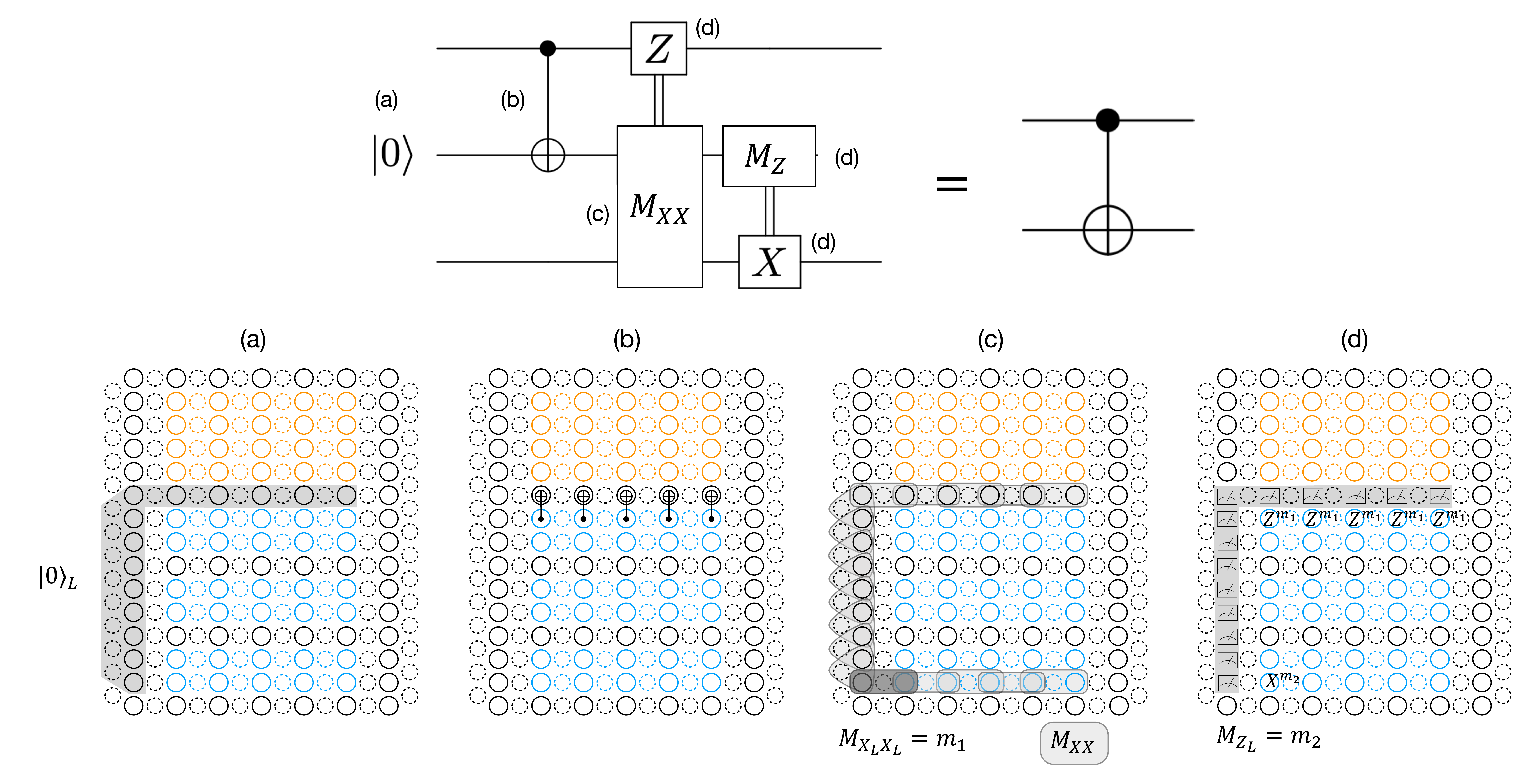}}
	\caption{(Top) Circuit identity used to teleport a logical CNOT gate in the processor.
		(Bottom) Physical mapping of the top circuit onto the processor.
		It proceeds in four steps: (a) Preparation of a logical ancilla in the $\ket{0}_L$ state.
		This is achieved by preparing all data qubits in the $\ket{0}$ state, followed by $d$ rounds of stabilizer measurements, decoding and correction.
		(b) A logical CNOT gate is realized using a local, transversal implementation.
		(c) A logical $X_L X_L$ measurement is realized by measuring $d$ times the $XX$ operator of the two physical qubits at the border between the logical qubits (in dark grey), as well as the stabilizers of the code (to ensure fault-tolerance).
		In terms of lattice surgery, this operation corresponds to a merge of the two logical qubits.
		The logical measurement outcome $m_1$ is obtained after decoding and subsequent correction.
		(d) To complete the gate teleportation, the logical $Z_L$ operator is measured on the logical ancilla qubit.
		This is achieved by measuring the value of all the $Z$ operators of the data qubits and the logical measurement outcome $m_2$ is obtained by taking the product of all measurement outcomes.
		Logical Pauli corrections, $Z_L^{m_1}$ on the control and $X_L^{m_2}$ may then be applied, or tracked in classical software to change the sign of subsequent logical measurement outcomes on these qubits.
	}\label{fig:LSCNOT}
\end{figure*}

\paragraph{State preparation, measurement, and Pauli gates ---}
The fault-tolerant preparation of $\ket{+}_L$ and $\ket{0}_L$ are realized by preparing the product states $\ket{+}^{\otimes d}$ and $\ket{0}^{\otimes d}$, respectively, followed by $d$ rounds of stabilizer measurements, decoding and correction.
This logical state preparation scheme is standard for stabilizer codes; see \latin{e.g.\@}~\cite{ClelandPRA2012SurfacecodesTowards,PreskillPRA2013Faulttolerantquantum}.
The fault-tolerant measurement of $X_L$, denoted $\mathcal{M}_{X_L}$, is realized by measuring all the data qubits of the code in the $X$ basis, followed by a majority vote on the measurement outcomes.
Note that in principle, the value of the logical operator $X_L$ can be read out on any data qubit of the code, but this is not fault-tolerant.
The measurement of the logical $Z_L$ operator, denoted $\mathcal{M}_{Z_L}$, is realized by measuring all the data qubits in the $Z$ basis and by taking the product of all the measurement outcomes.
The logical $X_L$ gate is implemented by performing a single physical $X$ gate on any of the data qubits, and the logical $Z_L$ gate is implemented transversally by applying a $Z$ gate to all of the data qubits.
Note that the $Z_L$ measurement and $X_L$ gate are sensitive to any single qubit $X$ error, but this is not a problem as cat qubits have exponentially rare bit-flip errors.

\paragraph{${\text{C}X}_L$ gate ---}
The logical ${\text{C}X}_L$ gate can be implemented transversally on the repetition code.
However, implementing a logical CNOT gate in such a way between an arbitrary pair of logical qubits in the processor requires long distance interactions between the physical qubits of the processor, which is not a realistic feature of a superconducting quantum processor.

Instead, the logical CNOT gate can be realized using a circuit based on lattice surgery involving only nearest-neighbor interactions~\cite{GottesmanCSaF1999FaultTolerantQuantum}.
Here, we only provide a superficial overview of how lattice surgery works to illustrate how it may be adapted to the case of a repetition code (instead of the more standard case of a surface code).
A detailed analysis of lattice surgery protocols may be found in~\cite{VanMeterNJoP2012Surfacecodequantum, Gidney2019Lowoverheadquantum, OppenQ2018LatticeSurgeryTwist}.
The implementation proceeds in four (logical) steps, as shown in \autoref{fig:LSCNOT}.

First, physical ancillary routing qubits are used to prepare a logical ancilla qubit in the $\ket{0}_L$ state that extends from the logical control qubit to the logical target qubit.
Note that, to implement a CNOT gate, which is a `$Z-X$ type' interaction, the value of the logical $Z_L$ operator on the control and of the logical $X_L$ operator on the target needs to be accessed.
For our QEC scheme, the $Z_L$ operator has support on all of the physical qubits while the $X_L$ operator can in principle be accessed on any of the data qubits, \latin{e.g.\@} at the edge of the repetition code.
For this reason, the logical ancilla prepared in the $\ket{0}_L$ state has an L shape, as depicted in \autoref{fig:LSCNOT}(a).

\begin{figure*}[t!]
	\resizebox{\textwidth}{!}{\includegraphics[width=\linewidth]{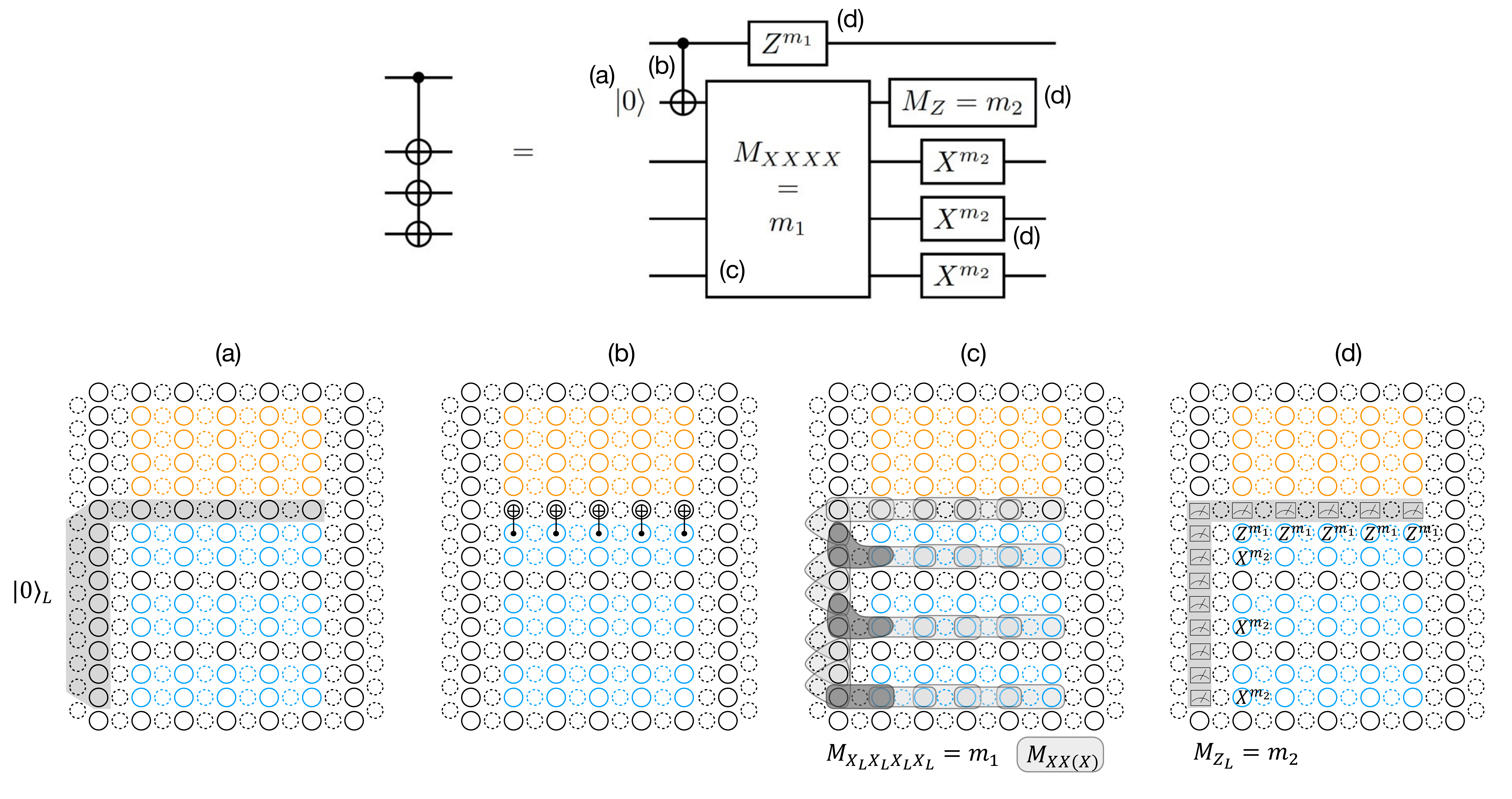}}
	\caption{(Top) Circuit identity used to teleport a logical $\text{C}X^k_L$ gate in the processor, with $k=3$.
		(Bottom) Physical mapping of the top circuit onto the processor.
		The four steps are identical to that of \autoref{fig:LSCNOT}, at the exception of the step (c) that consists of merging the logical qubits.
		This is achieved using weight-3 physical measurement of the $XXX$ operator between two data qubits of the logical ancilla and the last data qubit of the logical target qubit.
		The logical measurement outcome $m_1$ is obtained by taking the product of the operators in dark grey (interpreted after subsequent decoding).
	}\label{fig:LSCXX}
\end{figure*}

Second, a (transversal) CNOT gate is performed between the logical control qubit and the logical ancilla, mapping the state of the control qubit $\ket{\psi}_L = \alpha\ket{0}_L + \beta\ket{1}_L$ to $\alpha\ket{00}_L + \beta\ket{11}_L$.
Here, transversal means that physical CNOT gates are realized between pairs of physical qubits of the logical control qubit and the data qubits of the adjacent logical ancilla in the bus, as depicted in \autoref{fig:LSCNOT}(b).
Note however that the two logical qubits (control and ancilla) do not have the same distance, as the logical ancilla extends into the side bus.
Because the logical $X_L$ operator can be accessed on any of the physical qubits, however, a logical CNOT can be implemented using only physical CNOT gates on the part of the logical qubit that is in the horizontal bus
(in terms of lattice surgery on the surface code, this corresponds to a `smooth split' operation).
However, in the case of a repetition code, one needs to use an additional ancillary logical qubit to implement this smooth split~\footnote{Indeed, on a surface code, the smooth split operation consists of splitting the logical qubit in two logical qubits in the `logical $X_L$' direction; \latin{i.e.\@}, the $X$-distance $d_X$ of the two children logical qubits is half that of the parent qubit and (after correction) the new logical $X_L^i$ operators are related to the old one as $X_L^1\otimes X_L^2 = X_L^{\text{old}}$.
For repetition cat codes, the role of the $X$-distance $d_X$ is played by the average photon number in the cat qubits, and the support of the logical $X_L$ operator is a single physical qubit.
Therefore, unlike in the surface code case, one cannot `split' a string of physical $X$ operators to produce two new logical qubits.
Rather, preparing a logical ancilla qubit in the $\ket{0}_L$ state and applying a logical CNOT gate has the same effect.}.

Then, the multi-qubit Pauli $X_L \otimes X_L$ operator of the logical ancilla and the logical target is measured using a local $XX$ measurement between the edges of the two logical qubits, as depicted in dark grey in \autoref{fig:LSCNOT}(c)
(in terms of lattice surgery, one could also interpret this operation as a `rough merge').
For fault-tolerance, the measurement is repeated $d$ times and the outcome is interpreted after decoding the full syndrome history.

The final step consists of disentangling the logical ancilla qubit from the logical control and logical target qubits using a logical $Z_L$ measurement (\autoref{fig:LSCNOT}(d)).
To complete the teleportation, a (transversal) logical $Z_L$ operation is applied to the control qubit if the logical $X_L \otimes X_L$ measurement produces the outcome $-1$ (with probability $0.5$), and a logical $X_L$ operation is applied to the logical target qubit if the logical $Z_L$ measurement on the ancilla qubit produces the value $-1$ (with probability $0.5$).
In practice, these operations do not need to be performed but may be tracked in software (which is why we do not consider here that it is an additional time step).

As noted previously (\latin{e.g.\@}~\cite{VanMeterNJoP2012Surfacecodequantum,LitinskiQ2019GameSurfaceCodes}), an attractive feature of using lattice surgery to implement logical multi-qubit measurements is that decoding algorithms used to interpret the error syndromes extend naturally over these protocols~\cite{CampbellPRR2022Circuitlevelprotocol,CampbellPQ2022UniversalQuantumComputing,BrandaoPQ2022BuildingFaultTolerant}.
In its most general form, one may use lattice surgery to implement all the (single-qubit) Clifford gates using a classical tracking algorithm~\cite{OppenQ2018LatticeSurgeryTwist}, by tracking the so-called edges operators (the correspondence between the physical qubits operators and the logical qubit operators).
While this strategy is arguably most efficient in general, it also requires the ability to measure arbitrary logical Pauli operators, such as $X_L Y_L$, for instance.
Lattice surgery protocols that achieve these measurements are more involved, \latin{e.g.\@} using twist defects~\cite{OppenQ2018LatticeSurgeryTwist, Gidney2019Lowoverheadquantum} and the analysis of the associated logical failure rates under circuit-level noise has only been recently realized~\cite{CampbellPRR2022Circuitlevelprotocol,CampbellPQ2022UniversalQuantumComputing}.
In the context of Shor's algorithm however, the overhead of the algorithm is widely dominated by the fault-tolerant implementation of the modular arithmetic circuits and table-lookup, composed of Toffoli and (sometimes multi-target) CNOT gates.
Lattice surgery implementations of these gates only require $X_L X_L$ type measurements, which may be realized using regular stabilizer measurements only.
Thus, we do not attempt to perform circuit-level simulations of lattice surgery protocols to derive the exact error models of these gates.
Rather, we assume the same logical error rate per code cycle $\epsilon_L$ as in the case of the memory \autoref{eq:epsL_RepCode}, and take into account the fact that the logical gates take $\bigO{d}$ code cycles to complete.

\paragraph{$\text{C}X^k_L$ gate ---}
A major advantage of lattice surgery protocols is that they can be straightforwardly extended to multi-qubit Pauli measurements~\cite{Gidney2019Lowoverheadquantum, OppenQ2018LatticeSurgeryTwist}.
As an example, we illustrate in \autoref{fig:LSCXX} how the logical multi-target $\text{C}X^k_L$ may be implemented on our processor.
The only non-trivial difference with the case of the ${\text{C}X}_L$ gate is the measurement of the multi-qubit Pauli operator $X_L X_L \ldots X_L$.
This may be achieved as depicted in \autoref{fig:LSCXX}(c) using physical weight-3 measurements of the $XXX$ operator between three cat qubits.

\subsection{Logical Toffoli}\label{appendix:code:toffoli}
We conclude this Appendix by detailing the implementation of the last logical gate, the Toffoli gate.
In this work, we consider the teleportation based implementation~\cite{ChuangPRA2000Methodologyquantumlogic,BrandaoPQ2022BuildingFaultTolerant}.
Note that other proposals to implement a logical Toffoli gate directly on repetition cat codes using pieceable fault-tolerance or code concatenation have been proposed~\cite{MirrahimiPRA2021Errorratesresource}, but the teleportation implementation leads to a lower overhead and is compatible with the physical layout of the processor we consider here.

\subsubsection{Measurement based magic state preparation}
\begin{figure*}[t!]
	\includegraphics[width=0.8\linewidth]{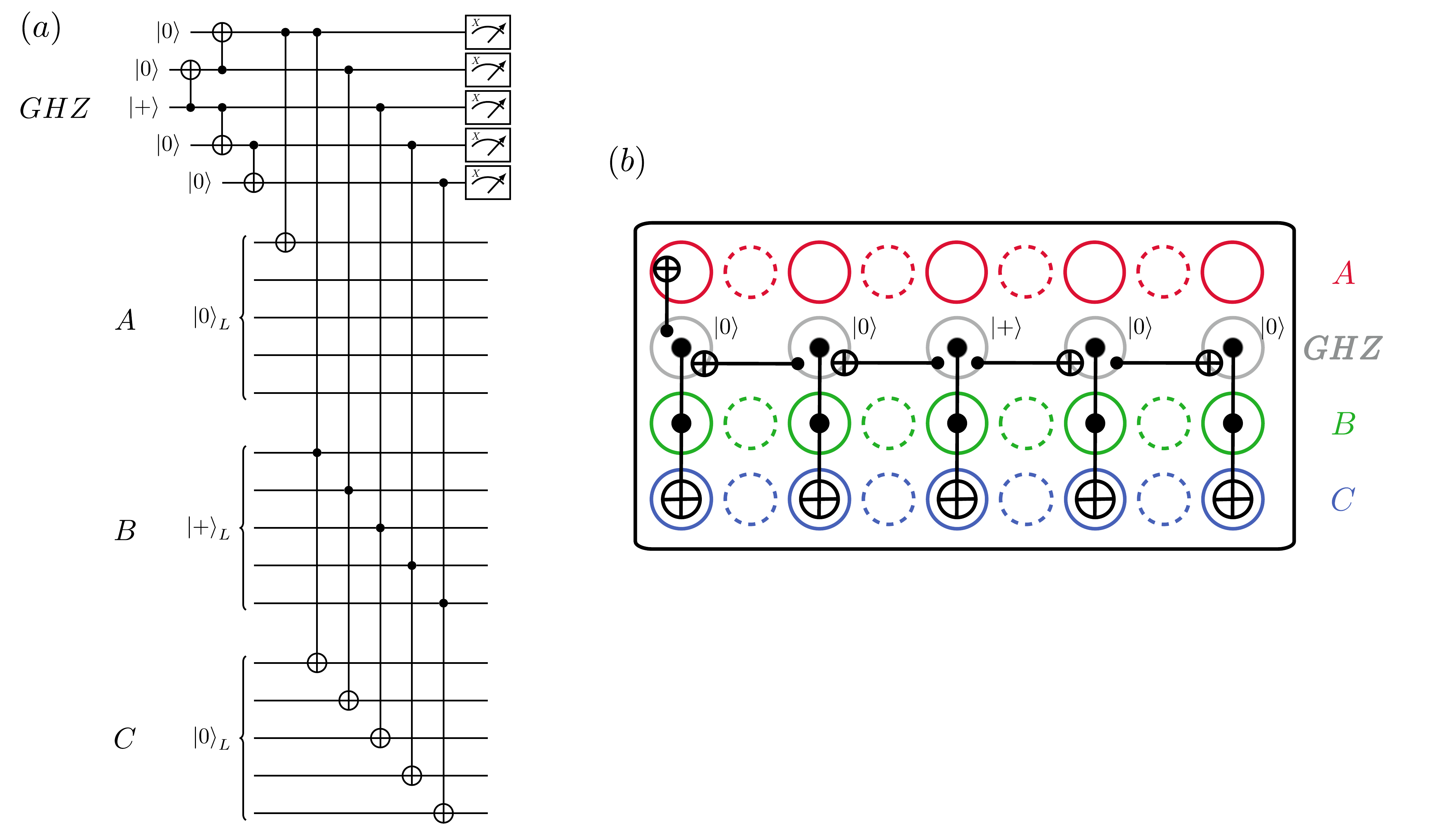}
	\caption{(a) $X_A \text{CNOT}_{B,C}$ measurement circuit for a distance $d=5$ code.
		An auxiliary $\ket{\text{GHZ}}$ state is prepared on the first block and is connected to the qubit A by a $\text{CNOT}$ gate to perform the $X_A$ stabilizer measurement, and to qubit B and C by transversal Toffoli gates to implement the $\text{CNOT}_{B,C}$ stabilizer measurement.
		The stabilizer measurement result is given by computing the parity of the $X$ measurements of the $\ket{\text{GHZ}}$ state.
		Starting from the state $\ket{0,+,0}$, this achieves the preparation of $\ket{\text{CC}X}$ or $Z_A \ket{\text{CC}X}$ if the measurement outcome is $+1$ or $-1$ respectively.
		(b) Mapping of the circuit on the physical layout of the processor.
		Note that all interactions are local~\cite{BrandaoPQ2022BuildingFaultTolerant}.
	}\label{fig:CCX_MagicState}
\end{figure*}

The teleportation of a Toffoli gate consumes the (logical) magic state $\ket{\text{CC}X} = \frac{1}{2} (\ket{000} + \ket{010} + \ket{100} + \ket{111})$~\cite{Shor1996Faulttolerantquantum}.
Labelling the three logical qubits A, B, C, note that $\ket{\text{CC}X}$ is the $+1$ eigenstate of $X_A \text{CNOT}_{B,C}$, $X_B \text{CNOT}_{A,C}$ and $Z_C {\text{C}Z}_{A,B}$ and can thus be obtained by starting from the logical state $\ket{000}_L$ and by performing a QND measurement of these operators.
However, starting from a well-chosen state which can be both prepared easily and is already a $+1$ eigenstate of two of the stabilizers, \latin{e.g.\@} $\ket{0,+,0}_L$, it is sufficient to measure only the remaining stabilizer, here $X_A \text{CNOT}_{B,C}$ to achieve the preparation of $\ket{\text{CC}X}$.
This stabilizer may be measured using the circuit depicted in \autoref{fig:CCX_MagicState}(a).
In this circuit, the Toffoli gates are used to measure $\text{CNOT}_{B,C}$ and a $\text{CNOT}$ gate to measure $X_A$.
In principle, a single physical ancilla qubit could be used to perform the measurement.
However, for connectivity issues, it is convenient to use instead a block of ancilla qubits prepared in the $\ket{\text{GHZ}} = (\ket{0}^{\otimes d} + \ket{1}^{\otimes d})/\sqrt{2}$ state, in order to apply transversal Toffoli gates.
The block of ancilla is then measured in the $X$ basis and the $X_A \text{CNOT}_{B,C}$ stabilizer measurement result is given by the product of the $X$ measurements.
This state is prepared with a circuit of depth $(d+3)/2$ as shown in \autoref{fig:CCX_MagicState}(a).
Following~\cite{BrandaoPQ2022BuildingFaultTolerant}, \autoref{fig:CCX_MagicState}(b) shows the 2D layout to perform the QND measurement.
If the measurement outcome is $-1$, the prepared state is $Z_A\ket{\text{CC}X}$ and a correction needs to be applied (or tracked in the software).
If the QND measurement is only performed once, a single physical $Z$ error on the $\ket{\text{GHZ}}$ state would lead to a logical $Z_A$ error.
To make this circuit fault-tolerant, the measurement needs to be repeated.
In the absence of errors, all measurement outcomes should be identical.
However, when it is repeated, a single $Z$ error on block $A$ flips the results of all the following measurements and an error on block $C$ randomly flips all the following measurements as well.
To prevent this, error detection or correction needs to be performed between each round of measurement by measuring the $XX$ stabilizers.
Below, we discuss two different fault-tolerant schemes to perform this magic state preparation.

\subsubsection{Heralded scheme}
\begin{figure}[h]
	\includegraphics[width=\linewidth]{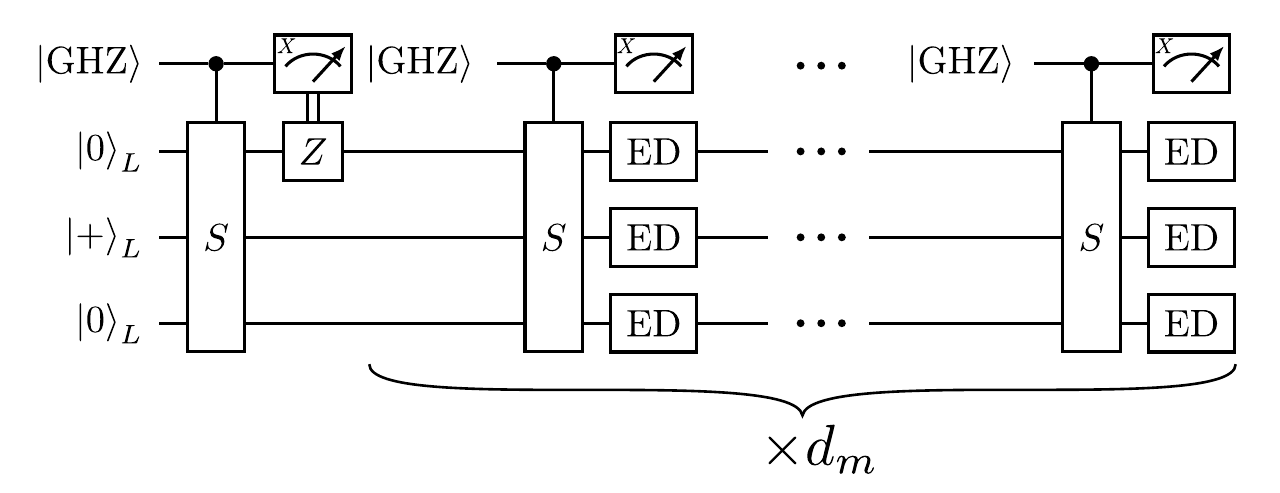}
	\caption{Heralded scheme of the magic state preparation.
		The first QND measurement ($S = X_A \text{CNOT}_{B,C}$) is followed by a $Z$ correction if the measurement outcome is $-1$.
		Then, the block `QND measurement + error detection (ED)' is repeated $d_m$ times and the preparation is rejected if the measurements do not all give $+1$.
		The error detection round is performed between the rounds of QND measurements to prevent errors from propagating to the GHZ measurement.
		$d_m$ is optimized to minimize the space-time overhead.
	}\label{fig:CCX_MagicStateFull}
\end{figure}

Following the magic state preparation described in~\cite{BrandaoPQ2022BuildingFaultTolerant}, the circuit of the logical Toffoli magic state preparation may be made fault-tolerant by post-selecting on the measurement outcomes.
After the first QND measurement, a $Z$ correction is applied to the qubit A if the measurement outcome is $-1$.
The following QND measurements are followed by a single round of error detection and the number of repetitions of the block `QND measurement + error detection' is denoted $d_m$.
The state preparation is rejected if the $d_m$ QND measurements do not give the same result $+1$.
If any error is detected during the error detection rounds, the state preparation is also rejected.
The success probability is called the acceptance probability.
The corresponding circuit is represented on \autoref{fig:CCX_MagicStateFull}.
Note that a slightly different error model, described in \autoref{tab:error_models_ccx}, is used compared to one described above for error correction.
Indeed, in the case of error correction, we considered the recent proposed scheme~\cite{MirrahimiIp2022Highperformancerepetition} where the fast, noisy CNOT gates are used to measure the stabilizers of the repetition code.
Here, in the context of a post-selected scheme, we consider the case where the CNOT gates are performed slower, in order to maximize the CNOT gate fidelity, so as not to lower the acceptance probability.
The CNOT gate is now performed in a time $89/(\alpha^2 \kappa_2)$ which depends on the number of photons, and the associated bit-flip error is given by $0.02 e^{-2 \alpha^2}$~\cite{BrandaoPQ2022BuildingFaultTolerant}.
Two scenarii can cause a logical error in this circuit: either all QND measurements are flipped, or a logical error happens on the blocks $A,B$ or $C$ without being detected by the error detection rounds.
When $d_m$ increases, the probability that each QND measurement fails decreases exponentially, while the probability of a logical error on the blocks $A,B$ or $C$ increases linearly, thus an optimal number of measurement repetitions $d_m$ that minimized the logical error can be found.

\begin{table}[t!]
	\begin{tabular}{ccc}
&                                                                                 &$\kappa_1/\kappa_2 = 10^{-5}$\\ \hline
	               &                                                                                 &                                             \\
	C$X$           &                                                                                 &$T_{\text{C}X} = 89/(\kappa_2 \alpha^2)$     \\ \hline
	$1-\mathcal{F}$&                                                                                 &$3.5 \times 10^{-3}$                         \\
	$Z_1$          &$\alpha^2 \kappa_1 T_{\text{C}X} + \frac{\pi^2}{64\alpha^2\kappa_2T_{\text{C}X}}$&$2.6\times 10^{-3}$                          \\
	$Z_2 = Z_1Z_2$ &$\tfrac{1}{2}\alpha^2 \kappa_1 T_{\text{C}X}$                                    &$4.4\times 10^{-4}$                          \\ \hline
	               &                                                                                 &                                             \\
	CC$X$          &                                                                                 &$T_{\text{CC}X} = 89/(\kappa_2 \alpha^2)$    \\ \hline
	$1-\mathcal{F}$&                                                                                 &$4.6 \times 10^{-3}$                         \\
	$Z_1=Z_2$      &$\alpha^2 \kappa_1 T^* + \frac{\pi^2}{128\alpha^2\kappa_2T^*}$                   &$1.8\times 10^{-3}$                          \\
	$Z_3$          &$\frac{5}{8}\alpha^2 \kappa_1 T^*$                                               &$1.1\times 10^{-4}$                          \\
	$Z_1Z_2$       &$\frac{\pi^2}{128\alpha^2\kappa_2T^*}$                                           &$8.8\times 10^{-4}$                          \\
	$Z_1Z_3=Z_2Z_3$&$\frac{1}{8} \alpha^2 \kappa_1 T^*$                                              &$2.3\times 10^{-5}$                          \\
	$=Z_1Z_2Z_3$   &                                                                                 &                                             \\ \hline
	\end{tabular}
	\caption{Pauli phase-flip error models used for the magic state preparation.
		For the dissipative implementations considered in this work, we summarize the analytical errors models derived in~\cite{BrandaoPQ2022BuildingFaultTolerant}.
		The equalities in the left column, \latin{e.g.\@} $Z_2=Z_1Z_2$, means that these errors occur with equal probability.
	}\label{tab:error_models_ccx}
\end{table}

However, when $d_m$ increases, the acceptance probability of the magic state preparation decreases exponentially.
To optimize these parameters, the space-time overhead; \latin{i.e.\@}, the number of qubits multiply by the depth of the circuit to generate one magic state, must be minimized.
The space-time overhead to generate one magic state is evaluated as
\begin{equation}
\text{space-time overhead} = \frac{\text{number of qubits} \times \text{circuit depth}}{\text{acceptance probability}}.
\end{equation}

To ensure that at least $(d+1)/2$ physical errors are needed to create a logical error, the QND measurement has to be repeated at least $d_m = (d-1)/2$ times.
However at this parameter value, the magic state preparation errors are dominated by the $\ket{\text{GHZ}}$ measurement errors by several orders of magnitude as noticed in~\cite[Fig.\,15c]{BrandaoPQ2022BuildingFaultTolerant}.
By increasing $d_m$, these measurement errors diminish and even if the acceptance rate also diminishes, it is beneficial for achieving a lower space-time overhead for the same logical phase-flip error.

\begin{figure}[b]
	\includegraphics[width=\linewidth]{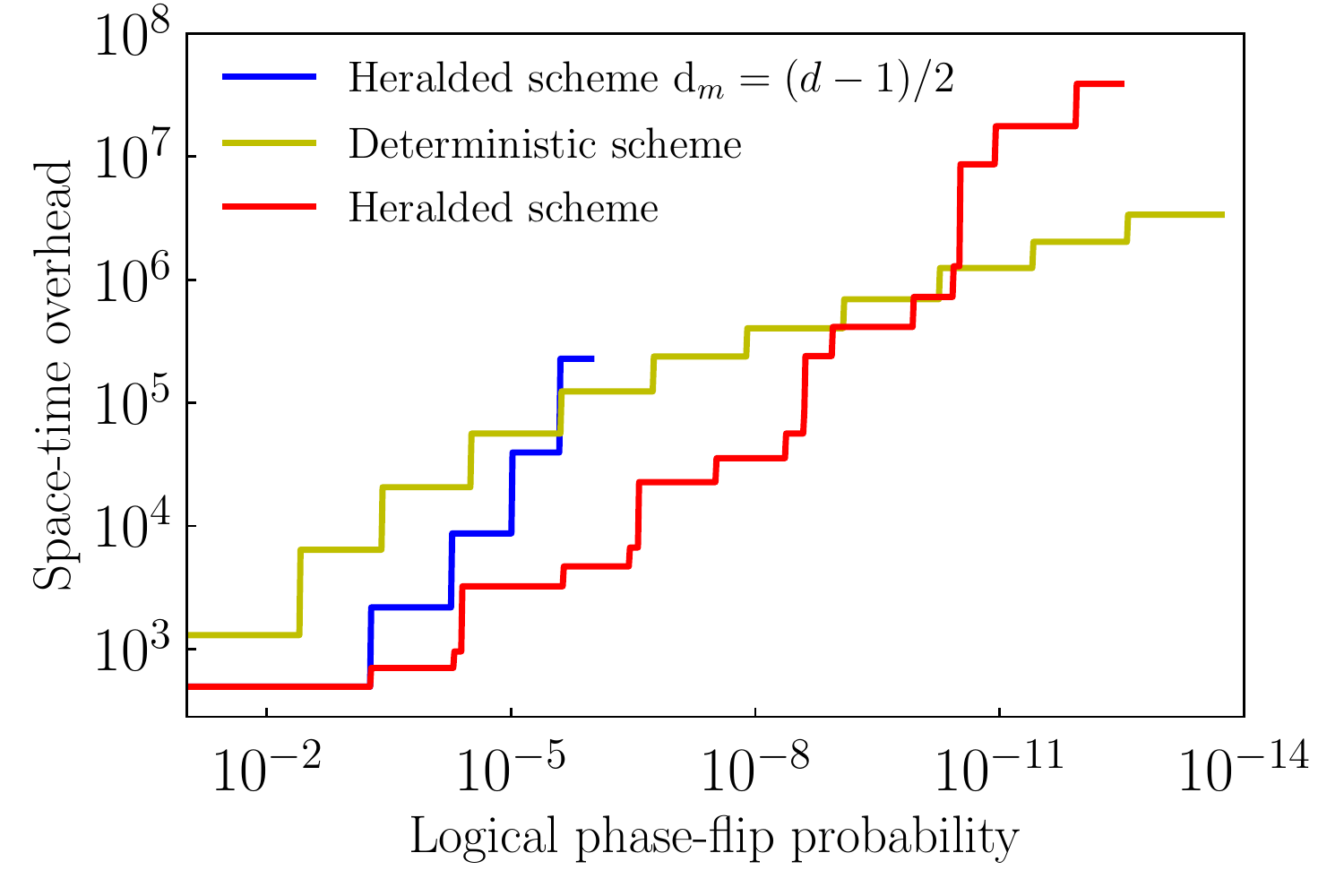}
	\caption{Space-time overhead as a function of the logical error probability needed for the Toffoli magic state prepared.
		The space-time overhead is calculated with $\text{space-time overhead} = \text{number of qubits} \times \text{circuit depth}/\text{acceptance probability}$.
		For the heralded and deterministic scheme, the distance of the code and $d_m$ are chosen to minimize the space-time overhead.
		For comparison, the proposition in~\cite{BrandaoPQ2022BuildingFaultTolerant} where $d_m = (d-1)/2$ is represented (in blue).
	}\label{fig:SpaceTimeOverhead}
\end{figure}

\autoref{fig:SpaceTimeOverhead} presents the space-time overhead needed as a function of the logical error targeted.
The space-time overhead shown for a certain logical error is optimized in terms of distance $d$ and number of repeated QND measurement $d_m$ (in red).
The proposition in~\cite{BrandaoPQ2022BuildingFaultTolerant} where $d_m = (d-1)/2$ is also represented (in blue).

\subsubsection{Deterministic scheme}
\begin{figure}[h]
	\includegraphics[width=\linewidth]{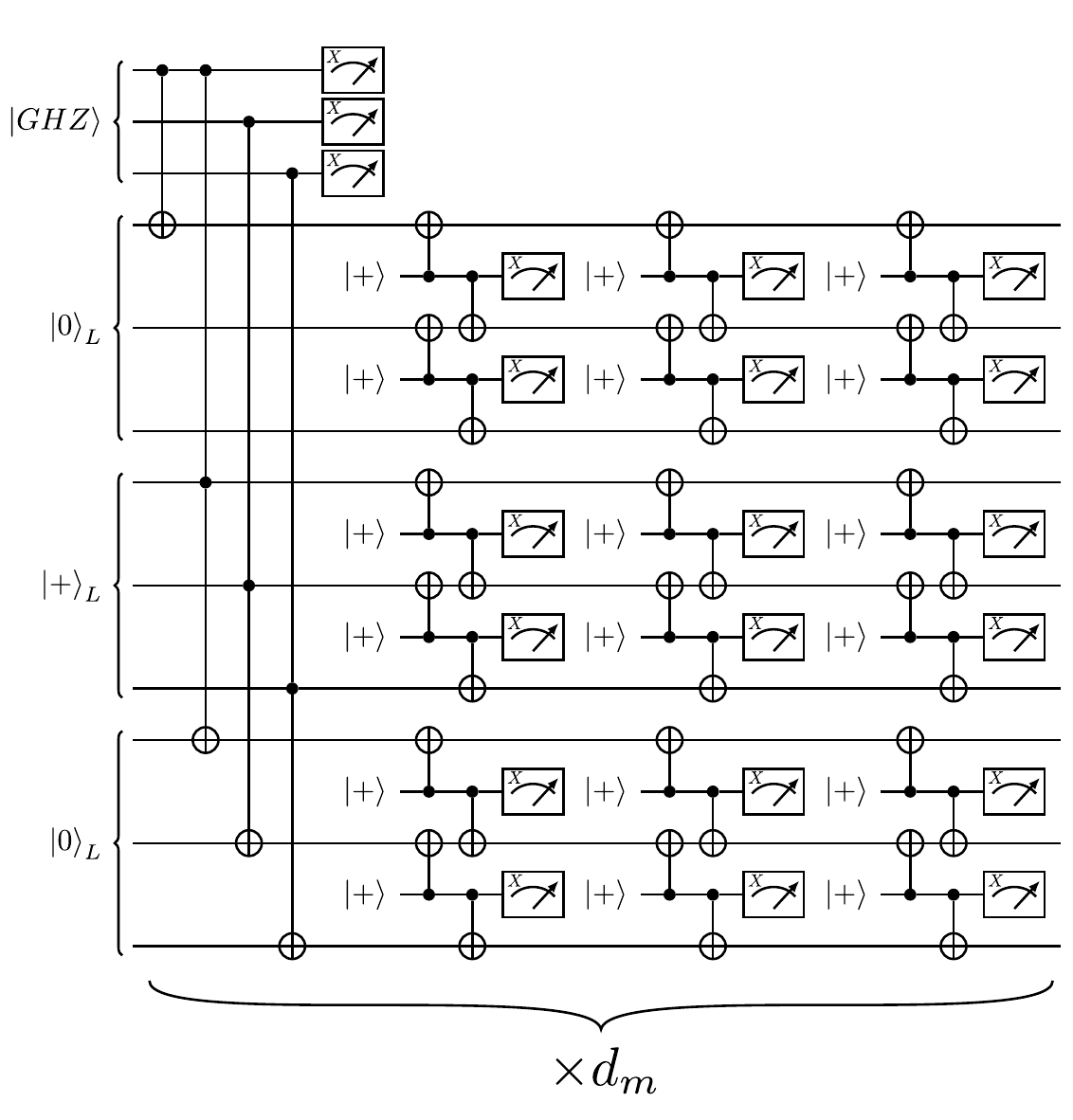}
	\caption{Deterministic scheme of the magic state preparation for $d=3$.
		For simplicity, only one block is shown, however this circuit is repeated $d_m$ times.
		$d$ rounds of error correction are performed between the rounds of QND measurements to correct errors before the next GHZ measurement.
		A majority vote between the QND measurements then determines the result of the measurement.
		The number of repetitions $d_m$ is optimized to minimize the space-time overhead.
	}\label{fig:FM}
\end{figure}

This section presents a deterministic scheme (with acceptance probability equals to one).
This scheme is similar to the one before but now includes a fault-tolerant error correction between each round of measurement (and not only error detection), and a majority vote between the $d_m$ outcomes of the stabilizer measurements (\latin{c.f.\@} \autoref{fig:FM}).
Note that error correction is needed at each round because in case a $Z$ error is not corrected on the block $C$ before the next stabilizer measurement, its outcome is random.
Further note that $d$ rounds of error correction are needed to make it fault-tolerant.
Here again, we neglect the classical processing time of the decoding of the syndrome and assume that the correction on the data qubits is perfect and instantaneous.
As in the previous case, when $d_m$ increases, the probability that the majority vote between the QND measurements fails decreases exponentially, while the probability of a logical error on the blocks $A,B$ or $C$ increases, such that there exists a finite optimal number of repetitions.

\autoref{fig:SpaceTimeOverhead} (yellow curve) shows the space-time overhead needed as a function of the logical error targeted for the deterministic scheme.
Like for the heralded scheme, the space-time overhead shown for a certain logical error is optimized in terms of distance $d$ and number of repeated QND measurement $d_m$.
This scheme needs a larger default overhead because of its very large depth, but has a better scaling as a function of the logical error targeted and thus become better than the heralded scheme if the phase-flip logical error targeted is below approximately $10^{-10}$.

\subsubsection{Magic state preparation total logical error}
As for the logical memory error, the bit-flip error is calculated independently of the logical phase-flip error.
The former is calculated with the CNOT and Toffoli gates physical bit-flip errors which are dominant.
The number of photon is chosen so that the logical bit-flip error rate equals the logical phase-flip error rate, and the results are shown in \autoref{tableau:preparation_etat_toffoli}.

\begin{table}
\begin{tabular}{c|cc|c|cr|c}
{$i$} & {$d$} & {$\alpha^2$} & {error prob.\@}        & {steps} & {time}                   & {accept.\@ prob.\@ (\si{\percent})} \\ \hline
 0    &  3    &  3.75        & $1.05 \times 10^{-3}$  & 23           & \SI{54.7}{\micro\second} & $84$    \\
 1    &  3    &  3.93        & $1.02 \times 10^{-4}$  & 29           & \SI{65.8}{\micro\second} & $74.5$  \\
 2    &  3    &  5.32        & $8.14 \times 10^{-5}$  & 35           & \SI{58.7}{\micro\second} & $66$    \\
 3    &  5    &  7.15        & $4.62 \times 10^{-6}$  & 46           & \SI{57.4}{\micro\second} & $45.6$  \\
 4    &  5    &  8.18        & $7.00 \times 10^{-7}$  & 53           & \SI{57.8}{\micro\second} & $36.2$  \\
 5    &  5    & 8.38         & $5.36 \times 10^{-7}$  & 60           & \SI{63.9}{\micro\second} & $28.8$  \\
 6    &  7    & 9.71         & $6.14 \times 10^{-8}$  & 73           & \SI{67.1}{\micro\second} & $14.8$  \\
 7    &  7    & 10.76        & $8.40 \times 10^{-9}$  & 81           & \SI{67.2}{\micro\second} & $10.5$  \\
 8    &  7    & 11.06        & $5.16 \times 10^{-9}$  & 89           & \SI{71.8}{\micro\second} & $7.27$  \\
 9    &  9    & 11.64        & $2.28 \times 10^{-9}$  & 104          & \SI{79.7}{\micro\second} & $2.62$  \\
10    &  9    & 12.83        & $2.30 \times 10^{-10}$ & 113          & \SI{78.6}{\micro\second} & $1.54$  \\
11    &  9    & 13.44        & $7.36 \times 10^{-11}$ & 122          & \SI{81  }{\micro\second} & $0.975$ \\
12    & 19    & 17.35        & $7.90 \times 10^{-12}$ & 9576         & \SI{4.92}{\milli\second} & $100$   \\
13    & 21    & 18.94        & $5.40 \times 10^{-13}$ & 14112        & \SI{6.65}{\milli\second} & $100$   \\
14    & 23    & 20.53        & $3.74 \times 10^{-14}$ & 21344        & \SI{9.27}{\milli\second} & $100$   \\
\end{tabular}
\caption{Parameters for the preparation of the Toffoli states.
	$i$ is an index identifying each parameter set, the one used in \autoref{tableau:resources}.
	$d$ and $\alpha^2$ are respectively the code distance and the average number of photons used during the preparation of the magic state.
	The fourth column describes the error probability during the process.
	Fifth column counts the depth of the magic state preparation, in number of physical gates, and the following one the corresponding time.
	Last column is the acceptance probability.
	The heralded scheme is used for $i \leq 11$ while for $i \geq 12$ the deterministic scheme is favored.
	The simulation techniques used to obtain the values presented in this table are described in the text.
}\label{tableau:preparation_etat_toffoli}
\end{table}

\subsubsection{Heralded scheme simulation technique}
In order to evaluate the logical failure of a circuit from the physical errors, the circuit-level error model is used, which includes every possible location for an error to occur.
Monte Carlo simulations are generally used to sample the logical failure rate.
However, sampling a logical failure probability of $10^{-10}$ with an acceptance probability of $10^{-2}$ would require thousands of years of CPU run-time to reach a relative standard deviation of the order of the percent.
We will now present two methods which have been used in order to reach reasonable computational time.

Stratified sampling consists of partitioning the population to sample in distinct subgroups, which are sampled independently.
Partitioning the population according to the number of errors happening in the circuit is an efficient way to reduce the time required for the Monte Carlo simulations.
\begin{multline}
	\mathbb{P}(\text{failure}) =
		\sum_{n=0}^{n=N} \mathbb{P}(\text{failure}\ |\ n\ \text{errors in the circuit}) \\
		\times \mathbb{P}(n \text{ errors in the circuit})
\end{multline}
where $N$ denotes the number of possible locations for an error to occur in the circuit.
$\mathbb{P}(n \text{ errors in the circuit})$ is a binomial law and becomes negligible for large $n$.
Due to the fault-tolerant construction of the scheme, $\mathbb{P}(\text{failure}\ |\ n\ \text{errors in the circuit} < n_{FT} )=0$ where $n_{FT} = \min(\frac{d+1}{2},d_m+1)$.
Thus, only a handful of $n$ contributes to the logical error probability.
$\mathbb{P}(n\ \text{errors in the circuit})$ can be calculated analytically and the only remaining terms to simulate with Monte Carlo simulations are $\mathbb{P}(\text{failure}\ |\ n\ \text{errors in the circuit})$ for a few relevant values of $n$.

However, this method alone cannot reduce sufficiently the computational time for large $d_m$.
We introduce a second type of stratified sampling, which combined with the first one, is sufficient to reach tractable computations.
Because the scheme uses post-selection, most of the trajectories which could cause a logical failure because they contain more $Z$ errors than $(d+1)/2$ are rejected because these errors are detected by the error detection rounds.
But we can finely choose the trajectories which can both cause a logical error while being undetected by syndrome measurements; \latin{i.e.\@}, a stratified sampling on the location where errors can happen.
If $\mathcal{C}$ designates the set of restricted locations where errors need to happen in order to create a logical failure without being detected, the logical failure is then given by
\begin{multline}\label{eqn:reduced}
	\mathbb{P}(\text{failure}) =
		\mathbb{P}(\text{failure}\ | \ n\ \text{errors} \in \mathcal{C}) \\
		\times \mathbb{P}(n\ \text{errors} \in \mathcal{C})
\end{multline}

\subsubsection{Deterministic scheme simulation technique}
The circuit-level error model is here intractable due to the depth of the circuit considered.
Once again, two logical error sources can be distinguished.
Either the majority vote of the repeated measurements fails, or a logical error happens on the logical qubits $A, B$ or $C$.
The latter has the same logical error as the memory.
After each correction applied to the logical qubits $A, B$ or $C$, residual errors can propagate to the QND measurement.
In order to perform a circuit level simulation of the failure probability of the QND measurement, the residual errors from the logical qubits $A, B$ or $C$ were sampled with circuit-level simulation, and added as input of the QND measurement simulation.
Because $d$ rounds of error correction are performed between each round of QND measurement, the failure probability of the different QND measurements are independent and can be added in order to calculate the failure probability of the majority vote.

\subsubsection{Teleportation}
\begin{figure}[h]
	\resizebox{0.95\linewidth}{!}{\begin{quantikz}[row sep=0.75em, column sep = 0.75em]
                                    &\ctrl{3}&\qw     &\qw      &\qw     &\qw     &\qw        &\ctrl{2}   &\ctrl{1}   &\qw        &\qw        &\gate{Z}   & \qw \\
                                    &\qw     &\ctrl{3}&\qw      &\qw     &\qw     &\ctrl{1}   &\qw        &\gate{Z}   &\qw        &\gate{Z}   &\qw        & \qw \\
                                    &\qw     &\qw     &\targ{}  &\qw     &\qw     &\targ{}    &\targ{}    &\qw        &\gate{X}   &\qw        &\qw        & \qw \\
\lstick[wires=3]{$\ket{\text{CC}X}$}&\targ{} &\qw     &\qw      &\qw     &\meter{}&\cwbend{-1}&\cw        &\cw        &\cwbend{-1}&\cwbend{-2}&           &     \\
                                    &\qw     &\targ{} &\qw      &\qw     &\meter{}&\cw        &\cwbend{-2}&\cw        &\cwbend{-1}&\cw        &\cwbend{-4}&     \\
                                    &\qw     &\qw     &\ctrl{-3}&\gate{H}&\meter{}&\cw        &\cw        &\cwbend{-4}&\cw        &\cwbend{-2}&\cwbend{-1}&
\end{quantikz}
 }
	\caption{Teleportation of a Toffoli gate.
		From top, the two first qubits are the controls and the third is the target.
		$\ket{\text{CC}X}$ is the magic state for the Toffoli gate, defined as the result of applying a Toffoli gate on $\ket{+}\ket{+}\ket{0}$, and obtained via the magic state preparation described in the text.
	}\label{fig:toffoli_teleport}
\end{figure}
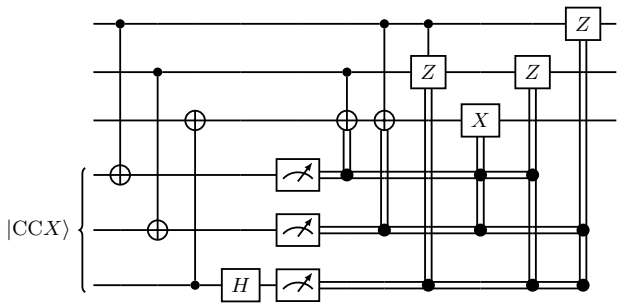

The teleportation of Toffoli gate, from the magic state $\ket{\text{CC}X}$, is done using the circuit represented in \autoref{fig:toffoli_teleport}.
Note that an auto-corrected Toffoli state would allow to also perform the corrections off-line, such as in~\cite{Fowler2019Flexiblelayoutsurface}.

\section{Resources for different key sizes}\label{appendix:tableau}
Estimation of resources required for solving elliptic curve discrete logarithm problem for different sizes are reported in \autoref{tableau:resources}.
The optimal parameters that we found are also given.
A description of symbols can be found in the caption of \autoref{tableau:resources}.
The optimization is run by an exhaustive search on the different parameters to minimize the product $\alpha^2 n_{\text{qubits}} t_{\text{exp}}$.

\begin{table*}
\begin{tabular}{cc|ccccc|cccccc}
	{$n$}                         &{$n_e$}                       &{$w_e$}                       &{$w_m$}                       &{$\alpha^2$}                  &{$d$}                         &{$i$}                         &{$\#\text{factories}$}        &{factories qubits}            &{$n_{\text{qubits}}$}         &{$t$}                         &{$t_{\text{exp}}$}            &{logical qubits}              \\ \hline
	8                             &16                            &9                             &2                             &12                            &7                             &4                             &5                             &388                           &2817                          &\SI{1}{\second}               &\SI{1}{\second}               &85                            \\
	16                            &32                            &11                            &4                             &14                            &9                             &5                             &6                             &463                           &5961                          &\SI{9}{\second}               &\SI{10}{\second}              &159                           \\
	32                            &64                            &13                            &4                             &15                            &9                             &7                             &16                            &1537                          &12050                         &\SI{55}{\second}              &\SI{1}{\minute}               &305                           \\
	64                            &128                           &15                            &4                             &17                            &11                            &7                             &13                            &1252                          &25346                         &\SI{8}{\minute}               &\SI{9}{\minute}               &595                           \\
	128                           &256                           &17                            &5                             &18                            &13                            &10                            &87                            &10026                         &64543                         &\SI{1}{hours}                 &\SI{1}{hours}                 &1173                          \\
	256                           &512                           &18                            &6                             &19                            &13                            &12                            &84                            &18101                         &126133                        &\SI{7}{hours}                 &\SI{9}{hours}                 &2326                          \\
	512                           &1024                          &20                            &7                             &21                            &15                            &12                            &73                            &15736                         &258739                        &\SI{3}{days}                  &\SI{3}{days}                  &4632                          \\
\end{tabular}
\caption{Resources and parameters for discrete logarithm computation.
	$n$ is the number of qubits to represent the coordinates of an elliptic curve point.
	$n_e$ is the number of qubits dedicated for the two registers containing the factors of the elliptic curve multiplication; $n_e = 2n$ with Shor's algorithm assuming the order of the cyclic group being of the same order of magnitude as $p$.
	$w_e$ and $w_m$ are the windows size for respectively the elliptic curve multiplication and the integer multiplication.
	$\alpha^2$ is the average number of photons used for each cat qubit.
	$d$ is the code distance and $i$ the index of the parameters of magic states factory, which are presented in \autoref{tableau:preparation_etat_toffoli}.
	Next detailed parameters are the number of magic states factories ($\#\text{factories}$) and number of qubits used in all of them.
	$n_\text{qubits}$ is the total number of (physical) cat qubits.
	$t$ and $t_{\text{exp}}$ are respectively the time for one run and the average time to successfully compute the discrete logarithm.
	Last parameter is the number of logical qubits involved in the computation.
}\label{tableau:resources}
\end{table*}

Here, $n$ (number of bits in $p$) is the parameter that controls the size of the problem (and $n_e=2n$).
$w_e$, $w_m$, $\alpha^2$, $d$ and $i$ are the parameters on which the optimization runs.
The number of factories is taken such that the average generation rate of magic states matches the teleportation time, and the number of qubits used for magic states factories is derived from the chosen layout; see \autoref{appendix:repetition_code:layout} for more details.
The number of logical qubits is computed by considering the place in the algorithm that consumes the largest number of ancillary states, as detailed in \autoref{appendix:arithmetic:qubit_count}.
The number of physical qubits is deduced from it by taking into account the code distance and routing qubits (and it also includes factories qubits).
The computation time is deduced from the logical algorithm, the repetition code implementation of logical gates and the time to implement physical gates.
The average time is deduced from the run-time and the success probability (the number of Bernoulli trials to get one success is given by a geometric distribution).

Similarly, the optimal parameters and resources for factoring RSA integers are detailed in \autoref{tableau:resources_rsa}.
The implementation of Shor's algorithm for factorization used to evaluate the resources is detailed in~\cite{SangouardPRL2021Factoring2048bit} (but with the architecture described here), and the code producing the table is available at~\cite{code}.
\begin{table*}
\begin{tabular}{cc|cccccc|cccccc}
	{$n$}                         &{$n_e$}                       &{$m$}                         &{$w_e$}                       &{$w_m$}                       &{$\alpha^2$}                  &{$d$}                         &{$i$}                         &{$\#\text{factories}$}        &{factories qubits}            &{$n_{\text{qubits}}$}         &{$t$}                         &{$t_{\text{exp}}$}            &{logical qubits}              \\ \hline
	6                             &6                             &5                             &2                             &2                             &10                            &5                             &2                             &4                             &229                           &986                           &\SI{62}{\milli\second}        &\SI{243}{\milli\second}       &33                            \\
	8                             &9                             &8                             &2                             &2                             &11                            &5                             &3                             &6                             &463                           &1476                          &\SI{159}{\milli\second}       &\SI{331}{\milli\second}       &45                            \\
	16                            &21                            &10                            &3                             &3                             &12                            &7                             &4                             &5                             &388                           &2551                          &\SI{992}{\milli\second}       &\SI{1}{\second}               &76                            \\
	128                           &189                           &18                            &4                             &3                             &16                            &11                            &7                             &13                            &1252                          &18650                         &\SI{3}{\minute}               &\SI{3}{\minute}               &430                           \\
	256                           &381                           &20                            &4                             &4                             &17                            &11                            &8                             &20                            &1917                          &35083                         &\SI{15}{\minute}              &\SI{19}{\minute}              &819                           \\
	512                           &765                           &23                            &4                             &4                             &19                            &13                            &10                            &87                            &10026                         &84073                         &\SI{2}{hours}                 &\SI{2}{hours}                 &1593                          \\
	829                           &1242                          &24                            &5                             &4                             &19                            &13                            &12                            &84                            &18101                         &136456                        &\SI{6}{hours}                 &\SI{8}{hours}                 &2548                          \\
	1024                          &1493                          &26                            &5                             &4                             &20                            &15                            &12                            &73                            &15736                         &180269                        &\SI{12}{hours}                &\SI{13}{hours}                &3137                          \\
	2048                          &3029                          &28                            &5                             &5                             &21                            &15                            &13                            &98                            &23075                         &349133                        &\SI{3}{days}                  &\SI{4}{days}                  &6214                          \\
\end{tabular}
\caption{Resources and parameters for integer factorization.
	The corresponding algorithm is detailed in~\cite{SangouardPRL2021Factoring2048bit} and the code generating this table is available at~\cite{code}.
	$n$ is the number of qubits to represent the coordinates of an elliptic curve point.
	$n_e$ is the number of qubits dedicated for the two registers containing the exponent of the modular exponentiation; $n_e \approx 1.5 n$ with Ekerå's algorithm.
	$m$ is the number of qubits added for the coset representation of integers.
	$w_e$ and $w_m$ are the windows size for respectively the exponentiation and the multiplications.
	$\alpha^2$ is the average number of photons used for each cat qubit.
	$d$ is the code distance and $i$ the index of the parameters of magic states factory, which are presented in \autoref{tableau:preparation_etat_toffoli}.
	Next detailed parameters are the number of magic states factories ($\#\text{factories}$) and number of qubits used in all of them.
	$n_\text{qubits}$ is the total number of (physical) cat qubits.
	$t$ and $t_{\text{exp}}$ are respectively the time for one run and the average time to successfully compute the discrete logarithm.
	Last parameter is the number of logical qubits involved in the computation.
}\label{tableau:resources_rsa}
\end{table*}

As stated in the main text, we find that computing a discrete logarithm on a 256-bit elliptic curve requires \num{126133} cat qubits, which corresponds to a distance-13 repetition code and cat qubits of size $\alpha^2 = 19$ photons.
This number was obtained under the assumption that the exponential suppression of bit-flip errors holds up until cat sizes of $19$ photons, an assumption that will be crucial to demonstrate experimentally.
Note that recent theoretical works~\cite{SavonaPRA2022Quantumerrorcorrection,Quijandria2022Quantumerrorcorrection,Jiang2022Autonomousquantumerror} have shown that, due to the better scaling of exponential suppression of bit flips in squeezed cats, the extremely low bit-flip error rates required to run large-scale algorithms on repetition cat codes could be attained with much fewer photons.
Alternatively, under similar assumptions than those used in this work, we estimate that a $(d_X, d_Z) = (3, 37)$ rectangular (rotated) surface code built on cat qubits of size $\alpha^2 = 10$ photons yields the same logical error rate as a distance-13 repetition code on $\alpha^2=19$ photons, which corresponds roughly to a $\approx \times 10$ overhead.

\section{Possible improvements}\label{appendix:ameliorations}
In this work, we favored solutions allowing for simple explanations while keeping the number of logical qubits as low as possible.
However, variations of algorithm implementation are possible and some of them may significantly reduce the computation time with a limited overhead in terms of qubit number (or even a reduction of qubit number if the run-time can be shortened enough to reduce the code distance, or use a less demanding magic state preparation).
They would have to be analyzed in detail to facilitate an experimental implementation of discrete logarithm computation with cat qubits.
In this appendix, we discuss techniques which might be relevant along this line.
In particular, we first discuss ways to achieve parallelization.
We then discuss implementations of Shor's algorithms using alternatives additions.
More generally, the repetition code can be seen as an extremely asymmetric surface code, and most of the work done in the past years to improve quantum computing on surface code can be adapted to the repetition code.

\subsection{Parallelization}
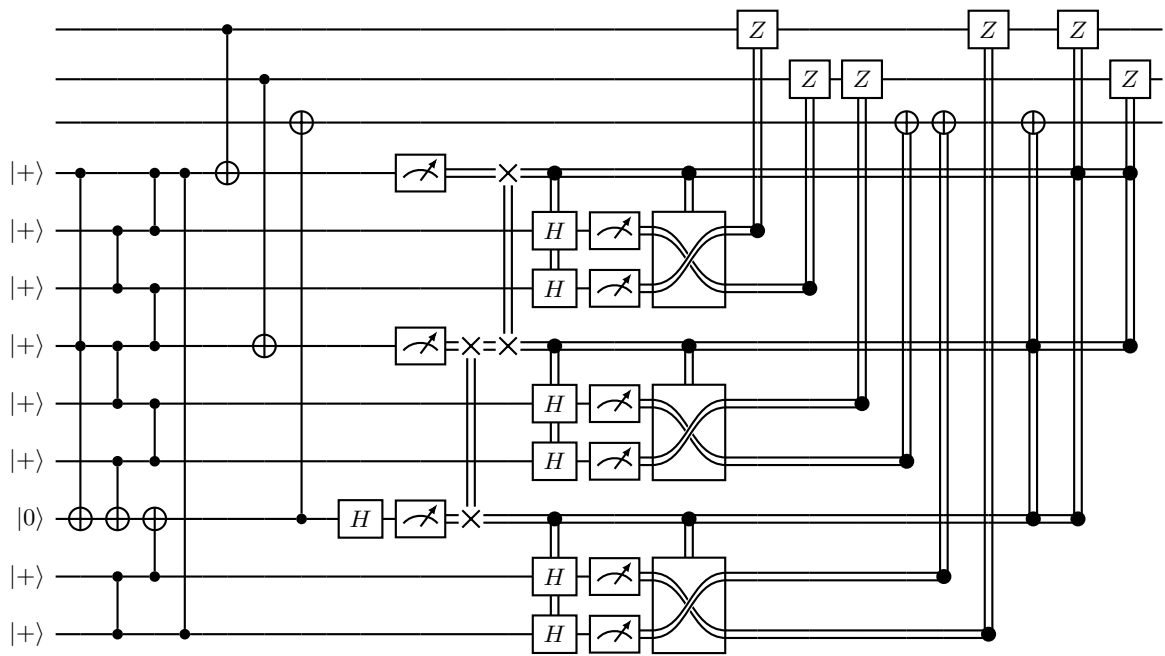
\begin{figure*}
\begin{quantikz}[row sep=0.5em, column sep=0.5em]
                        &\qw      & \qw        & \qw       & \qw      & \qw & \ctrl{3} & \qw      & \qw       & \qw & \qw      & \qw      & \qw              & \qw              & \qw        & \qw       & \qw        & \gate{Z}    & \qw         & \qw         & \qw         & \qw         & \gate{Z}     & \qw         & \gate{Z}    & \qw         & \qw \\
                        &\qw      & \qw        & \qw       & \qw      & \qw & \qw      & \ctrl{5} & \qw       & \qw & \qw      & \qw      & \qw              & \qw              & \qw        & \qw       & \qw        & \qw         & \gate{Z}    & \gate{Z}    & \qw         & \qw         & \qw          & \qw         & \qw         & \gate{Z}    & \qw \\
                        &\qw      & \qw        & \qw       & \qw      & \qw & \qw      & \qw      & \targ{}   & \qw & \qw      & \qw      & \qw              & \qw              & \qw        & \qw       & \qw        & \qw         & \qw         & \qw         & \targ{}     & \targ{}     & \qw          & \targ{}     & \qw         & \qw         & \qw \\
\lstick{$\ket{+}$}      &\ctrl{6} & \qw        & \ctrl{1}  & \ctrl{8} & \qw & \targ{}  & \qw      & \qw       & \qw & \qw      & \meter{} & \cw              & \ctargX{}\vcw{3} & \cwbend{2} & \cw       & \cwbend{1} & \cw         & \cw         & \cw         & \cw         & \cw         & \cw          & \cw         & \cwbend{-3} & \cwbend{-2} &     \\
\lstick{$\ket{+}$}      &\qw      & \ctrl{1}   & \ctrl{0}  & \qw      & \qw & \qw      & \qw      & \qw       & \qw & \qw      & \qw      & \qw              & \qw              & \gate{H}   & \meter{}  & \cwap{}    & \cwbend{-4} &             &             &             &             &              &             &             &             &     \\
\lstick{$\ket{+}$}      &\qw      & \ctrl{0}   & \ctrl{1}  & \qw      & \qw & \qw      & \qw      & \qw       & \qw & \qw      & \qw      & \qw              & \qw              & \gate{H}   & \meter{}  &            & \cw         & \cwbend{-4} &             &             &             &              &             &             &             &     \\
\lstick{$\ket{+}$}      &\ctrl{}  & \ctrl{1}   & \ctrl{0}  & \qw      & \qw & \qw      & \targ{}  & \qw       & \qw & \qw      & \meter{} & \ctargX{}\vcw{3} & \ctargX{}        & \cwbend{2} & \cw       & \cwbend{1} & \cw         & \cw         & \cw         & \cw         & \cw         & \cw          & \cwbend{-4} & \cw         & \cwbend{-3} &     \\
\lstick{$\ket{+}$}      &\qw      & \ctrl{0}   & \ctrl{1}  & \qw      & \qw & \qw      & \qw      & \qw       & \qw & \qw      & \qw      & \qw              & \qw              & \gate{H}   & \meter{}  & \cwap{}    & \cw         & \cw         & \cwbend{-6} &             &             &              &             &             &             &     \\
\lstick{$\ket{+}$}      &\qw      & \ctrl{1}   & \ctrl{0}  & \qw      & \qw & \qw      & \qw      & \qw       & \qw & \qw      & \qw      & \qw              & \qw              & \gate{H}   & \meter{}  &            & \cw         & \cw         & \cw         & \cwbend{-6} &             &              &             &             &             &     \\
\lstick{$\ket{0}$}      &\targ{}  & \targ{}    & \targ{}   & \qw      & \qw & \qw      & \qw      & \ctrl{-7} & \qw & \gate{H} & \meter{} & \ctargX{}        & \cw              & \cwbend{2} & \cw       & \cwbend{1} & \cw         & \cw         & \cw         & \cw         & \cw         & \cw          & \cwbend{-3} & \cwbend{-6} &             &     \\
\lstick{$\ket{+}$}      &\qw      & \ctrl{1}   & \ctrl{-1} & \qw      & \qw & \qw      & \qw      & \qw       & \qw & \qw      & \qw      & \qw              & \qw              & \gate{H}   & \meter{}  & \cwap{}    & \cw         & \cw         & \cw         & \cw         & \cwbend{-8} &              &             &             &             &     \\
\lstick{$\ket{+}$}      &\qw      & \ctrl{0}   & \qw       & \ctrl{0} & \qw & \qw      & \qw      & \qw       & \qw & \qw      & \qw      & \qw              & \qw              & \gate{H}   & \meter{}  &            & \cw         & \cw         & \cw         & \cw         & \cw         & \cwbend{-11} &             &             &             &
\end{quantikz}
 \caption{Toffoli teleportation with an autocorrected magic state, directly adapted from~\cite{Fowler2019Flexiblelayoutsurface}.
	The only non Clifford resource is the standard Toffoli magic state (same as in \autoref{fig:toffoli_teleport}) that can be obtained by applying a Toffoli gate on $\ket{+}\ket{+}\ket{0}$ or as detailed in \autoref{appendix:code:toffoli}.
}\label{fig:toffoli_teleport_auto}
\end{figure*}

For simplicity, we did not include any parallelization in this work.
In principle however, with its ``left'' and ``right' buses, the layout presented in \autoref{fig:architecture} would allow essentially two 2-qubit gates to be performed simultaneously (this is slightly restrictive as a bus can be used for implementing 2-qubit gates simultaneously in case there is no overlap between the routing qubits required to connect the different pairs).
For an augmented parallelization, wider busses can be used on the sides of the processor, or better, a 2D layout can be adopted by arranging several columns of logical qubits, similarly to what is usually considered for surface code (either square or rectangular)~\cite{Fowler2019Flexiblelayoutsurface,LitinskiQ2019GameSurfaceCodes,CampbellPQ2022UniversalQuantumComputing}.

A key element to reduce run-times is to use Fowler's time optimal quantum computing~\cite{Fowler2013Timeoptimalquantum} to trade between time and space through the use of teleportations.
When combined with auto-corrected magic state~\cite{Fowler2019Flexiblelayoutsurface}, the Clifford correction following the gate teleportation can be implemented off-line (remaining Pauli corrections can be followed by the control computer instead of being applied) and the space overhead of Fowler's original trick is reduced.
We present in \autoref{fig:toffoli_teleport_auto} an implementation of the Toffoli teleportation directly derived from~\cite{Fowler2019Flexiblelayoutsurface}.
Note that the first Toffoli gate applied on $\ket{+}\ket{+}\ket{0}$ corresponds to the preparation of a Toffoli magic state (the same consumed by \autoref{fig:toffoli_teleport}) and can be obtained according to \autoref{appendix:code:toffoli}.

\subsection{Alternative additions}

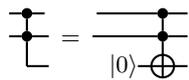
\begin{figure}[h]
\begin{quantikz}[column sep=0.5em,row sep=0.5em]
&\ctrl{1}&\qw\midstick[3,brackets=none]{=}&\qw    &\ctrl{1}&\qw\\
&\ctrl{1}&\qw                             &\qw    &\ctrl{1}&\qw\\
&        &\qw                             &\ket{0}&\targ{} &\qw
\end{quantikz}
 \caption{AND gate.}\label{fig:and}
\end{figure}

\begin{figure}[h]
\begin{quantikz}[column sep=0.5em,row sep=0.5em]
&\ctrl{1}&\qw& &&\qw  &\ctrl{1}&\qw               & &&\qw     &\qw     &\ctrl{1}   &\qw\\
&\ctrl{1}&\qw&=&&\qw  &\ctrl{1}&\qw               &=&&\qw     &\qw     &\gate{Z}   &\qw \\
&\qw     &   & &&\qw  &\targ{} &\push{\ket{0}} \qw& &&\gate{H}&\meter{}&\cwbend{-1}&
\end{quantikz}
 \caption{AND uncomputation, and its measurement-based uncomputation.}\label{fig:deand}
\end{figure}
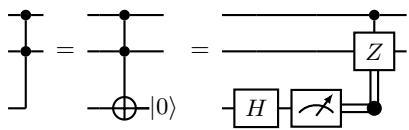

In the description of the algorithm presented in the previous sections, we considered addition circuits between two quantum registers derived from Cuccaro's adder~\cite{Moulton2004newquantumripple}.
It uses only one ancillary qubit and is based on Toffoli gates, the non-Clifford gate available on our platform.
Alternative adders exist, like Gidney adder~\cite{GidneyQ2018Halvingcostquantum}.
It replaces half (or a third for controlled additions) of Toffoli gates (and associated magic state preparation and teleportation) by a measurement-based circuit using only Clifford gates.
The idea is to store the result of an AND gate in a freshly initialized qubit (an AND gate being a Toffoli gate targeting an ancillary qubit initialized in $\ket{0}$, see \autoref{fig:and}), and replace the Toffoli required to reset the ancillary qubit to $\ket{0}$ by a measurement-based uncomputation~\cite{JonesPRA2013Lowoverheadconstructions,GidneyQ2018Halvingcostquantum,NevenPRX2018EncodingElectronicSpectra}, as shown in \autoref{fig:deand}.
Gidney adder requires more ancilliary qubits but there might be unused qubits in the algorithm depending on where the addition is needed~\footnote{The precise advantage brought by Gidney's adder also depends on the exact parameters of the algorithm.
	If Toffoli magic states are available when the addition have to be performed and if the teleportation is cheaper that measurement-based uncomputing, it might still be more interesting to use Cuccaro's adder.
	Also note that the Toffoli magic states are typically slightly noisier than Clifford gates, and for this reason reducing the number of Toffoli gates can be beneficial.
	Only a careful resource evaluation can really decide which adder circuit to use for each addition of the algorithm.
}.
In the following figures, we present the circuits exploiting the AND gate and its uncomputation which would replace Fig.\,\ref{fig:add_mod:add}, \ref{fig:add_mod:compair_uncompute}, \ref{fig:mult:add_ctrl_nomod}, \ref{fig:mult:add}, and \ref{fig:controlled_add}.

\begin{figure}[h]
\resizebox{0.95\linewidth}{!}{\begin{quantikz}[column sep=0.3em,row sep=0.3em]
\lstick{$\ket{x_0}$}&\ctrl{1}&\qw     &\qw     &\qw     &[1em]\qw                                                                                             &\qw     &\qw     &\qw       &\qw     &\qw     &\qw     &\qw     &\qw     &\qw     &\qw     &\qw     &\qw     &\qw     &\qw     &\qw     &\qw     &\qw     &[1em]\qw&\qw     &\qw     &\qw     &\ctrl{1}&\ctrl{1}&\rstick{$\ket{x_0}$}\qw \\
\lstick{$\ket{y_0}$}&\ctrl{1}&\qw     &\qw     &\qw     &\qw                                                                                                  &\qw     &\qw     &\qw       &\qw     &\qw     &\qw     &\qw     &\qw     &\qw     &\qw     &\qw     &\qw     &\qw     &\qw     &\qw     &\qw     &\qw     &\qw     &\qw     &\qw     &\qw     &\ctrl{1}&\targ{} &\rstick{$\ket{z_0}$}\qw \\
                    &        &\ctrl{2}&\qw     &\ctrl{3}&\qw                                                                                                  &\qw     &\qw     &\qw       &\qw     &\qw     &\qw     &\qw     &\qw     &\qw     &\qw     &\qw     &\qw     &\qw     &\qw     &\qw     &\qw     &\qw     &\ctrl{3}&\qw     &\ctrl{1}&\qw     &\qw     &        &                        \\
\lstick{$\ket{x_1}$}&\qw     &\targ{} &\ctrl{1}&\qw     &\qw                                                                                                  &\qw     &\qw     &\qw       &\qw     &\qw     &\qw     &\qw     &\qw     &\qw     &\qw     &\qw     &\qw     &\qw     &\qw     &\qw     &\qw     &\qw     &\qw     &\ctrl{1}&\targ{} &\ctrl{1}&\qw     &\qw     &\rstick{$\ket{x_1}$}\qw \\
\lstick{$\ket{y_1}$}&\qw     &\targ{} &\ctrl{1}&\qw     &\qw                                                                                                  &\qw     &\qw     &\qw       &\qw     &\qw     &\qw     &\qw     &\qw     &\qw     &\qw     &\qw     &\qw     &\qw     &\qw     &\qw     &\qw     &\qw     &\qw     &\ctrl{1}&\qw     &\targ{} &\qw     &\qw     &\rstick{$\ket{z_1}$}\qw \\
                    &        &        &        &\targ{} &\ctrl{2}\gategroup[wires=4,steps=18,style={dashed,rounded corners,inner xsep=0.5em,inner ysep=0em}]{}&\qw     &\ctrl{3}&\qw       &\qw     &\qw     &\qw     &\qw     &\qw     &\qw     &\qw     &\qw     &\qw     &\qw     &\ctrl{3}&\qw     &\ctrl{1}&\qw     &\targ{} &\qw     &        &        &        &        &                        \\
\lstick{$\ket{x_2}$}&\qw     &\qw     &\qw     &\qw     &\targ{}                                                                                              &\ctrl{1}&\qw     &\qw       &\qw     &\qw     &\qw     &\qw     &\qw     &\qw     &\qw     &\qw     &\qw     &\qw     &\qw     &\ctrl{1}&\targ{} &\ctrl{1}&\qw     &\qw     &\qw     &\qw     &\qw     &\qw     &\rstick{$\ket{x_2}$}\qw \\
\lstick{$\ket{y_2}$}&\qw     &\qw     &\qw     &\qw     &\targ{}                                                                                              &\ctrl{1}&\qw     &\qw       &\qw     &\qw     &\qw     &\qw     &\qw     &\qw     &\qw     &\qw     &\qw     &\qw     &\qw     &\ctrl{1}&\qw     &\targ{} &\qw     &\qw     &\qw     &\qw     &\qw     &\qw     &\rstick{$\ket{z_2}$}\qw \\
                    &        &        &        &        &                                                                                                     &        &\targ{} &\vqw{1}\qw&        &        &        &        &        &        &        &        &        &\vqw{1} &\targ{} &\qw     &        &        &        &        &        &        &        &        &                        \\[1ex]
                    &        &        &        &        &                                                                                                     &        &        &          &\ctrl{2}&\qw     &\ctrl{3}&\qw     &\qw     &\ctrl{3}&\qw     &\ctrl{1}&\qw     &\qw     &        &        &        &        &        &        &        &        &        &        &                        \\
\lstick{$\ket{x_3}$}&\qw     &\qw     &\qw     &\qw     &\qw                                                                                                  &\qw     &\qw     &\qw       &\targ{} &\ctrl{1}&\qw     &\qw     &\qw     &\qw     &\ctrl{1}&\targ{} &\ctrl{1}&\qw     &\qw     &\qw     &\qw     &\qw     &\qw     &\qw     &\qw     &\qw     &\qw     &\qw     &\rstick{$\ket{x_3}$}\qw \\
\lstick{$\ket{y_3}$}&\qw     &\qw     &\qw     &\qw     &\qw                                                                                                  &\qw     &\qw     &\qw       &\targ{} &\ctrl{1}&\qw     &\qw     &\qw     &\qw     &\ctrl{1}&\qw     &\targ{} &\qw     &\qw     &\qw     &\qw     &\qw     &\qw     &\qw     &\qw     &\qw     &\qw     &\qw     &\rstick{$\ket{z_3}$}\qw \\
                    &        &        &        &        &                                                                                                     &        &        &          &        &        &\targ{} &\ctrl{2}&\qw     &\targ{} &\qw     &        &        &        &        &        &        &        &        &        &        &        &        &        &                        \\
\lstick{$\ket{x_4}$}&\qw     &\qw     &\qw     &\qw     &\qw                                                                                                  &\qw     &\qw     &\qw       &\qw     &\qw     &\qw     &\qw     &\ctrl{1}&\qw     &\qw     &\qw     &\qw     &\qw     &\qw     &\qw     &\qw     &\qw     &\qw     &\qw     &\qw     &\qw     &\qw     &\qw     &\rstick{$\ket{x_4}$}\qw \\
\lstick{$\ket{y_4}$}&\qw     &\qw     &\qw     &\qw     &\qw                                                                                                  &\qw     &\qw     &\qw       &\qw     &\qw     &\qw     &\targ{} &\targ{} &\qw     &\qw     &\qw     &\qw     &\qw     &\qw     &\qw     &\qw     &\qw     &\qw     &\qw     &\qw     &\qw     &\qw     &\qw     &\rstick{$\ket{z_4}$}\qw
\end{quantikz} }
\caption{Gidney's addition circuit modulo $2^n$ from~\cite{GidneyQ2018Halvingcostquantum}, with $n$ the number of qubits of $x$ and $y$ (5 in the figure).
	The result is $z = x + y \mod{2^n}$.
	The boxed part is repeated for each qubit pair (except first and last).
	This scheme performs the same task as \autoref{fig:mult:add}.
}
\end{figure}
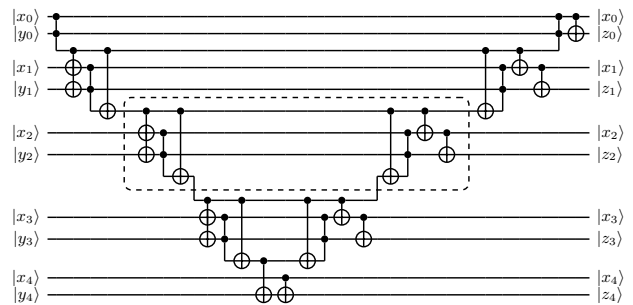

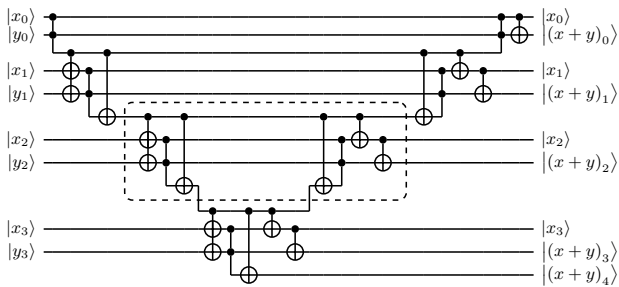
\begin{figure}[h]
\resizebox{0.95\linewidth}{!}{\begin{quantikz}[column sep=0.3em,row sep=0.3em]
\lstick{$\ket{x_0}$}&\ctrl{1} &\qw     &\qw     &\qw     &[1em]\qw                                                                                             &\qw     &\qw     &\qw       &\qw     &\qw     &\qw     &\qw     &\qw     &\qw    &\qw     &\qw     &\qw     &\qw     &[1em]\qw&\qw     &\qw     &\qw     &\ctrl{1}&\ctrl{1}&\rstick{$\ket{x_0}$}      \qw \\
\lstick{$\ket{y_0}$}&\ctrl{1} &\qw     &\qw     &\qw     &\qw                                                                                                  &\qw     &\qw     &\qw       &\qw     &\qw     &\qw     &\qw     &\qw     &\qw    &\qw     &\qw     &\qw     &\qw     &\qw     &\qw     &\qw     &\qw     &\ctrl{1}&\targ{} &\rstick{$\ket{{(x+y)}_0}$}\qw \\
                    &         &\ctrl{2}&\qw     &\ctrl{3}&\qw                                                                                                  &\qw     &\qw     &\qw       &\qw     &\qw     &\qw     &\qw     &\qw     &\qw    &\qw     &\qw     &\qw     &\qw     &\ctrl{3}&\qw     &\ctrl{1}&\qw     &\qw     &        &                              \\
\lstick{$\ket{x_1}$}&\qw      &\targ{} &\ctrl{1}&\qw     &\qw                                                                                                  &\qw     &\qw     &\qw       &\qw     &\qw     &\qw     &\qw     &\qw     &\qw    &\qw     &\qw     &\qw     &\qw     &\qw     &\ctrl{1}&\targ{} &\ctrl{1}&\qw     &\qw     &\rstick{$\ket{x_1}$}      \qw \\
\lstick{$\ket{y_1}$}&\qw      &\targ{} &\ctrl{1}&\qw     &\qw                                                                                                  &\qw     &\qw     &\qw       &\qw     &\qw     &\qw     &\qw     &\qw     &\qw    &\qw     &\qw     &\qw     &\qw     &\qw     &\ctrl{1}&\qw     &\targ{} &\qw     &\qw     &\rstick{$\ket{{(x+y)}_1}$}\qw \\
                    &         &        &        &\targ{} &\ctrl{2}\gategroup[wires=4,steps=14,style={dashed,rounded corners,inner xsep=0.5em,inner ysep=0em}]{}&\qw     &\ctrl{3}&\qw       &\qw     &\qw     &\qw     &\qw     &\qw     &\qw    &\ctrl{3}&\qw     &\ctrl{1}&\qw     &\targ{} &\qw     &        &        &        &        &                              \\
\lstick{$\ket{x_2}$}&\qw      &\qw     &\qw     &\qw     &\targ{}                                                                                              &\ctrl{1}&\qw     &\qw       &\qw     &\qw     &\qw     &\qw     &\qw     &\qw    &\qw     &\ctrl{1}&\targ{} &\ctrl{1}&\qw     &\qw     &\qw     &\qw     &\qw     &\qw     &\rstick{$\ket{x_2}$}      \qw \\
\lstick{$\ket{y_2}$}&\qw      &\qw     &\qw     &\qw     &\targ{}                                                                                              &\ctrl{1}&\qw     &\qw       &\qw     &\qw     &\qw     &\qw     &\qw     &\qw    &\qw     &\ctrl{1}&\qw     &\targ{} &\qw     &\qw     &\qw     &\qw     &\qw     &\qw     &\rstick{$\ket{{(x+y)}_2}$}\qw \\
                    &         &        &        &        &                                                                                                     &        &\targ{} &\vqw{1}\qw&        &        &        &        &        &\vqw{1}&\targ{} &\qw     &        &        &        &        &        &        &        &        &                              \\[1ex]
                    &         &        &        &        &                                                                                                     &        &        &          &\ctrl{2}&\qw     &\ctrl{3}&\ctrl{1}&\qw     &\qw    &        &        &        &        &        &        &        &        &        &        &                              \\
\lstick{$\ket{x_3}$}&\qw      &\qw     &\qw     &\qw     &\qw                                                                                                  &\qw     &\qw     &\qw       &\targ{} &\ctrl{1}&\qw     &\targ{} &\ctrl{1}&\qw    &\qw     &\qw     &\qw     &\qw     &\qw     &\qw     &\qw     &\qw     &\qw     &\qw     &\rstick{$\ket{x_3}$}      \qw \\
\lstick{$\ket{y_3}$}&\qw      &\qw     &\qw     &\qw     &\qw                                                                                                  &\qw     &\qw     &\qw       &\targ{} &\ctrl{1}&\qw     &\qw     &\targ{} &\qw    &\qw     &\qw     &\qw     &\qw     &\qw     &\qw     &\qw     &\qw     &\qw     &\qw     &\rstick{$\ket{{(x+y)}_3}$}\qw \\
                    &         &        &        &        &                                                                                                     &        &        &          &        &        &\targ{} &\qw     &\qw     &\qw    &\qw     &\qw     &\qw     &\qw     &\qw     &\qw     &\qw     &\qw     &\qw     &\qw     &\rstick{$\ket{{(x+y)}_4}$}\qw
\end{quantikz} }
\caption{Nonmodular adder circuit using AND gates.
	The boxed part is repeated for each qubit pair (except first and last).
	This scheme performs the same task as \autoref{fig:add_mod:add}.
}
\end{figure}

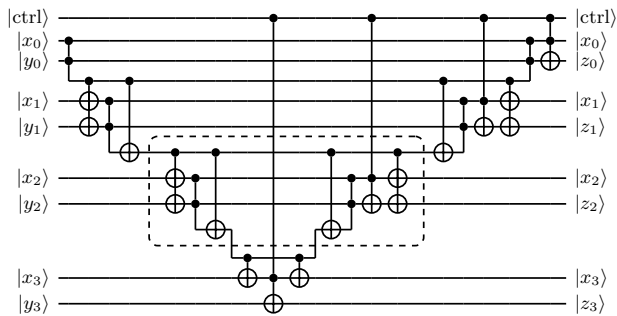
\begin{figure}[h]
\resizebox{0.95\linewidth}{!}{\begin{quantikz}[column sep=0.3em,row sep=0.3em]
\lstick{$\ket{\text{ctrl}}$}&\qw      &\qw     &\qw     &\qw     &[1em]\qw                                                                                             &\qw     &\qw     &\qw       &\qw     &\ctrl{11}&\qw     &\qw         &\qw     &\qw     &\ctrl{7}&\qw     &[1em]\qw&\qw     &\ctrl{4}&\qw     &\qw     &\ctrl{1}&\rstick{$\ket{\text{ctrl}}$}\qw \\[1ex]
\lstick{$\ket{x_0}$}        &\ctrl{1} &\qw     &\qw     &\qw     &\qw                                                                                                  &\qw     &\qw     &\qw       &\qw     &\qw      &\qw     &\qw         &\qw     &\qw     &\qw     &\qw     &\qw     &\qw     &\qw     &\qw     &\ctrl{1}&\ctrl{1}&\rstick{$\ket{x_0}$}        \qw \\
\lstick{$\ket{y_0}$}        &\ctrl{1} &\qw     &\qw     &\qw     &\qw                                                                                                  &\qw     &\qw     &\qw       &\qw     &\qw      &\qw     &\qw         &\qw     &\qw     &\qw     &\qw     &\qw     &\qw     &\qw     &\qw     &\ctrl{1}&\targ{} &\rstick{$\ket{z_0}$}        \qw \\
                            &         &\ctrl{2}&\qw     &\ctrl{3}&\qw                                                                                                  &\qw     &\qw     &\qw       &\qw     &\qw      &\qw     &\qw         &\qw     &\qw     &\qw     &\qw     &\ctrl{3}&\qw     &\qw     &\ctrl{2}&\qw     &        &                                \\
\lstick{$\ket{x_1}$}        &\qw      &\targ{} &\ctrl{1}&\qw     &\qw                                                                                                  &\qw     &\qw     &\qw       &\qw     &\qw      &\qw     &\qw         &\qw     &\qw     &\qw     &\qw     &\qw     &\ctrl{1}&\ctrl{1}&\targ{} &\qw     &\qw     &\rstick{$\ket{x_1}$}        \qw \\
\lstick{$\ket{y_1}$}        &\qw      &\targ{} &\ctrl{1}&\qw     &\qw                                                                                                  &\qw     &\qw     &\qw       &\qw     &\qw      &\qw     &\qw         &\qw     &\qw     &\qw     &\qw     &\qw     &\ctrl{1}&\targ{} &\targ{} &\qw     &\qw     &\rstick{$\ket{z_1}$}        \qw \\
                            &         &        &        &\targ{} &\ctrl{2}\gategroup[wires=4,steps=12,style={dashed,rounded corners,inner xsep=0.5em,inner ysep=0em}]{}&\qw     &\ctrl{3}&\qw       &\qw     &\qw      &\qw     &\qw         &\ctrl{3}&\qw     &\qw     &\ctrl{2}&\targ{} &\qw     &        &        &        &        &                                \\
\lstick{$\ket{x_2}$}        &\qw      &\qw     &\qw     &\qw     &\targ{}                                                                                              &\ctrl{1}&\qw     &\qw       &\qw     &\qw      &\qw     &\qw         &\qw     &\ctrl{1}&\ctrl{1}&\targ{} &\qw     &\qw     &\qw     &\qw     &\qw     &\qw     &\rstick{$\ket{x_2}$}        \qw \\
\lstick{$\ket{y_2}$}        &\qw      &\qw     &\qw     &\qw     &\targ{}                                                                                              &\ctrl{1}&\qw     &\qw       &\qw     &\qw      &\qw     &\qw         &\qw     &\ctrl{1}&\targ{} &\targ{} &\qw     &\qw     &\qw     &\qw     &\qw     &\qw     &\rstick{$\ket{z_2}$}        \qw \\
                            &         &        &        &        &                                                                                                     &        &\targ{} &\vqw{1}\qw&        &         &        &\vqw{1}     &\targ{} &\qw     &        &        &        &        &        &        &        &        &                                \\[1ex]
                            &         &        &        &        &                                                                                                     &        &        &          &\ctrl{1}&\qw      &\ctrl{1}&\qw         &        &        &        &        &        &        &        &        &        &        &                                \\
\lstick{$\ket{x_3}$}        &\qw      &\qw     &\qw     &\qw     &\qw                                                                                                  &\qw     &\qw     &\qw       &\targ{} &\ctrl{1} &\targ{} &\qw         &\qw     &\qw     &\qw     &\qw     &\qw     &\qw     &\qw     &\qw     &\qw     &\qw     &\rstick{$\ket{x_3}$}        \qw \\
\lstick{$\ket{y_3}$}        &\qw      &\qw     &\qw     &\qw     &\qw                                                                                                  &\qw     &\qw     &\qw       &\qw     &\targ{}  &\qw     &\qw         &\qw     &\qw     &\qw     &\qw     &\qw     &\qw     &\qw     &\qw     &\qw     &\qw     &\rstick{$\ket{z_3}$}        \qw
\end{quantikz} }
\caption{Controlled adder modulo $2^n$ circuit using AND gates, with $n$ the number of qubits of the inputs.
	The result is $z = y + x.\text{ctrl} \mod{2^n}$.
	The boxed part comes from~\cite{GidneyQ2018Halvingcostquantum} (the represented Toffoli gate should be implemented as such with the architecture studied here but can be replaced by a AND gate followed by a CNOT and the uncomputation of AND gate) and is repeated for each qubit pair (except first and last).
	This scheme performs the same task as \autoref{fig:controlled_add}.
}
\end{figure}

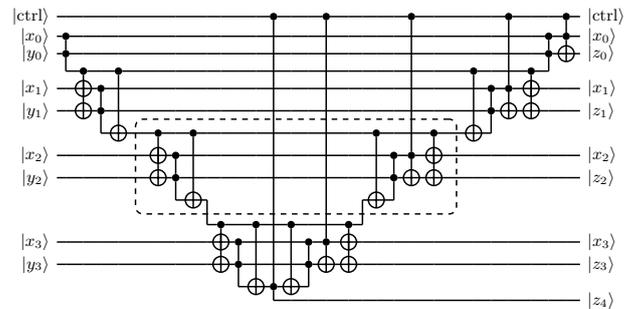
\begin{figure}[h]
\resizebox{0.95\linewidth}{!}{\begin{quantikz}[column sep=0.3em,row sep=0.3em]
\lstick{$\ket{\text{ctrl}}$}&\qw      &\qw     &\qw     &\qw     &[1em]\qw                                                                                             &\qw     &\qw     &\qw       &\qw     &\qw     &\qw     &\ctrl{13}&\qw     &\qw     &\ctrl{11}&\qw     &\qw    &\qw     &\qw     &\ctrl{7}&\qw     &[1em]\qw&\qw     &\ctrl{4}&\qw     &\qw     &\ctrl{1}&\rstick{$\ket{\text{ctrl}}$} \qw \\[1ex]
\lstick{$\ket{x_0}$}        &\ctrl{1} &\qw     &\qw     &\qw     &\qw                                                                                                  &\qw     &\qw     &\qw       &\qw     &\qw     &\qw     &\qw      &\qw     &\qw     &\qw      &\qw     &\qw    &\qw     &\qw     &\qw     &\qw     &\qw     &\qw     &\qw     &\qw     &\ctrl{1}&\ctrl{1}&\rstick{$\ket{x_0}$}         \qw \\
\lstick{$\ket{y_0}$}        &\ctrl{1} &\qw     &\qw     &\qw     &\qw                                                                                                  &\qw     &\qw     &\qw       &\qw     &\qw     &\qw     &\qw      &\qw     &\qw     &\qw      &\qw     &\qw    &\qw     &\qw     &\qw     &\qw     &\qw     &\qw     &\qw     &\qw     &\ctrl{1}&\targ{} &\rstick{$\ket{z_0}$}         \qw \\
                            &         &\ctrl{2}&\qw     &\ctrl{3}&\qw                                                                                                  &\qw     &\qw     &\qw       &\qw     &\qw     &\qw     &\qw      &\qw     &\qw     &\qw      &\qw     &\qw    &\qw     &\qw     &\qw     &\qw     &\ctrl{3}&\qw     &\qw     &\ctrl{2}&\qw     &        &                                 \\
\lstick{$\ket{x_1}$}        &\qw      &\targ{} &\ctrl{1}&\qw     &\qw                                                                                                  &\qw     &\qw     &\qw       &\qw     &\qw     &\qw     &\qw      &\qw     &\qw     &\qw      &\qw     &\qw    &\qw     &\qw     &\qw     &\qw     &\qw     &\ctrl{1}&\ctrl{1}&\targ{} &\qw     &\qw     &\rstick{$\ket{x_1}$}         \qw \\
\lstick{$\ket{y_1}$}        &\qw      &\targ{} &\ctrl{1}&\qw     &\qw                                                                                                  &\qw     &\qw     &\qw       &\qw     &\qw     &\qw     &\qw      &\qw     &\qw     &\qw      &\qw     &\qw    &\qw     &\qw     &\qw     &\qw     &\qw     &\ctrl{1}&\targ{} &\targ{} &\qw     &\qw     &\rstick{$\ket{z_1}$}         \qw \\
                            &         &        &        &\targ{} &\ctrl{2}\gategroup[wires=4,steps=17,style={dashed,rounded corners,inner xsep=0.5em,inner ysep=0em}]{}&\qw     &\ctrl{3}&\qw       &\qw     &\qw     &\qw     &\qw      &\qw     &\qw     &\qw      &\qw     &\qw    &\ctrl{3}&\qw     &\qw     &\ctrl{2}&\targ{} &\qw     &        &        &        &        &                                 \\
\lstick{$\ket{x_2}$}        &\qw      &\qw     &\qw     &\qw     &\targ{}                                                                                              &\ctrl{1}&\qw     &\qw       &\qw     &\qw     &\qw     &\qw      &\qw     &\qw     &\qw      &\qw     &\qw    &\qw     &\ctrl{1}&\ctrl{1}&\targ{} &\qw     &\qw     &\qw     &\qw     &\qw     &\qw     &\rstick{$\ket{x_2}$}         \qw \\
\lstick{$\ket{y_2}$}        &\qw      &\qw     &\qw     &\qw     &\targ{}                                                                                              &\ctrl{1}&\qw     &\qw       &\qw     &\qw     &\qw     &\qw      &\qw     &\qw     &\qw      &\qw     &\qw    &\qw     &\ctrl{1}&\targ{} &\targ{} &\qw     &\qw     &\qw     &\qw     &\qw     &\qw     &\rstick{$\ket{z_2}$}         \qw \\
                            &         &        &        &        &                                                                                                     &        &\targ{} &\vqw{1}\qw&        &        &        &         &        &        &         &        &\vqw{1}&\targ{} &\qw     &        &        &        &        &        &        &        &        &                                 \\[1ex]
                            &         &        &        &        &                                                                                                     &        &        &          &\ctrl{2}&\qw     &\ctrl{3}&\qw      &\ctrl{3}&\qw     &\qw      &\ctrl{2}&\qw    &        &        &        &        &        &        &        &        &        &        &                                 \\
\lstick{$\ket{x_3}$}        &\qw      &\qw     &\qw     &\qw     &\qw                                                                                                  &\qw     &\qw     &\qw       &\targ{} &\ctrl{1}&\qw     &\qw      &\qw     &\ctrl{1}&\ctrl{1} &\targ{} &\qw    &\qw     &\qw     &\qw     &\qw     &\qw     &\qw     &\qw     &\qw     &\qw     &\qw     &\rstick{$\ket{x_3}$}         \qw \\
\lstick{$\ket{y_3}$}        &\qw      &\qw     &\qw     &\qw     &\qw                                                                                                  &\qw     &\qw     &\qw       &\targ{} &\ctrl{1}&\qw     &\qw      &\qw     &\ctrl{1}&\targ{}  &\targ{} &\qw    &\qw     &\qw     &\qw     &\qw     &\qw     &\qw     &\qw     &\qw     &\qw     &\qw     &\rstick{$\ket{z_3}$}         \qw \\
                            &         &        &        &        &                                                                                                     &        &        &          &        &        &\targ{} &\ctrl{1} &\targ{} &\qw     &         &        &       &        &        &        &        &        &        &        &        &        &        &                                 \\
                            &         &        &        &        &                                                                                                     &        &        &          &        &        &        &         &\qw     &\qw     &\qw      &\qw     &\qw    &\qw     &\qw     &\qw     &\qw     &\qw     &\qw     &\qw     &\qw     &\qw     &\qw     &\rstick{$\ket{z_4}$}         \qw
\end{quantikz} }
\caption{Controlled nonmodular adder circuit using AND gates.
	The result is $z = y + x.\text{ctrl}$.
	The boxed part (still from~\cite{GidneyQ2018Halvingcostquantum}) is repeated for each qubit pair (except first and last).
	This scheme performs the same task as \autoref{fig:mult:add_ctrl_nomod}.
}
\end{figure}

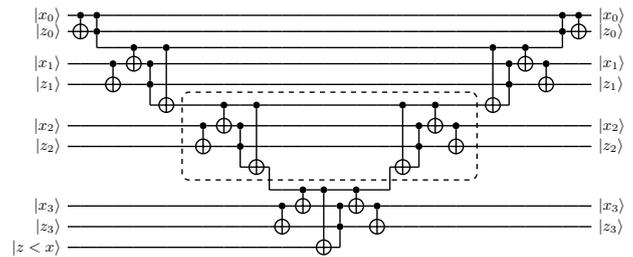
\begin{figure}[h]
\resizebox{0.95\linewidth}{!}{\begin{quantikz}[column sep=0.3em,row sep=0.3em]
\lstick{$\ket{x_0}$}&\ctrl{1}&\ctrl{1}&\qw     &\qw     &\qw     &\qw     &[1em]\qw                                                                                        &\qw     &\qw     &\qw     &\qw       &\qw     &\qw     &\qw     &\qw     &\qw     &\qw     &\qw    &\qw     &\qw     &\qw     &\qw     &[1em]\qw&\qw     &\qw     &\qw     &\ctrl{1}&\ctrl{1}&\rstick{$\ket{x_0}$}\qw\\
\lstick{$\ket{z_0}$}&\targ{} &\ctrl{1}&\qw     &\qw     &\qw     &\qw     &\qw                                                                                             &\qw     &\qw     &\qw     &\qw       &\qw     &\qw     &\qw     &\qw     &\qw     &\qw     &\qw    &\qw     &\qw     &\qw     &\qw     &\qw     &\qw     &\qw     &\qw     &\ctrl{1}&\targ{} &\rstick{$\ket{z_0}$}\qw\\
                    &        &        &\qw     &\ctrl{1}&\qw     &\ctrl{3}&\qw                                                                                             &\qw     &\qw     &\qw     &\qw       &\qw     &\qw     &\qw     &\qw     &\qw     &\qw     &\qw    &\qw     &\qw     &\qw     &\qw     &\ctrl{3}&\qw     &\ctrl{1}&\qw     &\qw     &        &                       \\
\lstick{$\ket{x_1}$}&\qw     &\qw     &\ctrl{1}&\targ{} &\ctrl{1}&\qw     &\qw                                                                                             &\qw     &\qw     &\qw     &\qw       &\qw     &\qw     &\qw     &\qw     &\qw     &\qw     &\qw    &\qw     &\qw     &\qw     &\qw     &\qw     &\ctrl{1}&\targ{} &\ctrl{1}&\qw     &\qw     &\rstick{$\ket{x_1}$}\qw\\
\lstick{$\ket{z_1}$}&\qw     &\qw     &\targ{} &\qw     &\ctrl{1}&\qw     &\qw                                                                                             &\qw     &\qw     &\qw     &\qw       &\qw     &\qw     &\qw     &\qw     &\qw     &\qw     &\qw    &\qw     &\qw     &\qw     &\qw     &\qw     &\ctrl{1}&\qw     &\targ{} &\qw     &\qw     &\rstick{$\ket{z_1}$}\qw\\
                    &        &        &        &        &        &\targ{} &\qw\gategroup[wires=4,steps=16,style={dashed,rounded corners,inner xsep=0.5em,inner ysep=0em}]{}&\ctrl{1}&\qw     &\ctrl{3}&\qw       &\qw     &\qw     &\qw     &\qw     &\qw     &\qw     &\qw    &\ctrl{3}&\qw     &\ctrl{1}&\qw     &\targ{} &\qw     &        &        &        &        &                       \\
\lstick{$\ket{x_2}$}&\qw     &\qw     &\qw     &\qw     &\qw     &\qw     &\ctrl{1}                                                                                        &\targ{} &\ctrl{1}&\qw     &\qw       &\qw     &\qw     &\qw     &\qw     &\qw     &\qw     &\qw    &\qw     &\ctrl{1}&\targ{} &\ctrl{1}&\qw     &\qw     &\qw     &\qw     &\qw     &\qw     &\rstick{$\ket{x_2}$}\qw\\
\lstick{$\ket{z_2}$}&\qw     &\qw     &\qw     &\qw     &\qw     &\qw     &\targ{}                                                                                         &\qw     &\ctrl{1}&\qw     &\qw       &\qw     &\qw     &\qw     &\qw     &\qw     &\qw     &\qw    &\qw     &\ctrl{1}&\qw     &\targ{} &\qw     &\qw     &\qw     &\qw     &\qw     &\qw     &\rstick{$\ket{z_2}$}\qw\\
                    &        &        &        &        &        &        &                                                                                                &        &        &\targ{} &\vqw{1}\qw&        &        &        &        &        &        &\vqw{1}&\targ{} &\qw     &        &        &        &        &        &        &        &        &                       \\[1ex]
                    &        &        &        &        &        &        &                                                                                                &        &        &        &          &\qw     &\ctrl{1}&\ctrl{3}&\qw     &\ctrl{1}&\qw     &\qw    &        &        &        &        &        &        &        &        &        &        &                       \\
\lstick{$\ket{x_3}$}&\qw     &\qw     &\qw     &\qw     &\qw     &\qw     &\qw                                                                                             &\qw     &\qw     &\qw     &\qw       &\ctrl{1}&\targ{} &\qw     &\ctrl{1}&\targ{} &\ctrl{1}&\qw    &\qw     &\qw     &\qw     &\qw     &\qw     &\qw     &\qw     &\qw     &\qw     &\qw     &\rstick{$\ket{x_3}$}\qw\\
\lstick{$\ket{z_3}$}&\qw     &\qw     &\qw     &\qw     &\qw     &\qw     &\qw                                                                                             &\qw     &\qw     &\qw     &\qw       &\targ{} &\qw     &\qw     &\ctrl{1}&\qw     &\targ{} &\qw    &\qw     &\qw     &\qw     &\qw     &\qw     &\qw     &\qw     &\qw     &\qw     &\qw     &\rstick{$\ket{z_3}$}\qw\\
\lstick{$\ket{z<x}$}&\qw     &\qw     &\qw     &\qw     &\qw     &\qw     &\qw                                                                                             &\qw     &\qw     &\qw     &\qw       &\qw     &\qw     &\targ{} &\qw     &        &        &       &        &        &        &        &        &        &        &        &        &        &
\end{quantikz} }
\caption{Uncomputation of comparison result based on a substraction using AND gates.
	The boxed part is repeated for each qubit pair (except first and last).
	This scheme performs the same task as \autoref{fig:add_mod:compair_uncompute}.
}
\end{figure}

Semiclassical additions used in this work are designed in the spirit of the circuit presented in~\cite[Fig.\,7]{Babbush2020CompilationFaultTolerant} (but using slightly less resources or achieving different tasks): the carries are successively computed and each one is stored in a new ancillary qubit.
In the circuits used in this work (Fig.\,\ref{fig:semi_classical_subtract}, \ref{fig:semiclassical_ctrl_add}, \ref{fig:semiclassical_comparison}, \ref{fig:modular_reduce_3}, and \ref{fig:semiclassical-add}), the Toffoli gates resetting a qubit to $\ket{0}$ could eventually be replaced by the measurement-based uncomputing discussed before (\autoref{fig:deand}).
Note that the addition from~\cite{SvoreQIC2017Factoringusing2n2} could be more appropriate to reduce the number of ancillary qubits at the cost of logarithmically longer circuits.

Note also that the additions mentioned before may be improved by using oblivious carry runway~\cite{Gidney2019Approximateencodedpermutations} to split the computation of one addition into several smaller additions that can be carried almost independently.
We did not study how this technique interferes with the circuits around the additions but the necessity to go back to the standard representation of integers might negate the advantage.

Note also that in order to further parallelize, look-ahead adders~\cite{Gidney2020Quantumblocklookahead} could be considered.
On the one hand, they require the simultaneous availability of a lot of magic states.
On the other hand, the relatively small size of Toffoli states factories suggest that look-ahead adders might be suitable in our case.
A thorough analysis including a precise noise model would be needed to conclude about the relevance of look-ahead adders.

\bibliography{article_chat_log,article_chat_logRepNotes.bib}

\begin{thebibliography}{96}
\bibitem [{pyt()}]{python}%
  \BibitemOpen
  \href@noop {} {}\bibinfo {note} {``\texttt{=}'' is the affectation operator,
  ``\texttt{==}'' check the equality, ``\texttt{//}'' is the integer division
  operator (quotient of Euclidean division) and ``\texttt{\%}'' returns the
  rest of the Euclidean division.}\BibitemShut {Stop}%
\end{thebibliography}

\end{document}